# Sustainable Satellite Communications in the 6G Era: A European View for Multi-Layer Systems and Space Safety

**Marko Höyhtyä[1], Senior Member, IEEE, Sandrine Boumard[1], Anastasia Yastrebova[1], Pertti Järvensivu[1], Markku Kiviranta[1], Senior Member, IEEE and Antti Anttonen[1], Senior Member, IEEE**

[1]VTT Technical Research Centre of Finland Ltd, Oulu, 90571 Finland

Corresponding author: Marko Höyhtyä (e-mail: firstname.surname@ vtt.fi).

This work was supported by the CF Space project and in part by Business Finland through the 6G_Sat project, grant 1537/31/2021.

**ABSTRACT** During the New Space era small countries are becoming important players in the space business. While space activities are rapidly increasing globally, it is important to make operations in a sustainable and safe way in order to preserve satellite services for future generations. Unfortunately, the sustainability aspect has been largely overlooked in the existing surveys on space technologies. As a result, in this survey paper, we discuss the multi-layer networking approaches in the 6G era specifically from the sustainability perspective. Moreover, our comprehensive survey includes aspects of some interesting industrial, proprietary, and standardization views. We review the most important regulations and international guidelines and revisit a three-dimensional architecture vision to support the sustainability target for a variety of application areas. We then classify and discuss space safety paradigms that are important sustainability enablers of future satellite communications. These include space traffic management, debris detection, environmental impacts, spectrum sharing, and cyber security aspects. The paper also discusses advances towards a planned European connectivity constellation that could become a third flagship infrastructure along with the Galileo and Copernicus systems. Finally, we define potential research directions into the 2030s.

**INDEX TERMS** Aerospace engineering, Low earth orbit satellites, Radio spectrum management, Spaceborne radar

## I. INTRODUCTION

Technology trends in satellite communications include on-board processing and the use of ever-higher frequencies to enable high-throughput satellites (HTS) to fulfill growing data demands. The two main disruptions currently driving the development and rapid growth of satellite communications (SatCom) are increasing satellite constellation size and the integration of satellite and terrestrial networks. The former aims to provide broadband services to currently underserved areas with improved performance. The latter is related to the evolution of mobile networks where different wireless and wired technologies converge. This creates a vast amount of new opportunities in different application fields such as public safety, digital health, logistics, and Internet services in developing countries. The annual space business related to the 5$^{th}$ generation (5G) and 6$^{th}$ generation (6G) of communication systems is expected to grow to more than €500B during the next two decades [1]–[3]. This is more than the current entire space business including scientific missions, earth observation (EO), and navigations.

At the same time, the whole space sector is in a transformation phase due to the so-called New Space Economy. A significant reduction in launch costs and easy and affordable access to space have attracted new innovative players to the space business [4] [5]. Low Earth Orbit (LEO) systems and small satellites in particular are increasing rapidly. The most typical orbit heights are above 500 km but there are significant efforts to also use very low Earth orbits (vLEO) to provide sensing and communications services. The so called Karman line, defining where the atmosphere ends and space begins, is above 80 km and orbiting objects can survive multiple perigee passages at altitudes around 80–90 km [6]. Small satellites in the range of 80-220 kg can be seen as a sweet spot [5] since they are large enough for payloads to support broadband communications [7]–[9] or synthetic aperture radar (SAR) imaging [10] [11].

### *A. MULTI-LAYER NETWORKS*

6G systems will be used to provide pervasive services worldwide to support both dense and less-dense areas. To achieve this goal, 6G systems will need to integrate terrestrial, airborne (drones, high-altitude platforms (HAPs)), and satellite communications in different orbits [12] [13]. This means that in contrast to traditional research and development (R&D) work, network analysis, planning and optimization will be updated from two dimensions to three dimensions (3D), where also the heights of communications nodes are taken into consideration [12]–[15]. In this way, 6G networks will be able to provide drastically higher performance to support passengers in ships and airplanes.

The initiatives spawned recently range from very high throughput geostationary orbit (GEO) and medium Earth orbit (MEO) systems to unmanned aerial vehicles (UAVs) [16]–[18] and small satellite systems dedicated to machine-to-machine (M2M) and Internet-of-Things (IoT) services [19]–[21]. Especially interesting are mega-constellations consisting of hundreds to thousands of small and medium-size satellites like those proprietary ones envisaged by OneWeb, Starlink, Orbcomm and Telesat to mention but a few. There is also ongoing active work in the 3rd Generation Partnership Project (3GPP) standardization to define non-terrestrial networks (NTN) with interoperable interfaces in order to have truly seamless connectivity in the future, described in detail in Section V.B.

### *B. SPACE SAFETY AND SUSTAINABILITY*

There are not only technical drivers in the development of multi-layer 6G networks. It is essential to develop services and technologies in a sustainable way to ensure high quality services also to coming generations. To mention a few examples: 1) According to International Telecommunication Union (ITU) only half of the world's population has access to broadband services above 256 kbits/s currently [22]. 2) The COVID-19 pandemic has shown that video communications provide means for people and businesses, including medical professionals, and their patients to remain in virtual contact, avoiding the need for travel while remaining socially, professionally, and commercially active [23].

A comprehensive analysis of the linkage between 6G and the United Nations Sustainable Development Goals (UN

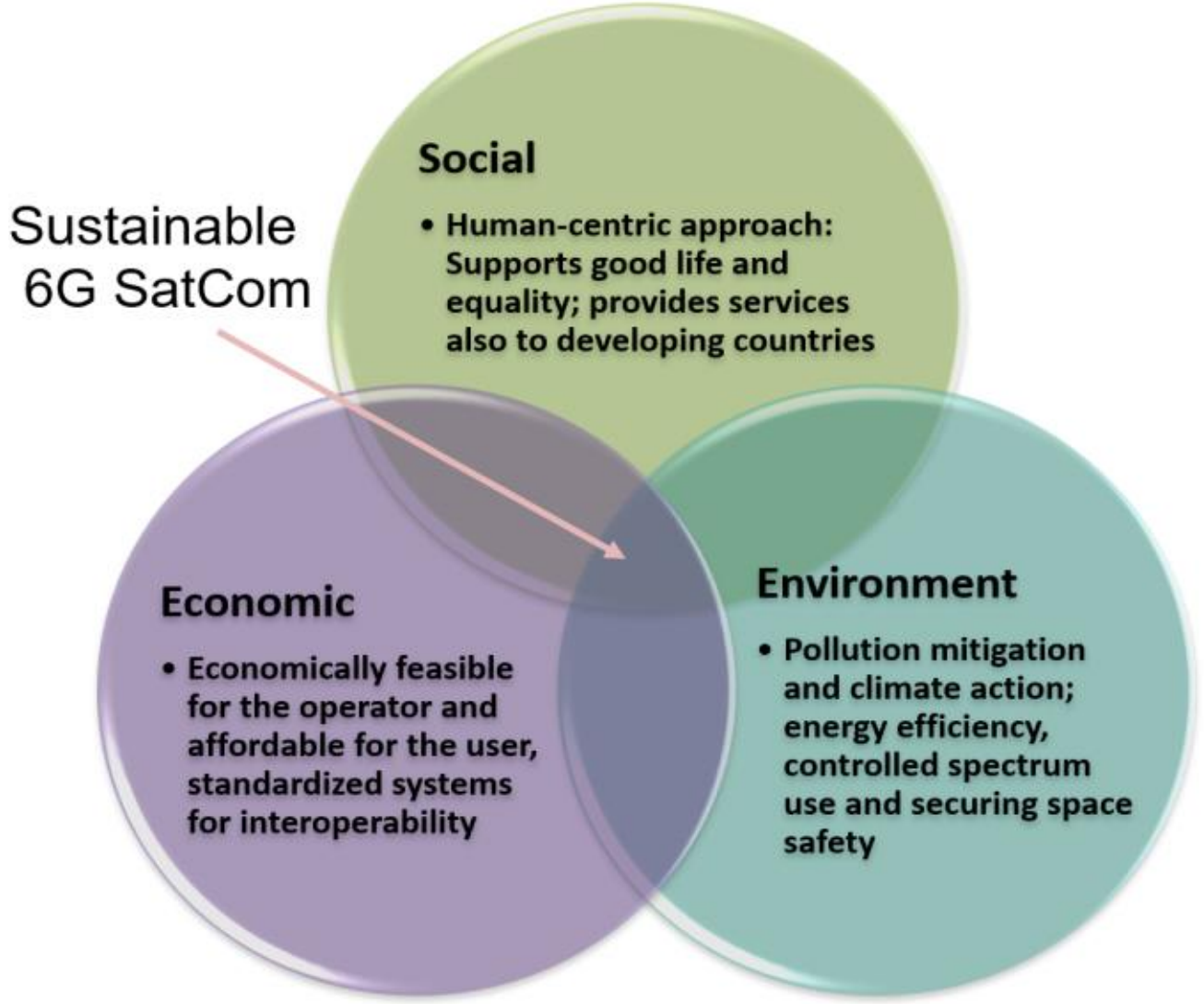

Figure 1. Sustainable 6G SatCom aspects.

SDGs) from technological, business and regulation perspectives has been provided in [24] [25]. A very good overview of how European Space Agency (ESA) programs support SDGs is given in [26]. For instance, satellite communication technologies provide e-learning in the Congo, tools for telemedicine and transmission of key medical data to and from remote locations, and means to gather and share data on arctic sea and climate conditions. Thus, it supports a multitude of SDGs including good health and wellbeing, climate action, quality education, sustainable cities and communities, reduced inequalities, and life on land by helping to protect terrestrial ecosystems.

Therefore, modern communication networks will be purposefully designed to be socially, economically, and environmentally sustainable, and they will provide the means to support equality globally. The main sustainability aspects, visualized in Figure 1, are:

*Social drivers*: The system should support reaching societal SDG goals and do that with a human-centric approach. This means supporting good life, equality of service, and access around different regions worldwide including developing countries in order to tackle the digital divide.

*Economic drivers*: It is not economically feasible to build terrestrial infrastructure to provide connectivity in Arctic areas or many other remote locations. There are market expectations and user needs to be fulfilled. The system to be built has to be economically feasible for the operators and affordable for the users. Satellites provide a cost-efficient platform in order to connect in remote locations. Standardization is also driving the development, making systems interoperable and easier to adopt.

*Environment drivers*: Environmental and ecological factors are driving development of future technological solutions. From the satellite system point of view drivers include avoidance of space debris creation, energy and spectrum efficiency, and pollution mitigation. It is good to note that it is not always ecologically feasible to build terrestrial systems in fragile areas such as the Arctic. The use of satellites to provide wanted services is therefore a good choice.

TABLE I. COMPARISON TO EXISTING REVIEW ARTICLES.

| Topic of the article | Ref. | Contributions given in the article |
|---|---|---|
| Review of small satellites | [7], [8] | Survey on small satellites and their capabilities and related transformation of the space business. |
| Unmanned aerial vehicle based communications | [16] | Comprehensive review of the use of airborne platforms to support wireless services. |
| Satellite communications in the New Space era | [28] | Comprehensive overview covering technical topics and development environments. |
| 5G and 6G satellite communications | [29]-[31], [34], [35] | Reviews describing 5G use cases, technologies, and standardization activities related to non-terrestrial networking. |
| 6G wireless systems visions | [12], [24], [25], [36]-[40] | Survey papers and white papers on 6G connectivity regarding main drivers, sustainability, requirements, technical building blocks and architecture visions. Focus on terrestrial aspects but also remote connectivity. |
| Dynamic spectrum sharing | [32], [33] | How to share spectrum between different networks such as satellite and terrestrial. Focus on database-assisted technology. |
| Sustainability and threats caused by constellations | [27], [41], [43] | Overview of sustainable space operations and emerging threats |
| Sustainable satellite communications in the 6G era | | Novelty: Industry views based on interviews. Sustainability and space safety aspects in multi-layer networks |

In the following, we list a couple of key points from the SatCom point of view.

- The number of satellites is increasing especially in LEO orbits. The sustainable use of space [27] ensures flight safety and mitigation of debris and creates the means to detect the debris to avoid collisions and enable space cleaning.
- The spectrum must be allocated and used in a way that does not endanger existing services. For example, some frequencies are actively used to gather information about the space environment and should not be interfered with communications signals.
- Integrated systems need to be designed from the beginning to be cyber secure in order to ensure reliable services and prevent data and systems to be accessed by unauthorized users.

### C. BACKGROUND AND CONTRIBUTIONS OF THIS PAPER

This paper has been prepared as an outcome of a project that aimed at creating a national roadmap for 6G systems. In the project, we have interviewed more than 20 mostly national organizations, including companies, ministries, and funding entities operating on SatCom aspects. In addition, during the roadmap work, we have been collaborating with ESA. Finland is an example of a relatively small country that has significantly increased space activities during the New Space era and has chosen to focus strongly on sustainability in development. Part of the material presented in this paper originates from the interviewed organizations. Insights and findings can be generalized to other countries as well. We aim to provide a comprehensive view that covers not only a single growing space country but also Europe in general. This provides a complementary view of the vast literature covering developments in North America and Asia.

Many recent survey and vision papers discuss technology developments related to integrated satellite-terrestrial networks [13], [28]–[35], and 6G topics [36]–[40]. In addition, there are papers discussing space sustainability and threats related to emerging constellations [27], [41]–[44]. However, there are gaps in the literature that we aim to fill. First, industrial interviews have not been used much to describe needs and use cases for SatCom. Second, regulations and space safety aspects related to multi-layer networks have not been covered in detail in previous papers. Third, there are no up-to-date surveys of proprietary systems and standardization with linkage to space safety aspects and related research directions. Thus, the novelty of our paper (elaborated in TABLE I) includes the following:

1. We define sustainable 6G SatCom systems and the three main elements related to them, i.e. social, economic and environmental drivers.
2. We revisit a multi-layered architecture from a sustainability perspective and extend our previous work presented in [13] to a survey with up-to-date information from extensive interviews and the latest 6G literature.
3. We discuss national and European-level developments related to a planned flagship connectivity initiative [45] that could complement Copernicus and Galileo programs and support European sovereignty.
4. We examine 6G SatCom systems from the sustainability perspective focusing on space safety aspects. We classify the space safety related development into seven subtopics and discuss related technical and regulatory issues.

The organization of the paper is described in Figure 2. First we will provide a system overview in Section II, defining use cases and the most relevant regulations. Then, we discuss multi-layer architecture and sustainability targets in Section III. We analyze and describe 6G related developments including large constellations in Section IV before taking a deeper look at the recent progress in technology enablers in Section V. Sustainability enablers and space safety aspects are reviewed in Section VI. Finally, research directions toward sustainable 6G SatCom are given in Section VII before concluding the paper.

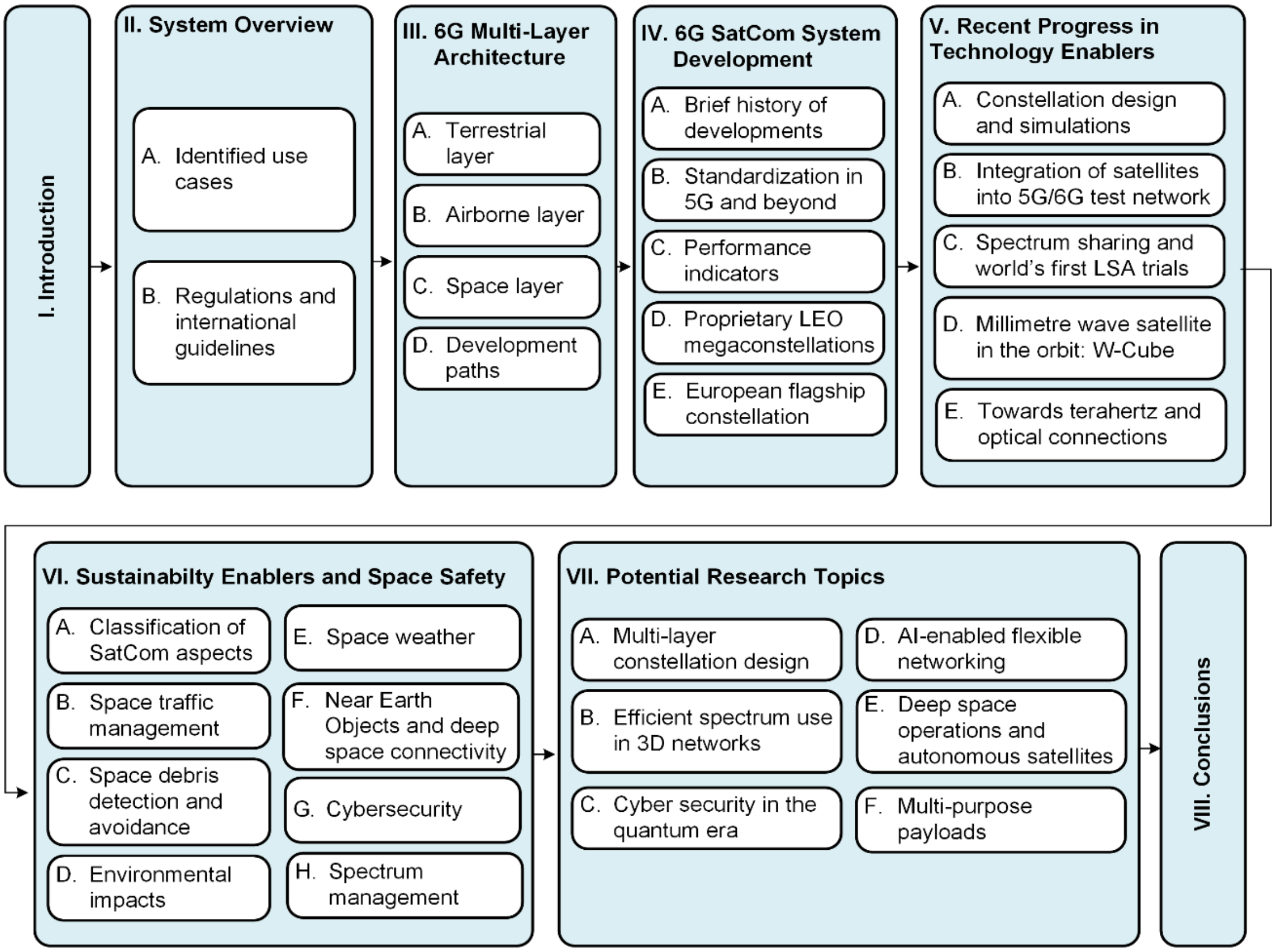

Figure 2. Structure of this paper.

## II. SYSTEM OVERVIEW

A high-level architecture of a future multi-layer system is depicted in Figure 3. It is a 3D network consisting of terrestrial communications, aerial platforms, and satellites at different orbits interconnected via high throughput inter-satellite links (ISLs), which can directly route data packets through space [13] [46]. The architecture has to accommodate the requirements of the targeted vertical industries and utilize the assets and infrastructure owned by multiple stakeholders. The user segment includes user terminals that may be fixed, or mobile ones deployed on platforms such as trains or airplanes i.e., located in multiple layers.

The main motivation for such integrated 3D networks lies in the fact that terrestrial networks will be mainly developed to cover urban areas but their coverage is poor in a harsh, remote environment [35]. Thus, satellites are needed in 6G systems to extend coverage to mountains, oceans and other less-populated regions. There are also significant differences between satellite and terrestrial systems regarding the transmission delay and number of users served. The development and use of integrated systems need to be done in a sustainable way, fulfilling the aspects described in Figure 1.

### A. IDENTIFIED USE CASES

There are numerous papers discussing 6G related use cases, some of which also discuss multi-layer networking. Our objective is to provide a consolidated Nordic view based on interviews with Finnish industry actors and administrations. In addition, we highlight the sustainability aspect throughout the descriptions. Obviously, similar use cases are also of high interest in other regions and we have aimed to create an internationally applicable view. Here we list the findings that are also depicted in Figure 4. The list is not meant to be exhaustive but it indicates potential application areas and shows where needs from the interviewed industrial and governmental organizations are.

The views are also in line with the Finnish recovery and resilience plan, part of Finland's sustainable growth program, where Pillar 2 focuses on digitalization and the data economy [47]. This pillar is further divided into three component areas: digital infrastructure, accelerating the data economy and digitalization, and digital security. The aim of the digital infrastructure is to have a high-speed reliable network and nation-wide coverage. Another target is to promote the digitalization of traffic, which will further support the attainment of transport emission reduction targets. This includes a future railway mobile communication system to enable a better and more punctual provision of trains. We have classified the use cases into five main categories.

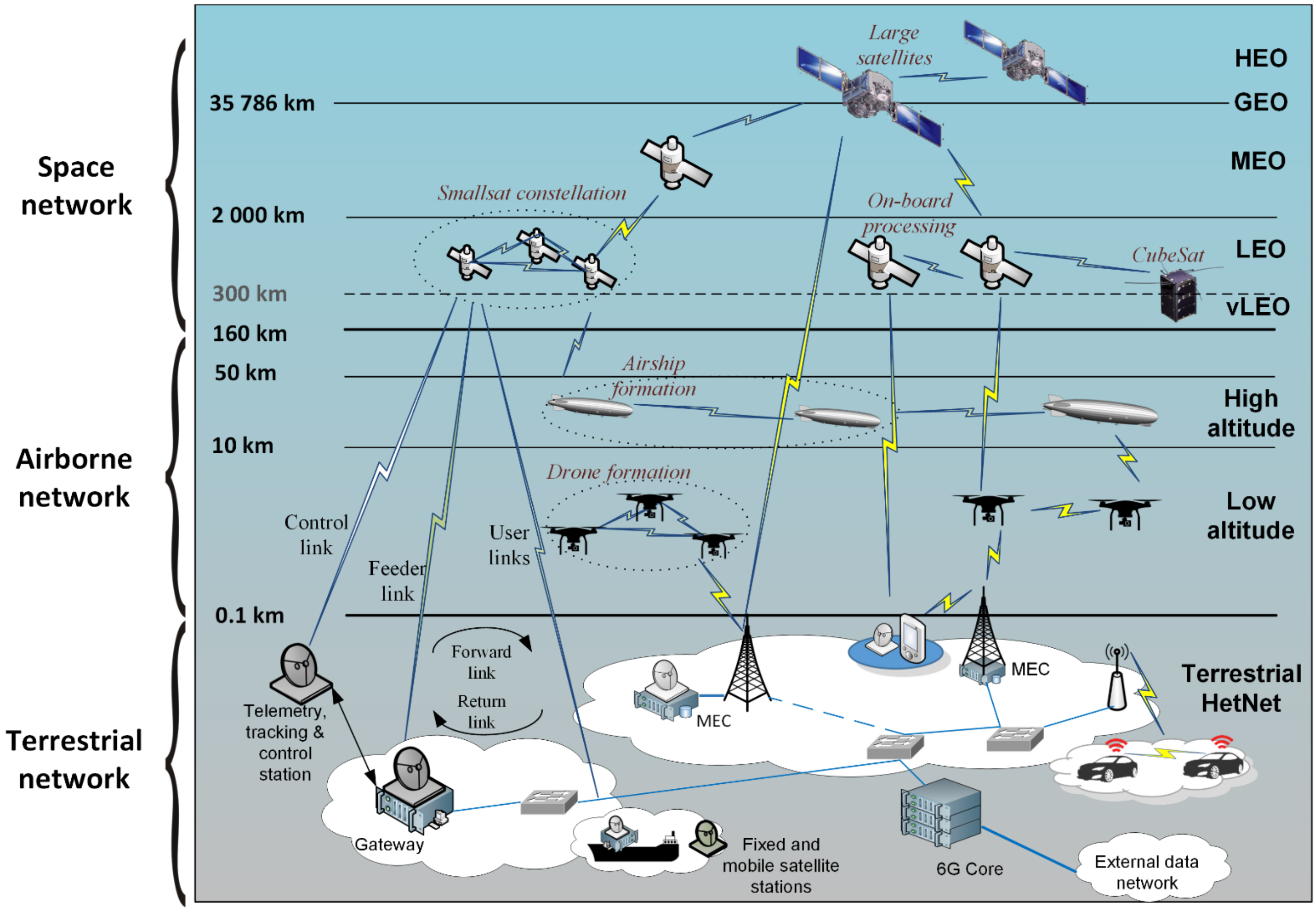

Figure 3. Multi-layer network architecture for 6G.

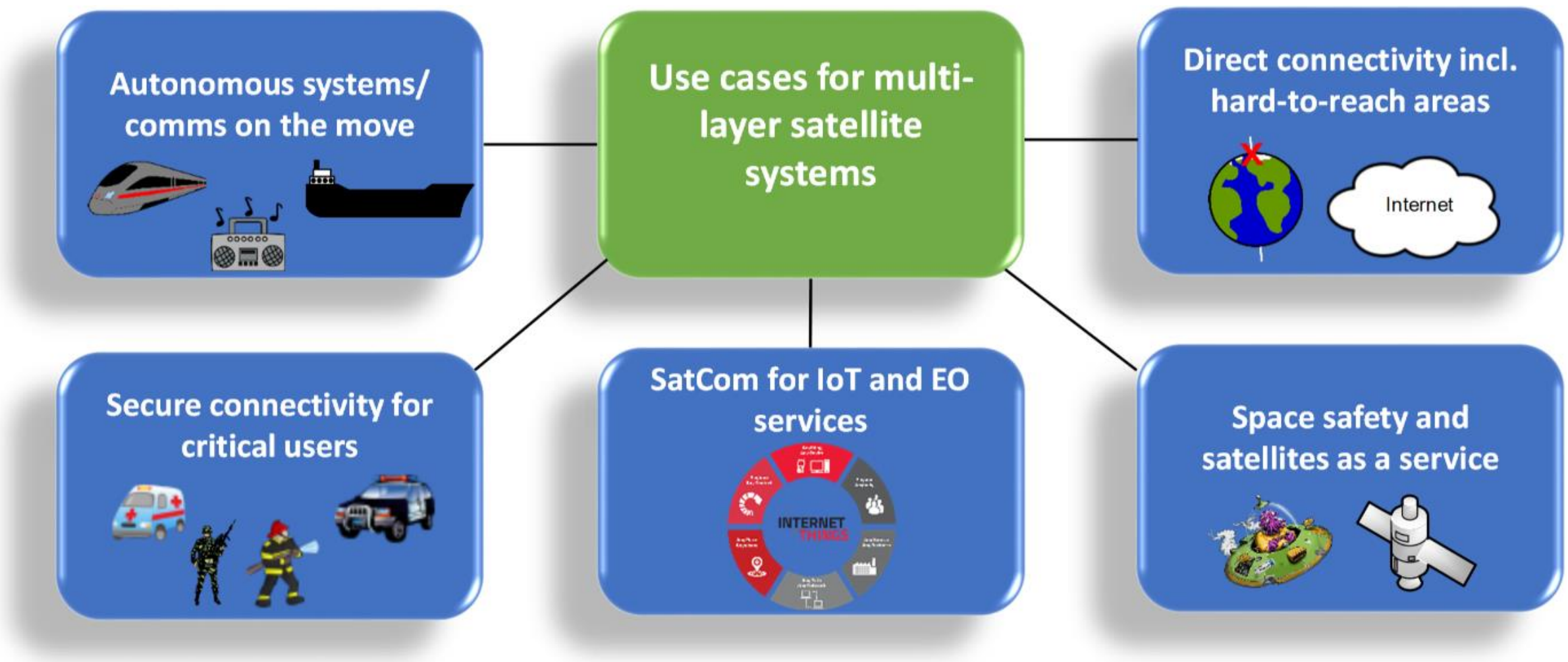

Figure 4. Use cases for future satellite communications.

### 1) AUTONOMOUS AND REMOTE-CONTROLLED SYSTEMS / COMMUNICATIONS ON THE MOVE

Communications on the move is a natural use case for future satellite-terrestrial systems. 6G technologies will need to provide robust high-capacity connections to airplanes, trains, cars, working machines, and maritime users across the globe. Satellites support operations by providing connection and enhancing situational awareness via remote sensing and navigation capabilities. Satellites can provide complementary redundant connection in a sustainable way, without the need to build extensive terrestrial infrastructure. Latency is still seen as a challenge for remote-controlled and autonomous systems such as cars, but upcoming LEO solutions are expected to provide significant improvement for latency reduction. Broadcasting to a large number of receivers will be provided by satellite and terrestrial infrastructure in the future. The broadcasting capability of satellites will also be used in sending software updates to devices and moving platforms over a large area.

The satellite connections and integrated systems are inherently needed in the maritime domain. Digitalization of maritime requires development, which is described in [48], [49]. The 3GPP standardization forum identifies the following main services for maritime communications (MarCom), covering the needs of humans as well as maritime systems: 1) Mobile broadband services for users at sea; 2) Machine type communication services inside a vessel, between vessels and between UEs at sea; 3) Maritime communication services between authorities and users at sea; 4) Interworking and harmonization (with very high frequency (VHF) and satellite systems). Additionally, other services such as collection and dissemination of sensor data (e.g., meteorological and hydrographic data) are possible. Based on the conducted interviews of MarCom end users it is essential to have integrated systems that can provide services globally also in areas where mobile cellular networks do not provide coverage. There is quite a good connection using the Finnish public safety network (VIRVE) towards Tallinn but commercial networks are not so good far from the shore.

### 2) SECURE CONNECTIVITY FOR CRITICAL USERS

A secure and resilient connectivity solution with sovereignty –also during crisis times– is an essential part of future development in Europe. There are plans to develop governmental SatCom (GovSatCom) solutions that can support broadband needs and provide services to governmental users in any location over the continent. GovSatCom is defined in [50] as a new service class between fully commercial and military applications, providing highly available secure connections. The air interface of the system could be based on 5G and later 6G in order to create a standardized system on top of which security functions are built. From the cybersecurity viewpoint, quantum solutions are foreseen.

Public safety authorities are increasingly using broadband services, new multimedia applications, and smart devices. Many new critical user sectors are emerging, including live broadcasting, critical business, security, remote control, and value transportation. The resulting connectivity needs cannot be supported by traditional narrowband systems. Instead, broadband mobile networks and satellite connections are needed as enabling technologies. For example, border guards and military personnel need rapidly deployable networks, (or so called "tactical bubbles" [51]) in remote locations. Coverage and capacity for integrated networks and easy-to-use solutions are mentioned in interviews as goals for adopting satellite systems for operational use. The use of NTN for public safety applications in Finland was studied in [52]. It was concluded that "Satellites could be used as backup connections especially in locations where terrestrial network is not available. This will create resiliency to the operations."

Usually, the military has the satellite system under their control. Also in Finland, ground stations, terminals, and personnel are internally managed; only satellites are not controlled by them. In addition to the mentioned tactical bubbles, military users need satellites as redundant connections that can provide services anywhere. Satellites can, e.g., connect hierarchically different parts during operations. For example, the facility area that handles the main resources can be connected to the battle zone where satellites and drone swarms provide redundant performance to the troops.

### 3) SATCOM FOR IOT AND EARTH OBSERVATION SERVICES

Internet of Things (IoT) and machine-type communication services include cheap, low-complexity sensors, and actuators that are able to generate and exchange data. Due to the large number of devices, traffic generated by them will have a significant impact on the network load. Satellites can help to offload the terrestrial IoT network traffic through backhauling or provide service continuity in cases where a terrestrial network cannot be reached. Several interviewed companies thought that this is an important future use case since IoT does not require so many satellites in a constellation in order to have reasonable business cases available. However, there are many old and new players and the IoT area is competitive. NTN-IoT interconnecting every point on Earth to service centers is interesting for many industries. There is ongoing development for narrowband IoT (NB-IoT) services over satellites in 3GPP standardization.

Although EO can be considered a separate business area from communications, the amount of EO data transmitted to dedicated ground stations is significant. Collected sensor data at the satellites may be so large, that some pre-processing is required before downlink transmission. In such a case, the data links form the limiting factor of the system.

The situation arises especially with LEO satellites, which are visible to ground stations for only a short period of time during which the download of data must be performed. Inter-satellite links (ISLs) to GEO, highly elliptical orbit (HEO) or other LEO satellites solve the problem, but then the EO satellites become a part of a satellite communication networks, justifying the inclusion in the communications business area. The number of EO satellites is constantly increasing due to an increased need to monitor the changing environment and resources of Earth.

#### *4) DIRECT CONNECTIVITY INCLUDING HARD-TO-REACH AREAS*

Direct 5G and 6G handheld connections could make emerging satellite services widely used by consumers and enable seamless integration with mobile cellular networks. The ability to connect without separate satellite equipment is definitely of high interest both to entertainment and authoritative applications. 3GPP is working on a global new radio (NR)-based solution enabling hand-held devices to be used both in terrestrial and non-terrestrial networks as well as mobility in between. This is a promising business opportunity for mobile phone manufacturers.

Roughly half of the world's population cannot access the Internet via broadband connections [53]. In addition, Internet services are very limited e.g., in Arctic areas. Building terrestrial infrastructure in remote locations and developing countries is not economically feasible. Thus, this is a good opportunity and driver for the development of satellite services since they can provide truly global connectivity. In more populated places satellites will most probably only complement terrestrial systems, not replace them. Satellites can still provide backhaul services to mobile networks, playing an important role in their development.

The Arctic area is very relevant for many European stakeholders. From the Finnish viewpoint the solutions that generally work in the Arctic also work in Finland. One of the main SDGs is to provide coverage to poorly connected areas. The developed solution should not only be technically capable, but it should also be easy to use and adaptable to different cultures. One company pointed out that in general they are looking for use cases that promise sustainable development from the climate point of view. It is vital to develop situational awareness solutions that can be used to detect environmental impacts.

#### *5) SPACE SAFETY AND SATELLITES AS A SERVICE*

One part of sustainable development is to take care of space safety related aspects. Space weather services provide information to the space infrastructure and helps in defining required protection for the coming satellites and their electronics. The number of satellites is increasing rapidly in the New Space era and both avoidance and detection of space debris are very important. Both ground-borne radars and space-borne radars can help in space debris management. It was found in a recent paper [54] that while high-density constellations intrinsically increase the risk of satellite collisions, they can also be used to mitigate the debris problem. Tight integration between space-based and ground-based radars might provide the best solution for real-time debris detection.

As future needs and technology developments cannot be predicted accurately there is a need to develop programmable satellites [55] and concepts such as satellite-as-a-service or payload-as-a-service. The former concept refers to providing the satellite capability to a customer whereas in the latter concept, a single satellite may carry several payloads for different customers. Still the development of the platform and operation of the satellite is done by the satellite operator, easing customer access to the satellite data.

The envisioned development requires the use of software-defined radio technology and software-defined satellite platforms. Then, the satellite or its payload could be reconfigured over the air to needs that are not even known yet to improve its longevity and sustainability. In addition, this could also make updates of technology releases possible so that regular updates of 3GPP releases would not mean launching new satellites into orbit. Satellite-as-a-service and payload-as-a-service reduce the need for large technical teams in different organizations, they can pay for the service only when needed and the satellites can be used by multiple customers for multiple missions. A recent example is Lockheed Martin's SmartSat concept, which allows satellite operators to quickly change missions while in orbit with the simplicity of starting, stopping, or uploading new applications [56].

**Lessons learned**: While researchers and research organizations typically provide futuristic visions about use cases that might require a totally new kind of performance from 6G networks, there are still many industrial players that seek mostly moderate improvements in reliability, scalability, and cost. Interviews showed that many organizations would be very willing to increase their use of satellite services and predicted that developments in performance, cost, and availability will spark innovations in many industrial fields. Concepts such as satellite-as-a-service may provide useful tools that will limit the exponential growth in the number of satellites.

### *B. REGULATIONS AND INTERNATIONAL GUIDELINES*

Regulation is one of the most important, or perhaps the most important aspect for ensuring sustainable operations. They provide an international framework to be followed by space nations and ways to control and monitor the actual operations. The following sections cover the identified key topics related to multi-layer networks. The relation of identified aspects to the three main sustainability pillars is presented in TABLE II.

TABLE II. REGULATIONS TO ENABLE SUSTAINABILITY

| | Relevant regulations and guidelines | Comments |
|---|---|---|
| **Social** | Cybersecurity<br>UAV regulations<br>Territorial aspects<br>Ground stations and data | People need to trust that new technology is safe and reliable. |
| **Economic** | Cybersecurity<br>Frequency management<br>Territorial aspects | Enables industry to operate internationally. |
| **Environment** | Space debris<br>UAV regulations<br>Frequency management | Keeping the environment safe and other services protected. |

*1) FREQUENCY MANAGEMENT AND LICENSES TO OPERATE*

Currently, different frequency bands are assigned to different users and service providers, and licenses are required to operate within those bands. Cellular communication systems, broadcasting services, satellites, etc., all have dedicated bands on which they may operate. These licensed systems possess characteristics that are distinctively different from each other and thus require dedicated bands for interference-free operation. In addition, there are unlicensed bands where several systems such as Wi-Fi, Bluetooth, and other short-range communication systems may operate according to given rules without any regulatory protection against interference.

Operators, both in terrestrial and satellite networks, need a license from a regulator to be able to provide wireless services. Regarding the space segment, the license specifies the allowed orbital positions in order to avoid physical collisions or interference with other satellites in the same frequency band. There are two key sustainability aspects in frequency management: (1) To ensure that coming services do not endanger existing services and (2) A frequency license enables good business and a basis for infrastructure investment. The interested reader may look at [32] for more details about the regulation process.

*2) SPACE DEBRIS RELATED ASPECTS*

There is an increasing number of satellites in orbit that are increasing the risk for collisions and the generation of debris that could put satellite services in danger. Space debris, also called orbital debris, are non-functional man-made objects orbiting the Earth or re-entering the atmosphere [57]. Space debris is an important aspect of space sustainability [27]. Most international guidelines are not legally binding. Such guidelines include the Inter-Agency Space Debris Coordination Committee (IADC) space debris mitigation guidelines, the UN Committee on the Peaceful Uses of Outer Space (COPUOS) space debris mitigation guidelines, and the UN COPUOS Guidelines for the Long-term Sustainability of Outer Space Activities (LTS Guidelines). The EU has also included space situational awareness (SSA) in its space program for 2021–2027. SSA includes space surveillance and tracking (SST), which tracks resident space objects (RSOs), including both active and inactive satellites and space debris. In addition to SSA, a more comprehensive target is prompt space traffic management capability similar to air traffic management. However, as things are more complicated in space, we are still quite far from an effective space traffic management system [58].

Companies have also realized the important economic aspects of space sustainability and debris mitigation. The space safety coalition is an ad-hoc coalition of companies, organizations, and other government and industry stakeholders that have endorsed a set of best practices for the sustainability of space operations to address gaps in current space legislations [59]. They promote the exchange of information between all actors, the careful selection of launch providers, the prioritization of space safety in the design and operation, best practices in design, and sustainability enhancing operations. Additionally, the World Economic Forum's Global Future Council on Space Technologies has developed, together with stakeholders, the concept of Space Sustainability Rating (SSR) [60]. The SSR will score space missions based on markers such as evidenced-based debris mitigation and alignment with international guidelines. The sustainability certifications for mission operators will start in 2022.

The importance of space debris mitigation has led many countries to incorporate some of the guidelines into legally binding national instruments [27]. In Finland, the Ministry of Economic Affairs and Employment takes care of the national space legislation and authorizes and registers any space activities [61]. The Finnish act on space activities of 2018 includes a section on environmental protection and space debris. The operator of the space activity "shall assess the environmental impacts of the activities on the earth, in the atmosphere and in outer space, and present a plan for measures to counter and reduce adverse environmental impacts." Additionally, "the operator shall seek to ensure that the space activities do not generate space debris. In particular, the operator shall restrict the generation of space debris during the normal operations of the space object, reduce the risks of in-orbit break-ups and in-orbit collisions and, after the space object has completed its mission, seek to move it into a less used orbit or into the atmosphere."

*3) TERRITORIAL ASPECTS RELATED TO 5G AND 6G NETWORKS*

A SatCom operation and communication service from high altitude platforms (HAPs) can easily cross borders i.e., cover areas in more than one country. In addition, there are also international water and land areas without territorial claims such as the Antarctic where service from satellites is provided. Regulations allowing fluent operation across borders are good from a social point of view, allowing end users to access services easily while traveling. In addition, it supports international business and makes the latest technology widely available.

An example of extraterritorial access is where a mobile network identifier (ID) is used for a local network on a ship or plane traveling in or over international waters and then satellites are used to connect to the outside world [62]. IDs are authorized by one administration in one country, but they could also be transmitted by a radio access network (RAN) in another country. 3GPP is currently developing guidelines for regulatory aspects related to regional operation including how to enable routing to a core network in a specific country. Regulations may require that the satellite ground station and/or base station and the core network all have to be in the same country as the user equipment (UE), unless countries have made specific agreements [62]. Finally, international regulations ensure safe and efficient maritime traffic management and air traffic management and other safety-critical applications. On the other hand, passenger communications must comply with regulations of the territory with sovereignty over the location they are in.

#### *4) UAV REGULATIONS*

Airborne platforms differ from satellites in the sense that they operate in the air space of different countries whereas space is international. UAV regulations include not only connectivity but also operational limitations on different locations and ethical constraints related to privacy protection [63]. When the UAV is small enough, its use is regulated by national aviation authorities but when they are larger than 150 kg, they are usually regulated similarly to manned aircraft. Typically, there are also no-fly zones defined to minimize risks to manned air flights, and UAVs can only be operated in line-of-sight conditions. Regulations make the UAV more acceptable and safe for people as well as help to protect critical services in the environment.

#### *5) GROUND STATIONS AND DATA*

An important item in the regulatory domain is work related to ground stations and data aspects. This work is actively ongoing in Finland to support practical ground station operators in their work. There are new operators willing to start operations to support both Earth observation and communications systems. Nordic locations are more favorable than southern Europe locations due to longer visibility times for satellites using polar orbits and consequently better data downlinks. However, there are regulatory aspects that need to be clear for commercial operators willing to support a number of operators from different countries. For example, the data that goes through the ground station is not necessarily visible at all to the operator. With whom can the operator collaborate? Regulations allow for efficient collaboration, transparency and consequently trust for people and businesses.

#### *6) CYBER SECURITY REQUIREMENTS AND GUIDELINES*

Cyber security is not meant to be a legal concept. However, regulators, such as Traficom's National Cyber Security Centre Finland, develop and monitor networks and services and also provide guidelines nationally. It is well stated in [64] as "Ensuring the security of society is a key task of the government authorities and the vital functions of our society must be secured in all situations." The main cyber aspects for multi-layer 6G networks include three attributes: (a) Availability i.e., communication services must be available for legitimate users to access at any time. Radio frequency interference can be seen as a denial-of-service attack when it is intentional; (b) Confidentiality, maintaining the secrecy of information by preventing access to systems and data from unauthorized users. The platforms, interfaces, and end-to-end connections should be well protected so that "hostile quarters cannot obtain the control of constellations" as one interviewee put it. (c) Integrity, i.e., protecting data from unauthorized alteration such as modification, deletion, or injection of wrong data. It is also good to protect the system against unintentional alteration such as data loss caused by a system malfunction.

TABLE III. SUSTAINABILITY IN MULTI-LAYER NETWORKS

| | **Sustainability targets for design** |
|---|---|
| **Terrestrial layer** | Multi-tenancy support and sharing of infrastructure, resource efficiency, ability to tailor the network for different needs. |
| **Airborne layer** | Selecting the right platform for temporal use, electricity, support for critical infrastructures, environmental monitoring |
| **Space layer** | Preventing space debris, coverage everywhere, efficient resource use, ability to update over the air, low latency |

**Lessons learned**: To develop a sustainable multi-layer space-air-ground system one needs to take into account multiple regulations and guidelines, which also partially limits technologies that can be applied in practice. When proposing new technologies and frequency bands to use, one also needs to conduct regulatory studies and discuss with administrations before those can be used in operational systems. Good regulations can be seen as a driver and enabler for sustainable operations.

## III. 6G MULTI-LAYER ARCHITECTURE

Considering now the multi-layer architecture given in Figure 3 we will describe each layer of a sustainable network. The main design targets are summarized in TABLE III.

### *A. TERRESTRIAL NETWORK LAYER: GROUND SEGMENT*

The terrestrial layer includes several radio access technologies such as cellular radios, WiFi, and IoT solutions to support fixed and mobile users. Car-to-car and ship-to-ship communications are also enabled by radios specifically developed for those purposes. The ground segment consists of a gateway and a core network operated by the network operator. System control, network access, and backhauling are done at the ground segment. A satellite operator uses telemetry, tracking & control (TT&C) stations to monitor the status of satellites and their subsystems, run updates, and update the configurations. These can be used to keep satellites at the desired orbits, update camera parameters, etc.

Small user terminals are handheld devices with small antennas and very small aperture terminals (VSATs) with dish or flat antennas installed e.g., on a ship deck. The UE of a future system will be a multi-radio terminal (any type of integrated communication device) including satellite access. The RAN is assumed to be software-defined networking (SDN) capable of multi-tenancy support. That is essential because, in practice, different RANs and transport networks are often managed by separate network operators. Network virtualization and slicing techniques enable different operators to share network resources with other (virtual) operators and provide end-to-end connectivity across operator boundaries. Efficiency in spectrum and energy use is targeted as well. The whole 3D network could be controlled by a centralized entity, an SDN controller, which has control over the network devices and global knowledge about the network state within an administrative region. The core network in the 6G architecture supports seamless cooperation between the terrestrial and non-terrestrial segments and enables QoS management of data transmission e.g., by dedicating part of the resources to higher-priority applications.

Furthermore, multi-access edge computing (MEC) provides localized computing and storage resources for applications as well as real-time information on local network conditions. The satellite can provide a reliable backhaul link for edge computing. Together, software networks permitting flexible control of network traffic with fine granularity and MEC enabling the provision of scalable distributed services and network functions create a highly elastic integrated satellite-airborne-terrestrial system. This means that the whole system can be tailored during the operation of the network to support different kind of services.

### B. AIRBORNE NETWORK LAYER

The airborne layer comprises UAVs that can be classified according to their altitude. Low altitude platforms (LAPs) are characterized by an altitude lying within the troposphere [65] and HAPs are between 10 km and 50 km (mostly concentrated around 20 km) in altitude [18]. UAV types include (1) balloons, (2) fixed-wing aircraft, and (3) rotary-wing aircraft. There can be WiFi, 5G, and 6G types of payloads providing connectivity to terrestrial users, and Earth observation sensors for remote-sensing purposes. Due to the short distance, there is no need to use different radio equipment, and standardized cellular equipment can be used to provide services from HAPs to cellular users. HAPs are used for various use cases and their implementation scenarios include dedicated, shared, and neutral hosts. It is important to select a suitable platform for operations to save cost and resources.

A major challenge for many operations is the ability of a HAP to maintain a stationary position due to windy conditions at high altitudes. An operating altitude between 17 and 22 km is often chosen for platforms because in most regions of the world, this represents a layer of relatively mild wind and turbulence above the jet stream. This altitude (> 17 km) is also above commercial air-traffic heights, which would otherwise prove to be a potentially prohibitive constraint.

Tethered aerostations, drones, and unmanned balloons are density neutral, floating at the desired altitude [66]. Propulsion is only used to maintain their position. Tethered balloons are generally LAPs, operated within a few hundred meters in altitude. Although tether limits the achievable height of the aerial systems, it also offers the means to feed electric power and communications cable to the platform. Thus, they can be used for long-duration missions.

The majority of drones or UAVs operate at a low altitude. They are versatile and easily deployable aerial platforms, which are increasingly used for different applications and purposes. According to [67] the following attributes make drones a desirable candidate to substitute or complement terrestrial networks: (1) a higher probability for line-of-sight (LoS) links to connect users on the ground and the ability to adjust locations to maintain high-quality links; (2) dynamic deployment capability according to needs. No need for site rental costs; (3) UAV-based swarm networks for ubiquitous connectivity to recover and expand communications fast and effectively. Sustainable UAVs aim to use electricity in the form of rechargeable batteries instead of fossil fuels. In addition, they can play a crucial role in monitoring the environment and critical infrastructure.

### C. SPACE NETWORK LAYER

The space layer comprises satellite constellations and large satellites operating in different orbits – and links between them. The satellite payloads can be transparent or regenerative. In the latter case, part of the base station functionalities are performed by the satellite (e.g., demodulating and re-modulating the signal) whereas in the traditional transparent case, the satellite acts as a simple repeater that amplifies the signal and makes a frequency conversion between the uplink (UL) and downlink (DL) frequencies.

There are satellites in LEO, vLEO, MEO, GEO and even HEO. Large satellites in high orbits are mostly built, launched, and operated by established companies within the space industry. GEO satellites are used for providing broadband access especially to remote areas, which otherwise cannot be served – either for technical or economic reasons. Modern high-throughput satellites (HTS) are able to provide sufficient data rates to most consumers. However, the availability of GEO broadband services becomes limited in polar areas above 60° latitude. Thus, satellite constellations at lower orbits are developed to also support

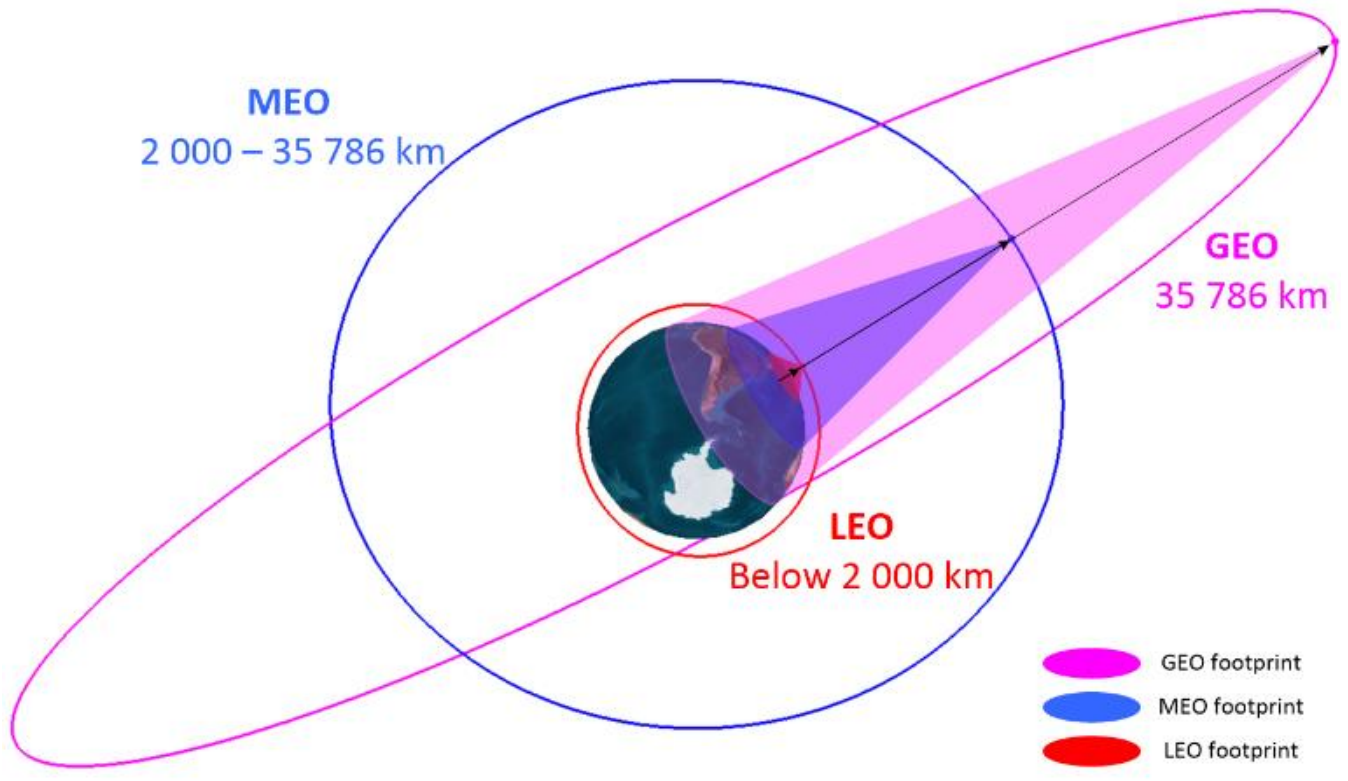

Figure 5. GEO, MEO and LEO orbits.

TABLE IV. COMPARISON OF SATELLITE ORBITS.

| Orbit | (v)LEO | MEO | GEO |
|---|---|---|---|
| **Typical orbit height (km)** | 160 – 1400 | 10000 – 20000 | 35786 |
| **Path loss at 17.7 GHz (dB)** | 161 – 180 | 197 – 203 | 208 |
| **Number of satellites for global coverage** | 40 – 200 | 10 – 30 | 3 |
| **Orbital period (h)** | 1.5 – 2 | 6 – 12 | 24 |
| **Pass time (min)** | 6 – 22 | 130 – 300 | - |
| **1-way latency, Zenith (ms)** | 0.5 – 5 | 33 – 67 | 119 |

Arctic environments.

Small satellite R&D efforts are growing rapidly with new players that focus on non-geostationary (NGSO) orbits and integrate products and services of different technology providers. The majority of planned small satellite missions around the world consider communications and the use of very small platforms such as CubeSats. In addition to providing coverage for remote areas, the space segment is required to be able to support low-latency services, and that is only possible with lower orbits. Satellites in a future system can be tailored on-the-fly with on-board processing capabilities. For example, space-hardened software-defined radios can enable on-board waveform-specific processing that can be upgraded during a satellite's lifetime [28]. While improving technical capabilities, it is also essential to be able to use radio resources very efficiently, which may mean narrow antenna beams and dynamic spectrum sharing.

The main orbits are depicted in Figure 5, showing footprints and the distance from the orbit to Earth. A comparison summary is given in TABLE IV. The number of satellites is related to coverage of the entire globe. However, more satellites might be needed to fulfill capacity requirements. The orbital period can be calculated with Kepler's third law in seconds as [9]

$$T = 2\pi\sqrt{(R_\mathrm{e}+h)^3/\mu}, \qquad (1)$$

where $\mu = 398600.5$ km$^3$/s$^2$ is the Earth's geocentric gravitational constant, $R_\mathrm{e} \approx 6378$ km is the Earth's radius, and $h$ is the orbit height defining the distance between the ground station and the satellite. Pass time or possible connection time from a specific location on the ground to a passing satellite from horizon to horizon is then

$$T_\mathrm{p} = \frac{T}{\pi}\arccos\left(\frac{R_\mathrm{e}}{R_\mathrm{e}+h}\right). \qquad (2)$$

The pass time also defines the maximum handover time from one satellite to another. Usually, the time is somewhat less than that since a safety margin is needed to guarantee connectivity.

### D. DEVELOPMENT PATHS

When we look specifically from the SatCom point of view, the following development paths can be seen when we move towards 6G [1], [13], [34], [68].

1) The networks will become multi-layered. The role of small satellites in LEO orbits is essential.
2) The on-board computer (OBC), the brain of the satellite, is evolving and its processing power increasing. This allows for the softwarization of the payload, which brings flexibility to the system and renders possible the dynamic adaptation of beams, power and frequency allocations, as well as reconfigurability of the payload itself. Reconfigurability improves the longevity and sustainability of the satellites. On the software side, isolation using partitioning, virtualization, or containers can prevent the on-board data processing from interfering with the basic operations of the satellite.
3) From the spectrum point of view, millimeter wave and terahertz technologies and optical links allow very high-rate data links, and spectrum sharing techniques will be used to reduce interference in the future.
4) Reconfigurable phased array antennas and multi-beam architectures are used to reduce power consumption and improve spectrum efficiency.

End-to-end cybersecurity is to be taken into account early in the design phase (security-by-design) to cover all interfaces, handovers, and the whole platform. Quantum technologies including post-quantum cryptography will be used for secure connectivity.

**Lessons learned**: Network planning needs to be updated to three dimensions. Consequently efficient resource use paradigms will have to cover vertical dimensions. The integrated architecture will be dynamic with multiple moving and potentially temporal components in it. Satellite-terrestrial infrastructure will form the backbone. Airborne platforms are mainly temporal additions for specific needs.

## IV. 6G SATCOM SYSTEMS DEVELOPMENT

### A. BRIEF HISTORY OF DEVELOPMENTS

Satellites have been studied and developed in parallel to mobile cellular networks during all generations. During 1G and 2G satellite networks were separate, proprietary systems providing services e.g., to remote areas [30]. During 3G the first step towards the convergence of satellite and terrestrial systems was made and the satellite air interface was made compatible with the terrestrial universal mobile telecommunication system (UMTS) infrastructure [69]. In addition to satellites, HAPs were also considered [70]. Satellites became more important in 4G and satellite systems were considered an essential part of achieving global roaming in places where a terrestrial network is impossible to be installed or too expensive [71]–[73]. 5G is making further progress and there are real promises of having wide-scale use of integrated systems in the future. In 5G, service continuity ensures a smooth handover, e.g., from terrestrial to NTN interface.

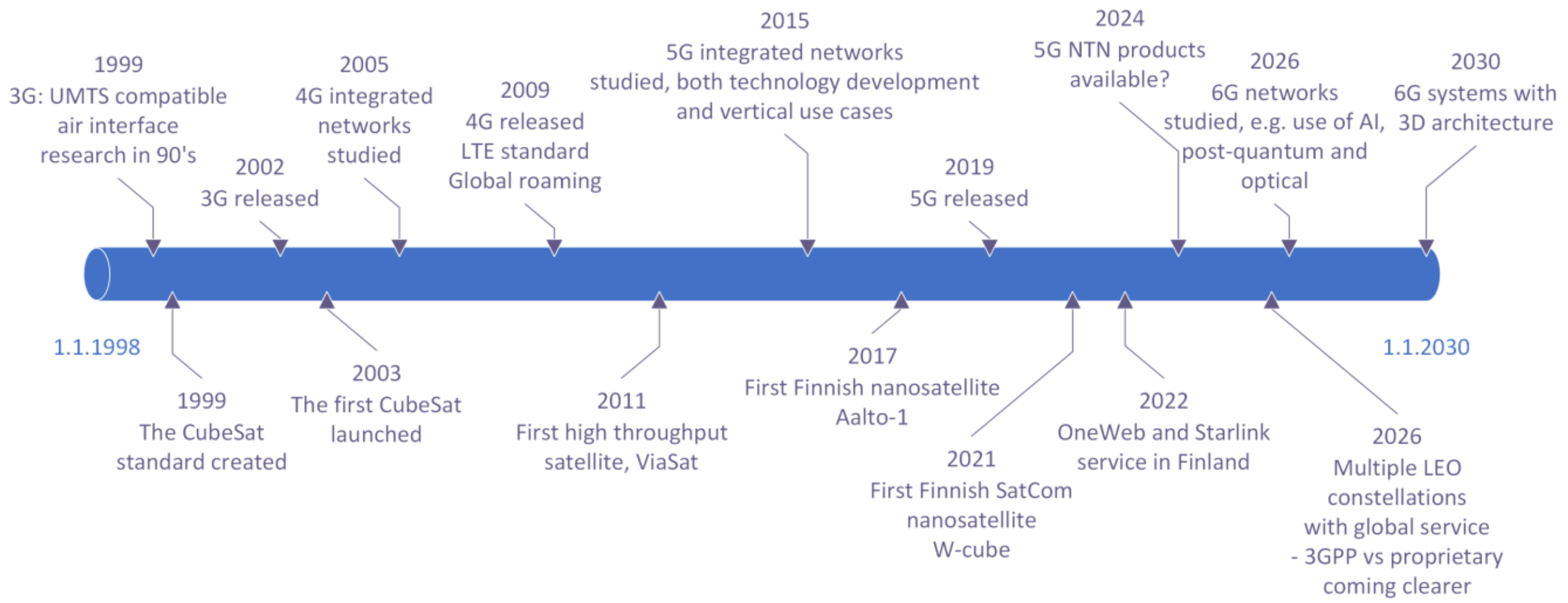

Figure 6. Timeline for 3GPP development and small satellites by 2030.

The timeline for 3GPP development since 3G and some milestones related to LEO constellations are shown in Figure 6. CubeSats have been in orbit for almost two decades. The first Finnish satellite was launched in 2017 [74] and the first telecommunications satellite was launched in the summer of 2021. The W-cube mission aims at modeling 75 GHz channels and their suitability for future satellite systems [75]. 5G NTN service will be available in a few years, and multiple LEO constellations with global service will exist before 2030 when 6G technology will become available.

In addition to 3GPP, other standards – such as the enhanced version of the second generation of Digital Video Broadcasting standard for satellites (DVB-S2X) — are still very relevant for HTS satellites [76] [77]. They form the basis for digital satellite transmission around the world.

### *B. STANDARDIZATION IN 5G AND BEYOND*

3GPP is the main standardization body for mobile networks. During the 5G standardization, an important action item has been to include non-terrestrial networks to support 5G use cases such as public safety and mobile autonomous systems. The consensus and general agreement on what satellite brings to achieving 5G requirements are [13], [78]:

- **Ubiquity**: Satellites provide high-speed capacity across the globe using the following enablers: capacity in-fill inside geographic gaps; overspill to satellite when terrestrial links are over capacity; global coverage; and backup for network fallback.
- **Mobility**: Satellite is the only technology capable of providing connectivity anywhere at sea, on land or in the air for moving platforms, aircraft, ships and trains, while requiring minimal terrestrial infrastructure for support.
- **Broadcast (Simultaneity):** A satellite can efficiently deliver rich multimedia and other content across multiple sites simultaneously, using broadcast/multicast streams with an information centric network and content-caching for local distribution.
- **Security/resilience**: Satellite networks can provide secure, highly reliable, rapid, and resilient deployment in challenging communication scenarios, such as in emergency responses. Rapid deployment essentially means that you can rapidly have a connection to the outside world even in cases where the terrestrial infrastructure is destroyed. Emergency personnel may bring with them a base station for local connections – and a satellite terminal to connect to the outside world to provide up-to-date situational awareness.

The same topics are still very relevant when going towards 6G. Excellent research covering NTN-related activities papers have recently been published. 5G use cases and scenarios defined in the Sat5G project are described in [29]. Non-terrestrial networking and standardization aspects are discussed in [30]. In addition, the importance and role of LEO satellite systems in the 6G era are studied in [31]. An up-to-date standardization status is summarized in TABLE V and documents therein including [62], [79]–[85]. We can analyze and see the following information from the table.

TABLE V. 3GPP NTN STANDARDIZATION AS OF AUG. 2022

| Technical spec. group | Release | Feature and study item | Objectives | Technical report / year |
|---|---|---|---|---|
| Radio Access Network (RAN) | Rel. 15 | Study on NR to support non-terrestrial networks | Channel model, deployment scenarios | TR 38.811 / 2018 |
| | Rel. 16 | Study on solutions for NR to support non-terrestrial networks | Necessary features | TR 38.821 / 2019 |
| | Rel. 17 | Solutions for NR to support non-terrestrial networks | Enhancements for LEO, GEO | n/a, completed 2021 |
| | Rel. 17 | Study on NB-IoT/eMTC support for NTN | Scenarios and changes to support IoT | TR 36.763 / 2021 |
| Service & System aspects (SA) | Rel. 16 | Study on using satellite access in 5G | Use cases and requirements | TR 22.822 / 2018 |
| | Rel. 17 | Integration of satellite access in 5G | Stage 1 requirements | n/a, requirements defined in 2018 |
| | Rel. 17 | Study on architecture aspects for using satellite access in 5G | Key issues for integrating satellites in 5G architecture | TR 23.737 / 2020 |
| | Rel. 17 | Integration of satellite systems in the 5G architecture | Normative specifications | n/a, completed 2020 |
| | Rel. 17 | Management and orchestration aspects with integrated satellite components in a 5G network | Business roles, service management and orchestration | TR 28.808 / 2020 |
| | Rel. 18 | Guidelines for extra-territorial 5G systems | Use cases, relevant features, operation over borders | TR 22.926, v18.0.0 in Dec. 2021 |
| Core Network and Terminals (CT) | Rel. 17 | CT aspects of 5GC architecture for satellite networks | Issues related to PLMN selection | TR 24.821 / 2021 |

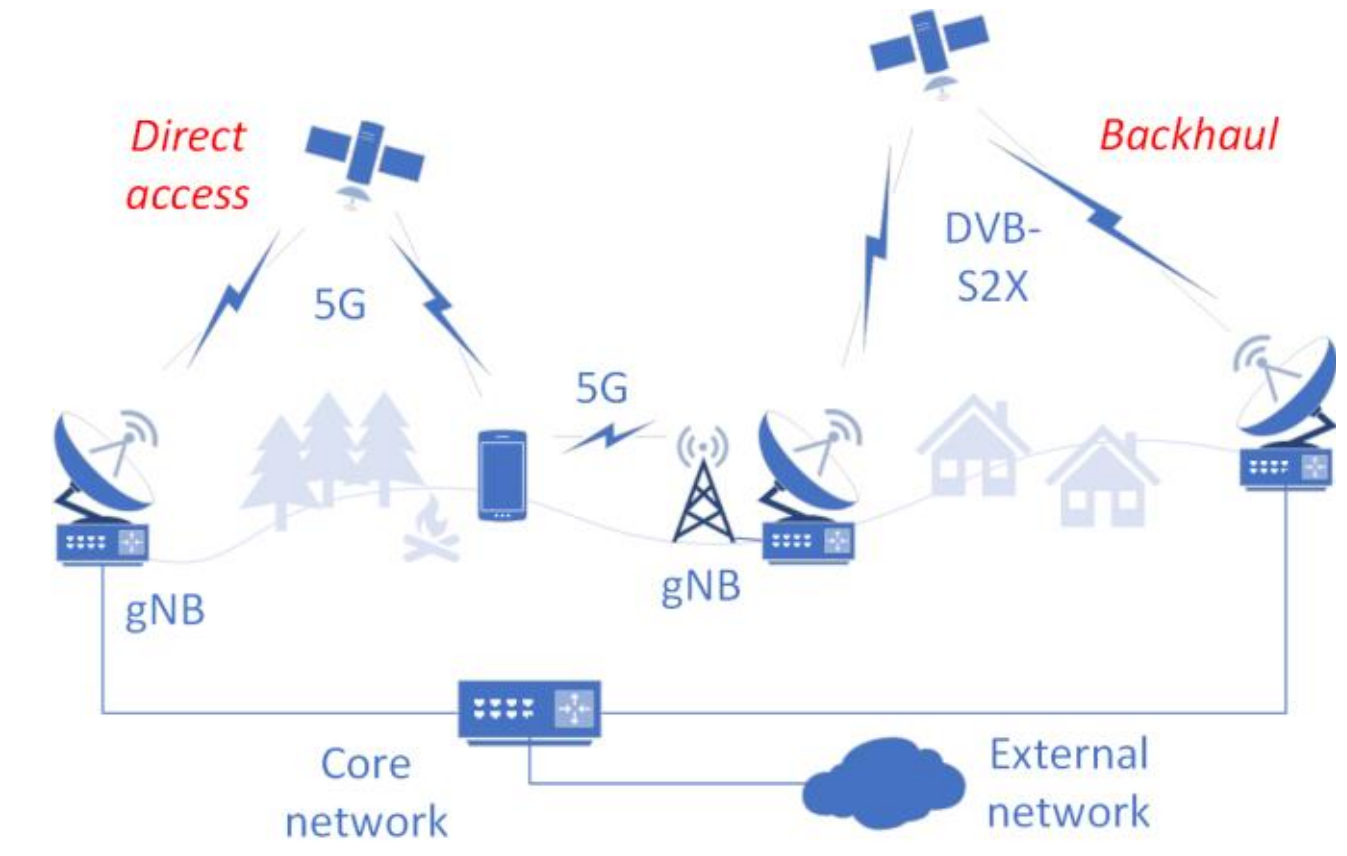

Figure 7. Direct and indirect (backhaul) access with some tentative technologies.

NTN networks have been actively advanced in the 3GPP forum since 2017. The work started by defining channel models and deployment scenarios in Rel. 15 and continued by identifying three main categories of use cases including service continuity, service ubiquity and service scalability. Rel. 16 continued providing a baseline for NR functionalities needed to support LEO and GEO satellites. The work has been developing technical solutions to cope with challenges such as long propagation delays and Doppler effects due to the movement of LEO satellites. Solutions include improved timing and uplink synchronization as well as Hybrid Automatic Repeat Request (HARQ).

The industry is working towards the deployment of satellites as part of the 5G and beyond networks in partnership with mobile network operators. This work has been done partly in Rel. 17 where both direct access and satellite backhaul and their QoS aspects have been studied as well as mobility management and core network architecture. 3GPP also approved a study item for enabling the operation of IoT in NTN networks, focusing on NB-IoT and enhanced MTC-over-satellite communications. Both LEO and GEO orbits have been considered to assume a transparent payload and sub-6 GHz frequency bands.

Rel. 17 specifications are expected to be completed during the second quarter of 2022 and the first NTN chips might become available within two years from that as shown also in Figure 6. Now the work is shifting towards Rel. 18 and 5G-Advanced finally leading to 6G systems.

The study items currently in Rel. 18 include public land mobile network (PLMN) related topics where new deployment scenarios are studied: (1) Terrestrial access and satellite access in the same PLMN, (2) PLMNs with shared satellite access networks, and (3) Mobility between PLMNs with terrestrial-only and satellite-only access. In addition, the regulatory guidelines for extra-territorial operations discussed in Section III. D are prepared.

Considering the architecture options for network integration, there are two main ways to do it as depicted in Figure 7. First, in direct access mode the end user is directly connected to the satellite as well as to the terrestrial base station. This enables access to satellite services anywhere with a typical mobile phone. Secondly, indirect access is basically the backhaul case where the end-user terminal is connected to the radio network using 3GPP or non-3GPP technology. The RAN is connected to the 5G core network via satellite. Indirect use enables connecting a local private network on the ground, ships, or aircraft to the outside world. In the backhaul case, the gNB could be located in an airborne platform such as a HAP.

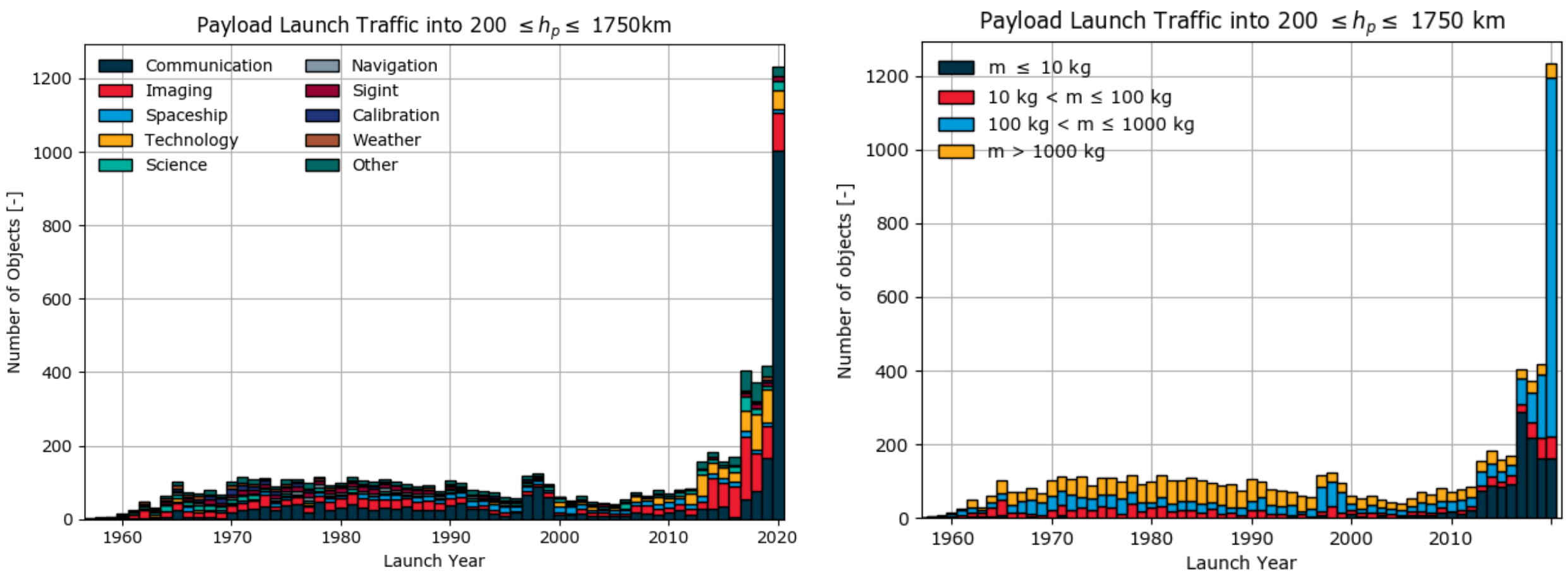

Figure 8. Evolution of launch traffic per mission type (left) and LEO launches categorized based on the launch mass (right) [86]. Reprinted with permission.

TABLE VI. PERFORMANCE INDICATORS OF THE MOBILE SYSTEMS.

| **Parameter** | **4G** | **5G** | **6G** |
|---|---|---|---|
| Peak data rate | 1 Gbps | 10 Gbps | 1 Tbps |
| Max spectral efficiency | 15 bps/Hz | 30 bps/Hz | 100 bps/Hz |
| Mobility support | up to 350 km/h | up to 500 km/h | up to 1000 km/h |
| End-to-end latency | 100 ms | 10 ms | 1 ms |
| Network architecture | Horizontal | Horizontal + NTN component studied | Three-dimensional (3D) with vertical |
| Positioning accuracy | Tens of meters in 2D | 10 cm on 2D | 1 cm with 3D |

## C. PERFORMANCE INDICATORS

The basic requirements for 6G connectivity given as performance indicators are shown in TABLE VI, with a comparison to 4G and 5G. In general, 6G systems [36]–[40] aim to offer:

- Extremely high data rates per device,
- Extremely low latency,
- Ultra-high reliable connectivity,
- A very large number of connected devices,
- Very low energy consumption with IoT devices,
- Global connectivity, and
- Connected intelligence with machine learning capability.

However, it should be noted that not all these happen simultaneously. Performance can be tailored according to application requirements. 6G networks will fuse digital, physical, and virtual worlds together and consequently increase the range of applications and services. Compared to previous generations, 6G will increase capacity and mobility support and aim for extremely low latencies. However, it is clearly evident that the integration of networks and the vertical dimension are taken into account in the network design and operations from the beginning, leading to the 3D architecture. Therefore, 6G systems can support the connectivity and positioning needs of future users and applications accurately and efficiently.

In discussions with ESA and the industry, it became evident that in the 6G era, researchers should not focus solely on the connectivity part of the space systems. Instead, telecommunication satellites, earth-imaging satellites, and navigation satellites may all be integrated to provide advanced services to end users, including localized services, situational awareness and Internet connectivity anywhere.

## D. PROPRIETARY LEO MEGACONSTELLATIONS AND MULTI-LAYER CONNECTIVITY

A major part of the planned small satellite missions globally considers communications, encouraged by the business visions presented earlier. There are many initiatives aiming to launch communications satellites into LEO orbits including mega-constellation initiatives from the USA, Asia, and Europe. For example, Starlink is already by far the largest satellite constellation ever built with over 2000 satellites in orbit.

A good state-of-the art analysis of the current situation is given in [86]. Figure 8 shows that the number of satellites launched into LEO has exploded and most of the launches include small communications satellites with 100+ kg mass. One thousand SatCom satellites were launched in 2020 alone. These launches are related to building mega-constellations. There is also a rapidly increasing number of nanosatellites with less than 10 kg mass in orbit.

TABLE VII. COMPARISON OF EXISTING AND PLANNED SATELLITE CONSTELLATIONS AS OF AUGUST 2022.

| Constellation | Frequency band | Altitude | Number launched/ authorised | Mass | User data rate/total capacity | Terminal type | Status |
|---|---|---|---|---|---|---|---|
| **SpaceX Starlink** | Ku-band<br>DL: 12 GHz<br>UL: 14 GHz<br>+ Ka-band | 550-570 km | 2748/4408 | 227 kg | 1/20 Gbps roughly 100 Mbps DL for users | VSAT, ESIM | > 100 000 users commercial in the Arctic 2022 |
| **OneWeb** | Ku-band<br>DL: 12 GHz<br>UL: 14 GHz<br>+ Ka | 1200 km | 428/648 | 150 kg | 10 Gbps per satellite | VSAT, ESIM | Arctic area coverage in 2022 |
| **Telesat Lightspeed** | Ka-band<br>DL: 20 GHz<br>UL: 29 GHz | 1015 km / 78 satellites 1325 km / 220 satellites | 2/292 both test satellites deorbited | 700 kg | Up to 7.5 Gbps for a single terminal, 20 Gbps for a hotspot | VSAT, ESIM | Plan to have commercial service in 2023 |
| **Kuiper Systems** | Ka-band<br>DL: 20 GHz<br>UL: 29 GHz | 590-630 km | -/3236 | - | 400 Mbps user data rates in terminal tests | VSAT, ESIM | First satellites in 2022, one-half to be launched by 2026 |
| **China SatNet** | Ka/Ku bands | 1145 km / 508-600 km | 1/12992 test satellite in 2018 | - | - | VSAT, ESIM | Plan to have 60 satellites in 2022 |
| **AST Space Mobile** | Terrestrial frequencies < 2 GHz | 700 km | 168 | 1500 kg | Initially 120 Mbit/s peak data rate for a cell | Commercial cellular handheld | Aiming first satellites to be launched in 2023 |
| **Lynk** | Terrestrial frequencies < 2 GHz | 500 km | 6 launched/ Plan up to 5000 | 25 kg | Narrowband transmission such as text messages | Commercial cellular handheld | FCC license files for up to 10 small satellites in May 2021 |
| **European Constellation plan: Secure Connectivity** | Not defined yet | Multiorbital LEO+MEO +GEO | - | - | Broadband services and governmental users targeted | - | Feasibility study finished in 2021. Estimated cost €6B. Potential new EU flagship |
| **Inmarsat Orchestra** | L-band<br>S-band<br>Ka-band | Multiorbital, GEO, LEO, and terrestrial 5G | 14 GEOs in orbit, 5 GEOs scheduled | 2000-6500 kg | L-band BGAN high data rate 600-700 kbps/user<br>EAN up to 100 Mbps/aircraft | VSAT, ESIM | 2 6th generation GEO to be launched in 2021-2022<br>Planning for 150-175 LEO satellites |
| **Iridium Next** | 1.6 GHz (TDD) | 780 km | 75/66 | 680 kg | Call 2400 bits/s. Up to 1100 simultaneous calls per satellite. | Proprietary handheld | Operational, 9 in-orbit spare satellites |
| **ViaSat-2** | Ka | GEO | 1 | 6400 kg | 260 Gbps total, up to 100 Mbps for the user | VSAT, ESIM | Operational since 2017 |
| **Sateliot** | Terrestrial frequencies > 2 GHz | LEO, ~550 km | 1/100 | 3U nanosat test, final satellites not confirmed | IoT | Cellular handheld, sensors | Plan to launch the whole constellation in 2022 |
| **Swarm Technologies** | 137-138 MHz<br>400-401 MHz | 300-585 km | 155/150 | 0.4 kg (smallest commercial satellites in space) | IoT | Proprietary system | SpaceX acquired Swarm in 2021, full constellation in 2022 |
| **SES 03b mPOWER** | Ka band | 8000 km + vLEO | 20/70 | 1700 kg (MEO) | max user rate 10 Gbps / total hundreds of Gbps | VSAT, ESIM | MEO satellites operational, plan to include 36 vLEO satellites |

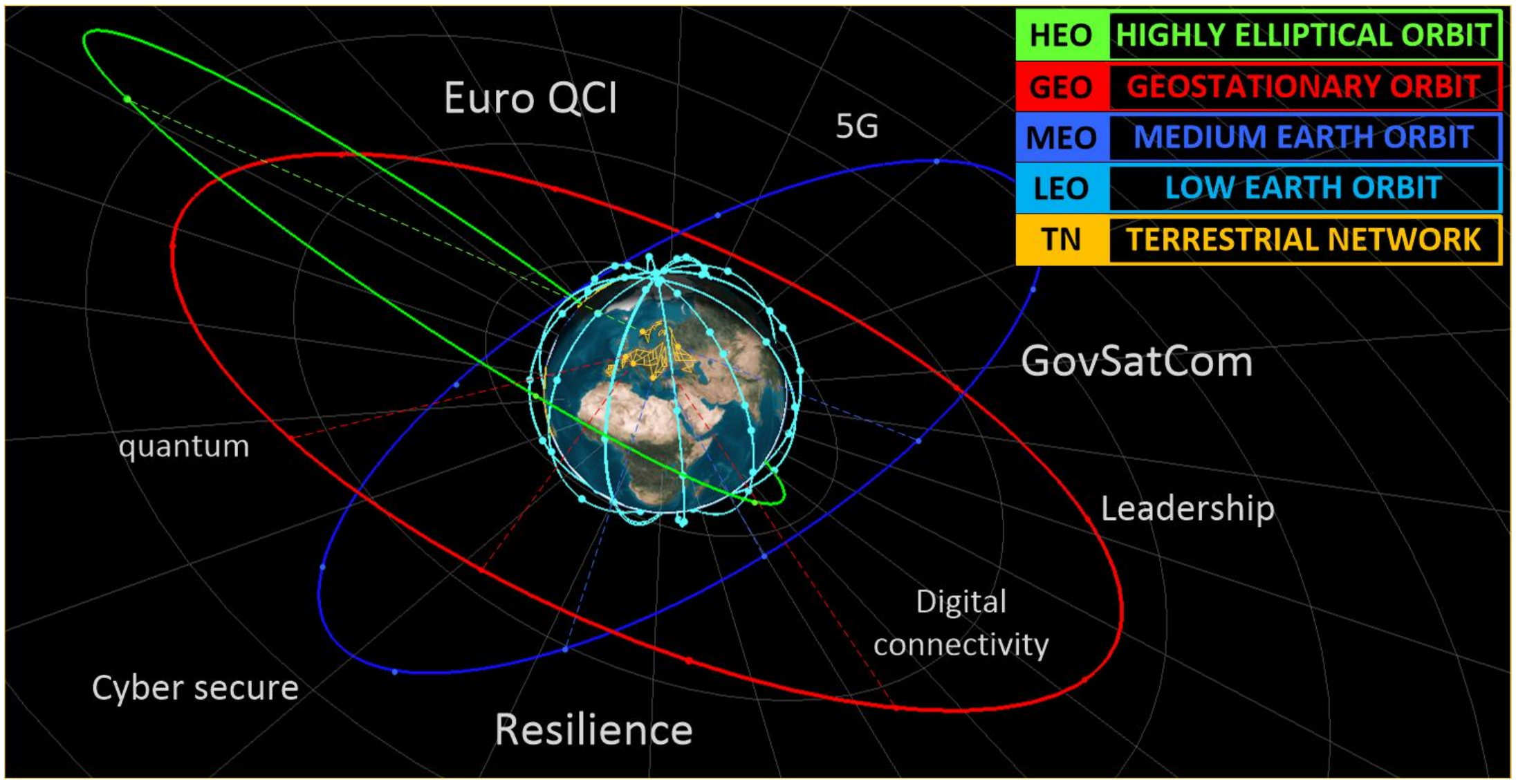

Figure 9. European secure connectivity initiative vision from DG DEFIS, redrawn by authors.

TABLE VII compares existing and planned LEO satellite constellations, including Iridium and Viasat-2 information as reference examples of traditional LEO and GEO systems. The information is gathered mostly from public webpages. In addition to Starlink, OneWeb is also already providing initial services and both systems plan to provide Arctic coverage, including Finland, in 2022. Other notable LEO constellation initiatives include Kuiper Systems, Telesat Lightspeed, AST Space Mobile, Lynk, and Chinese plans for their 13,000-satellite constellation. More information on initiatives on LEO orbits can be found in [87]–[90].

In addition to LEO, there are system developments related to other orbits and multi-layer systems including Inmarsat Orchestra [91] [92] and O3B [93]. Finally, there have been some failed attempts to build a constellation such as LeoSat even though it first raised a good amount of funding. Also OneWeb was very close to going bankrupt before the British government and an Indian investor supported the work. Already more than 400 satellites have been launched into orbit and the company seems commercially competitive.

The costs of the proposed constellations are quite high and there is clearly a strong belief in the business models. SpaceX Starlink, OneWeb, Telesat and Kuiper projects have estimated total costs ranging from $5B to $30B. AST Space mobile has a slightly lower price of around $2B but economic information about the Chinese state-sponsored ChinaSat plans is missing. Comparing the proposed systems, we can highlight the following findings:

- Iridium works with handhelds and low data rates, providing voice services.
- Lynk and SpaceMobile also aim for handhelds, even commercial mobile phones. Lynk has successfully demonstrated the ability for two-way communications between a satellite and a handheld for limited data rate applications such as text messaging.
- Many initiatives consider VSAT and the "cellular Internet" type of performance. Both fixed and mobile user terminals will be supported.
- Viasat-2 has a higher total capacity for a GEO satellite, but the total system capacity and maximum user throughput are clearly higher in megaconstellations.

TABLE VII provides an overview of the situation. However, there are a number of initiatives not presented in the table, such as the IoT satellite constellation being built by Kepler Communications [94]. Finally, there are also LEO constellations for Earth observation including Spire Global, Planet Labs and Iceye.

### E. EUROPEAN FLAGSHIP CONSTELLATION PLAN

The European Commission funded a feasibility study on a secure space-based connectivity system, fulfilling Europe's need to rapidly develop a space-based connectivity initiative [95] [96]. This could potentially become a third flagship space infrastructure besides Galileo and Copernicus. The initiative aims at implementing the most advanced satellite infrastructure for connectivity, providing the first broadband coverage to areas where terrestrial infrastructure cannot reach and later supporting services such as autonomous transport. The consortium doing the study includes European satellite manufacturers, operators and service providers, telco operators and launch service providers. Recently, a plan to invest €6 billion in the system was made public. It aims to provide worldwide uninterrupted access to secure and cost-effective satellite communication services [97].

The Directorate-General for Defence Industry and Space (DG DEFIS) has created a high-level figure of the system,

depicted in Figure 9. The planned system includes traditional large satellites at higher orbits as well as an LEO constellation and the 5G terrestrial infrastructure. The system would support European sovereignty, providing secure services to governmental users as well as broadband connection to consumers. The main objective is to create a broadband service for the whole of Europe so that the continent will have an ultra-secure connectivity system to guarantee operations during crises. EuroQCI, a secure quantum communication infrastructure, and standardized 5G technology are key components of the initiative.

There are ambitious goals set for the system since the initial service should be provided already in 2024 and the multi-layer network should be fully operational in 2027. To enable this, the European Commission and ESA are expected to fund several technology development projects in the coming years. The development is partly funded by ESA's 4S program (Space Systems for Safety and Security), which works on secure space systems to integrate them into seamless operations on Earth. Quoting the program webpage [98]: "There's no safety on Earth without safety in space."

**Lessons learned**: The importance of satellite communications is increasing throughout society, covering the needs of individual users, businesses and public safety organizations. Due to the estimated high business potential, there are a number of commercial players developing proprietary satellite constellations to support broadband and IoT needs from the sky. To make future systems interoperable with each other as well as with the terrestrial systems, standardization work is actively ongoing to create globally acceptable NTN solutions. Constellation developments are on-going in different continents and countries to ensure sovereignty also during times of crisis. It is not likely that all existing and new players will stay economically competitive in the future.

## V. RECENT PROGRESS IN TECHNOLOGY ENABLERS

### *A. CONSTELLATION DESIGN AND SIMULATIONS*

Satellite constellation design is a complex task that needs to take into account service requirements, coverage areas, and sustainability aspects. Traditionally, the constellation has been optimized for global coverage [99]–[102] or for regional coverage especially for EO constellations [103] using a single-layer system. There is no specific unified design approach for a local continuous coverage or surveillance mission over a region [104], and the design can aim at maximizing coverage while minimizing revisit time or achieving a revisit time target. Also, EO satellites need to download their measurement data to ground stations. Depending on the number of satellites in the constellation and the locations of ground stations it is also necessary to design proper datalinks and scheduling algorithms. Data downlink scheduling is part of a much wider problem that also involves scheduling the sensing and managing of on-board and terrestrial system limitations [105]. Finally, the described 3D architecture requires a multi-layer constellation design, which has been considered in [106], [107].

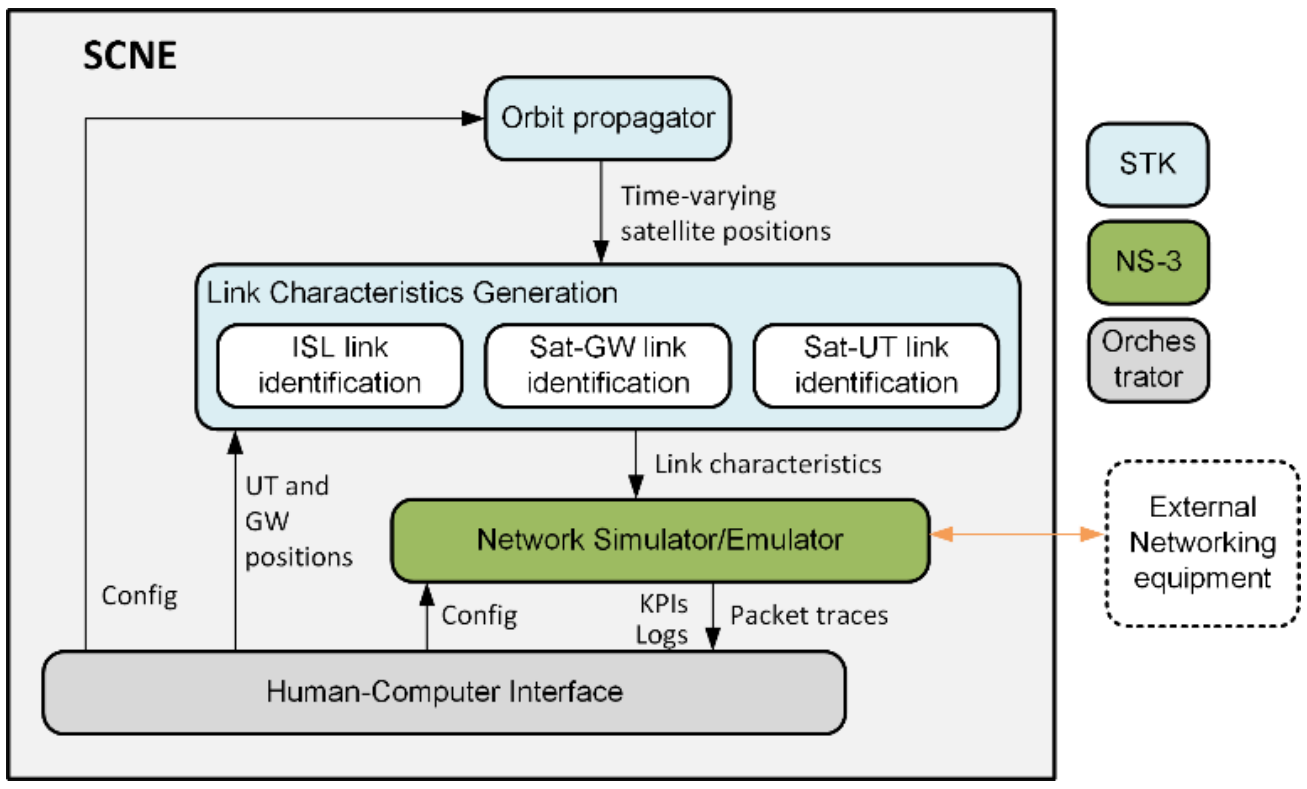

Figure 10. The SCNE simulator architecture.

Typically, constellation designs are made using advanced simulators and various constellation creation methods, such as the Walker constellation, which is a globally symmetrical configuration, or streets-of-coverage. The latter refers to the swath on the ground with continuous coverage. Deterministic and location-based models that have been applied to analyze satellite systems are typically restricted to support simulations. Recently, stochastic geometry analysis that abstracts generic networks into uniform binomial point processes has also been developed to support constellation design and analysis [108] [109]. Analytical methods can lead to very fast results. Since numerical models can lead to more accurate outcomes than analytical methods, they are still favored in constellation and orbit designs. There are also semi-analytic approaches proposed for lowering the computational burden related to satellite propagation [104] or probability of detection calculations [54] in the case of joint communication and sensing constellations.

Regarding the SatCom applications, it is not enough to be able to create a constellation that provides the required coverage globally or for the area of interest. The design needs to consider detailed physical layer protocols, network traffic characteristics such as end-to-end delays, throughput requirements, routing, and interference. In Finland, Magister Solutions Ltd has developed the Satellite Network Simulator 3 (SNS3) [110] as part of an ESA ARTES project to support network simulations. Magister has also created a system simulator for 5G NTN evaluations [111] to support 3GPP-based SatCom development. In addition, there are studies from the University of Oulu on the use of 5G NR-over-SatCom links [112].

There is no single simulation tool currently available on the market that can holistically cover all the needs. Traditional simulation tools handle either orbital or network aspects, with limited interfacing and coupling between the two. Thus, one approach is to combine different simulation tools intelligently e.g., by using Systems Tool Kit (STK) to create a constellation and then use MATLAB for detailed link level analysis in the designed constellation [113]. Another recent example of this approach is the Satellite Constellation Network Emulator (SCNE) project [114] done in collaboration between Airbus, Magister and VTT. The project has developed a novel co-simulation approach that enables the modelling of complete satellite constellations including orbital and network aspects. The objective of the project has been to enable the study and assessment of protocol performances used in large LEO constellations. The tool is particularly helpful in the study of routing protocols and consequently end-to-end quality of service aspects.

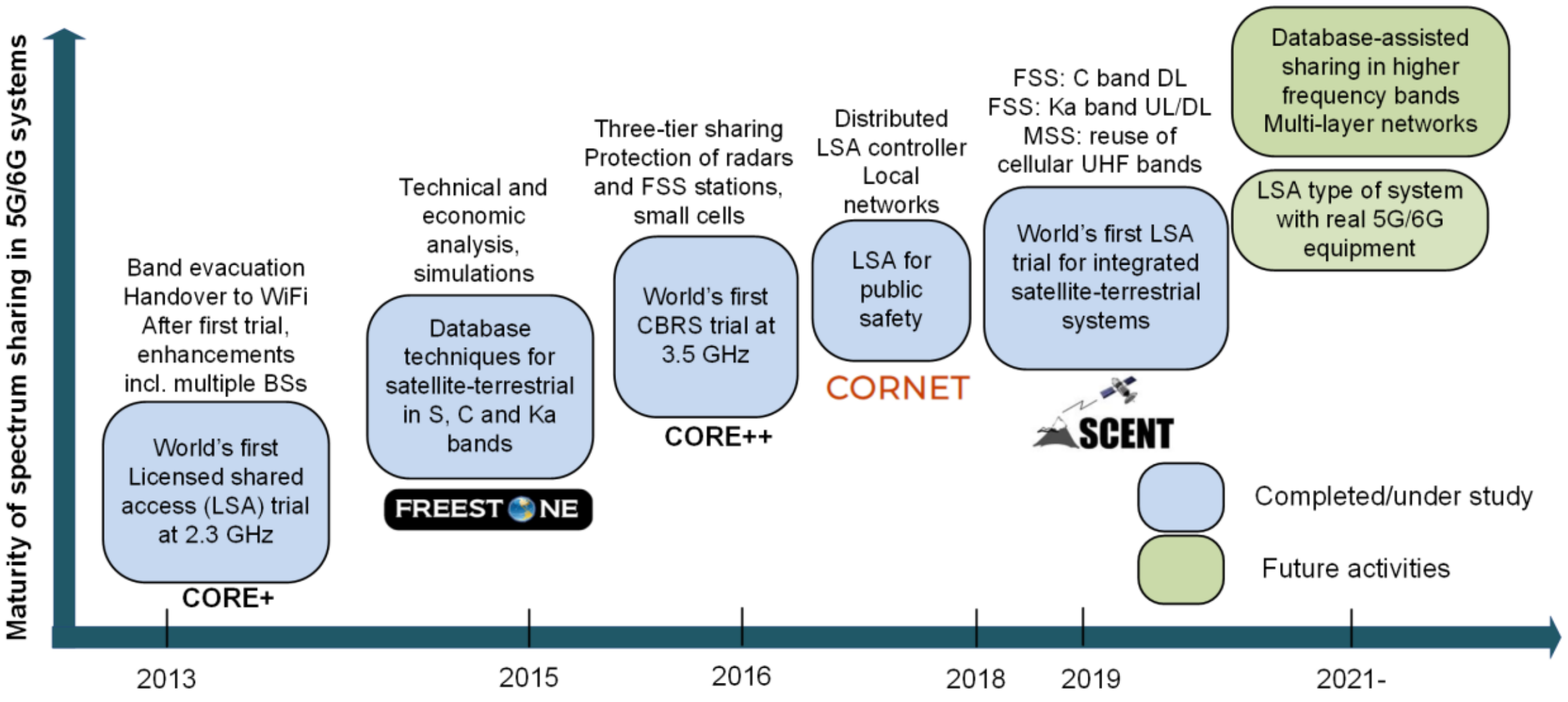

Figure 11. Evolution of database-assisted spectrum sharing towards 6G systems.

The high-level architecture of the tool is presented in Figure 10. It includes the following components. First, the human-computer interface (HCI) provides the means for a user to define the desired constellation and load the orbital and network scenarios. The HCI includes the STK Graphical User Interface (GUI) to define the constellations using a GUI or a scripting approach. Second, there is an orchestrator that is a plugin developed specifically for interwork between STK and ns-3. The orchestrator calculates the link characteristics of all the possible links based on STK accesses-defined restrictions between satellites, users, and gateways. Finally, the orchestrator transfers the link information to the ns-3, where the L2 and upper layer protocols are added and the end-to-end link performance is assessed.

### B. INTEGRATION OF SATELLITES INTO 5G/6G TEST NETWORKS

5G Test Network Finland is a multisite test environment and co-operation network [115]. Test networks provide the infrastructure to support 5G and 6G technology development, service research, and large-scale field trials. The majority of the projects have focused on terrestrial technology developments to support various verticals such as port automation and ultra-reliable low latency aspects [116] [117]. However, there are also drones and real satellite equipment (both LEO and GEO) included in the test network to support 3D network studies. The GEO connection in VTT's test network includes two different types of terminals that support high throughput broadband connections. The fixed terminal is located in Espoo and the nomadic one has been used in various locations. In addition, the LEO connection with lower latency supports 700 kbps DL and 300 kbps UL connections. We also plan to use megaconstellation-based connections in the near future when they become commercially available in Finland.

Part of the work has been developing concepts and conducting analysis and simulation studies for IoT satellites [118] [119]. Some areas such as autonomous shipping [120] or connected driving and road safety [121] [122], can be greatly enhanced with the integrated satellite-terrestrial networks. Satellites can provide both connectivity and positioning services. A practical connectivity solution of an autonomous ship includes both satellite and terrestrial communication systems. It may also include HAPs, e.g., along shipping routes in the Arctic.

Implementation activities in the test network have first focused on measuring the performance of the available solutions and then building proofs-of-concept for selected application areas. Regarding public safety networks, we implemented a private network, or "tactical bubble," to provide a local connection to authorities and used the GEO connection as a backhaul towards the core network [123].

### C. SPECTRUM SHARING AND THE WORLD'S FIRST TRIAL IN INTEGRATED SATELLITE-TERRESTRIAL NETWORKS

Adaptive communications research, which started in the sixties [124], led over years to cognitive radio (CR) studies aiming to share and use spectrum efficiently [125]. Spectrum coexistence studies in satellite communications were done at the same time [126] and the first ESA funded activity considering the application of CR techniques to SatCom provided results regarding suitable scenarios and frequency bands in [127]. In parallel, Business Finland funded test network projects started to develop practical solutions for licensed spectrum sharing starting from licensed shared access (LSA). In this approach the incumbent operators are required to provide a priori information for this database about their spectrum use over the area of interest, telling where, when, and which parts of the frequency bands are available. A limited number of users obtain the right to use the band while the LSA controller, using the information from the database called the LSA repository, ensures predictable QoS for all spectrum rights of use holders with proper power and frequency allocations.

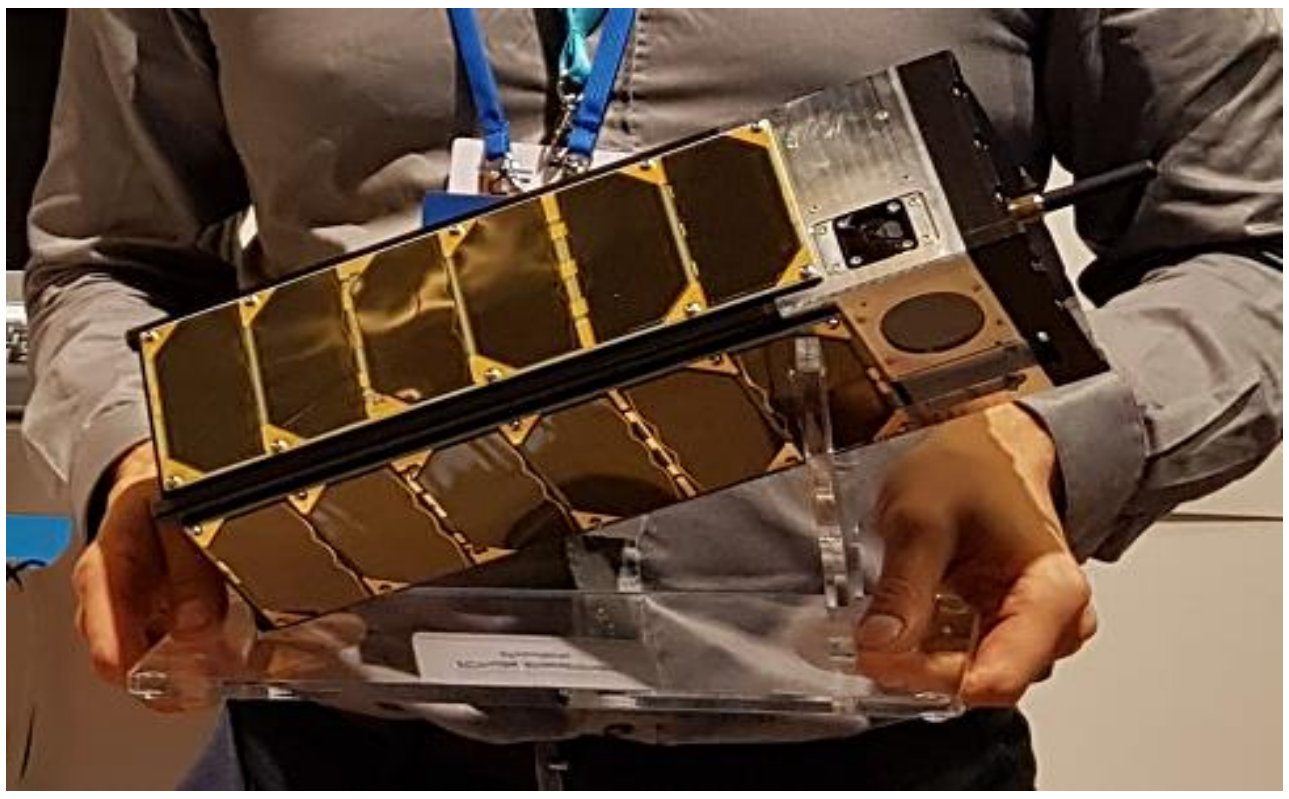

Figure 12. W-Cube nanosatellite.

The timeline for spectrum sharing developments from the database-assisted system point of view is given in Figure 11. The world's first LSA trial was done in 2013 in the CORE+ (cognitive radio trial environment) project at the 2.3 GHz band, showing the applicability of the system in practice [128]. Multiple enhancements were done during the following years adding more base stations and advanced technologies to the setup. Another spectrum sharing approach proposed in the 3.5 GHz band called spectrum access system (SAS) for citizens broadband radio service (CBRS), emerged in the USA [129]. While LSA is a two-tier model with primary and secondary users, the SAS model includes a third tier, called general authorized access (GAA) to facilitate opportunistic spectrum use. In order to protect FSS earth stations, the Federal Communications Commission (FCC) has adopted a rule that requires satellite operators to register their stations annually. The SAS obtains this information from the FCC database and uses the data when it grants or denies access to users willing to operate in the same band. The world's first CBRS trials were conducted in the CORE++ project [130].

The use of database-assisted technologies in SatCom continued in the ESA Freestone project [32]. Findings from the project provided technical and economic analysis for several frequency bands and application scenarios. Controlled sharing was shown to be an attractive option since, in some cases it can ensure the status of the satellite operator on the band instead of totally losing it to some other systems via new spectrum allocations. The applicability of LSA to public safety and to support local networks via the use of distributed architecture was studied in [131]. Finally, the recent ESA ASCENT project created a testbed and conducted field trials using real base stations and up to 1000 virtual base stations [132]. Trials considered spectrum sharing scenarios between satellite and cellular systems at the 5G pioneer bands at 3.4–3.8 GHz and 24.25–27.5 GHz, where a satellite system is operating in the downlink direction and a cellular system is accessing the same band.

As described in [133], spectrum discussions for 6G networks are currently in their infancy. Database-assisted technologies [134] can support local networks that will be more and more important in the future, and those networks can use satellite connections from any location to connect to the outside world. In addition, the development of joint communication and sensing especially at higher frequencies opens up further possibilities for 6G networks to sense and adapt their operations in real time.

### D. MILLIMETER WAVE SATELLITE IN ORBIT: W-CUBE

In order to maximize data rates for multilayer systems, millimeter wave technologies need to be applied in terrestrial and satellite links [135]–[137]. Frequency bands clearly above 10 GHz can provide larger bandwidths compared to conventionally used frequencies and enable the use of highly directional antennas. However, those frequencies are heavily attenuated due to effects such as atmospheric absorption and tropospheric scintillation. Thus, successful application of those frequencies requires the development of RF and antenna technology as well as channel modelling missions to really be able to evaluate performance over satellite links.

The ESA ARTES activity called W-Cube is using a 3U (where 1U is 10 cm x 10 cm x 11.35 cm) nanosatellite equipped with a beacon transmitter to measure and characterize the wireless channel in the 75 GHz band [138]. This opens up possibilities for the use of the high millimeter wave frequency range in communications satellites in the future. The first-ever W band satellite that was launched into orbit in June 2021 is depicted in Figure 12. The payload design includes an innovative concentric ring antenna for signal transmission. In addition to the main payload, the satellite broadcasts a Q band signal at 37.5 GHz to compare the information on measurements with previous models at lower frequencies. The satellite platform was designed, developed, and tested by the Kuva Space company, and the payload was developed by VTT.

One aim of the mission is to understand how weather phenomena affect signal propagation and polarization. To save battery power, the beacon signals are switched on only when they can be detected by measuring stations located in Austria and Finland. At other times, the satellite charges its batteries using the craft's solar panels. The satellite orbits the Earth approximately once every 1.5 hours and is visible to the ground station for about 10 minutes at a time. The signals from space have been successfully received at the ground station and the actual channel modeling work led by Joanneum Research from Austria is ongoing.

### E. TOWARDS TERAHERTZ AND OPTICAL COMMUNICATIONS

The quest for ever-higher frequencies is actively continuing especially in terrestrial short-range communications. For example, D-band (above 100 GHz) active electronically steerable antennas have been developed recently [139]. There are studies on terahertz technology, such as [140], in which the authors show that the capacity of the satellite-to-airplane THz link may reach speeds ranging from 50–150 Gbps, thus enabling cellular-equivalent data rates to the passengers and staff during the entire flight.

In small satellites, cost and power consumption are clearly limiting factors. Ground stations will use electronically steerable multi-beam antennas that are able to communicate with multiple LEO satellites simultaneously. Miniaturization and the creation of efficient antennas for small satellites is an important topic, and planar patch antennas are being actively developed. It is essential to develop antenna systems that can sustain wireless links or remote sensing requirements in a small, stowable package [141].

Optical communication is a disruptive technology that will enable ISLs and satellite formations and can lead to significant power savings compared to RF communications. According to [142] it can provide a safe and cyber-secure way to serve scalable 3D networks, enabling the shift from partitioned ground and space segments into a fully integrated system. Optical communications are limited in many areas due to clouds. There is actually an "optical belt" across the Sahara and the Middle East where satellite-to-ground connections are possible due to cloud-free availability [186].

**Lessons learned**: Analysis of recent research related to multi-layer networks shows that the research community is actively developing technologies for complex integrated multi-layer networks. Simulations enable the cost-efficient modeling of large systems, showing what could be possible in the future. Test networks enable the development and testing of practical applications. High-capacity demands may be fulfilled with very high-frequency bands and optical communications. Even though we have used many Finnish examples in the analysis, the topics can be generalized to other countries and areas since satellite communications is naturally an international business. The following potential research topics are targeted at the whole SatCom research community across the globe.

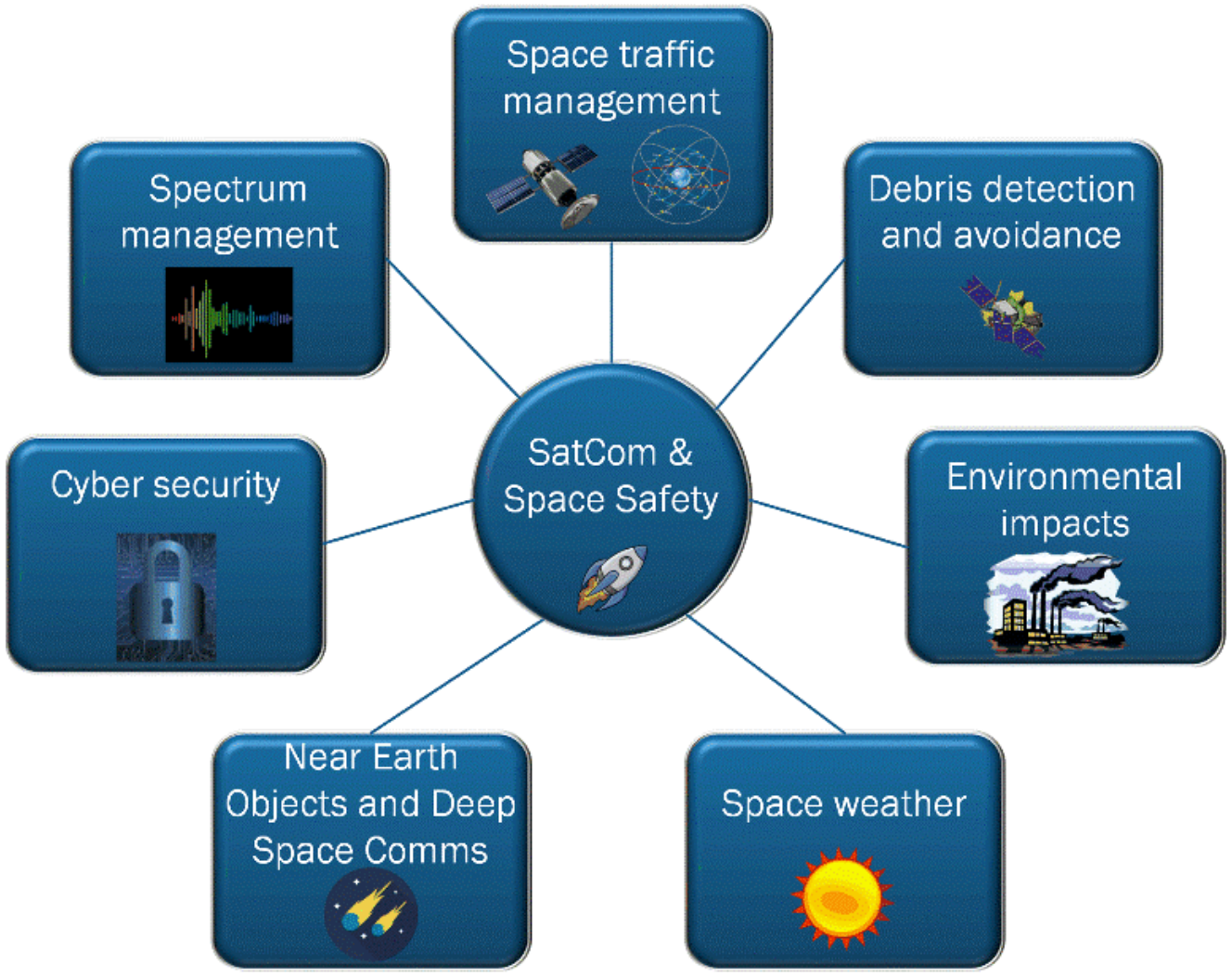

Figure 13. Space safety aspects related to 6G SatCom.

## VI. SUSTAINABILITY ENABLERS AND SPACE SAFETY

The described development is both enabling sustainable growth globally and inevitably creating challenges related to space congestion. Thus, while developing 6G SatCom systems, one needs to take care of space safety as well. The risks associated with space systems originate from hazardous characteristics of the system design and its operating environment and the hazardous effects of system failures [143]. The space system itself is comprised of hardware, software, and the human operator. ESA defines a "hazard" as a source of threat to safety. A hazard is therefore not an event but a characteristic of a system or its potentially dangerous environment. We adopt the same view and look at the situation from the SatCom perspective. Thus, we do not consider manned flights and related risks to onboard humans. Those topics are well covered in [144]. The topic of space safety has been visibly raised in many countries' agendas. For example, in [145] the UK Space Agency is welcoming all efforts – public and commercial – to preserve the space environment for future generations.

### A. CLASSIFICATION OF SATCOM RELATED ASPECTS

The emergence of a large number of small satellite systems in LEO orbits (including the flagship initiative) is creating challenges in keeping the systems safe and reliable. There should be no physical collisions to prevent a total loss of services and the creation of space debris. On the other hand, there is a need to manage radio system interference, both between satellite and terrestrial services and between different satellite systems. Current space safety procedures, including debris mitigation and collision avoidance are designed with lower traffic densities and the New Space era is causing clear challenges. We classify the space safety aspects related to SatCom into seven different themes, depicted in Figure 13. Even though the focus is on the communication systems the classification and discussion can be easily generalized to any space-based activity. Each topic is covered in detail in Table VIII and the following sections.

TABLE VIII. RESEARCH TOPICS FOR SPACE SAFETY AND SUSTAINABILITY IN THE 6G ERA.

| Space safety topic | Key considerations and selected research areas |
|---|---|
| **Space traffic management** | Automated collision avoidance procedures have been proposed for use with satellites in LEO constellations in order to make them react rapidly to threats. This ability requires the development of new ways to keep space object catalogs and their predictions up-to-date. For example, there might be a need to share data on collision avoidance actions in a global registry. |
| **Space debris detection and avoidance** | It is essential to avoid creating new debris and shield critical infra as well as possible against hits by very small pieces of debris. However, in order to avoid and remove the debris, reliable detection is essential. This could be achieved by efficiently combining space-based and ground-based detection capabilities. |
| **Environmental impacts** | Keeping the megaconstellations up-to-date with the continuous launching of new satellites and re-entrance of satellites at their end of life to the atmosphere should not cause too much harm to the atmosphere or land/ocean areas. The development of in-orbit servicing methods and innovative materials including how cleanly satellites burn when returning to Earth are clearly fruitful research topics. |
| **Space weather** | New Space Economy players entering the business with cheap small satellites should still consider high reliability and safety standards to avoid their equipment becoming debris due to the harsh space environment. How to ensure this while enabling flexible entrance to the space business? |
| **Near-Earth objects and deep space communications** | Life on Earth can be threatened by Near-Earth Objects such as asteroids. On the other hand, the space may provide useful resources to us as well as provide places to go for future generations. The development of multi-hop networks for deep space to support mining, space exploration, and the potential inhabitance of other planets is foreseen. |
| **Cyber security** | The protection of space infrastructure in space and on Earth covering multi-layer networks and their interfaces is very important. Sustainable and safe development requires effective coordination of actions among stakeholders and organizing training for space actors. |
| **Spectrum management** | Ensuring the availability of services and avoiding harmful interference will be ensured by regulated allocations. However, there is a need to develop the means to detect interference sources to prevent malicious actions. In addition, developing coordinated dynamic spectrum management approaches will enable more efficient use of limited resources. |

### B. SPACE TRAFFIC MANAGEMENT

Space traffic is increasing rapidly including both launchers and the number of satellites in orbit. According to the European Space Agency, there are currently 8840 satellites in space of which 6200 are still functioning [146]. Due to the miniaturization of thrusters in the satellites, small satellites also are capable of maneuvering themselves. NewSpace operators are making extensive use of low-thrust systems for both transit and station-keeping. One approach is to launch into low LEO orbit, transition to the higher operational altitude via low-thrust, and at end-of-life, deorbit the same way [147].

Space surveillance networks are keeping their catalogs of space objects (including satellites and tracked debris) up-to-date and also aim at predicting where they are going in order to prevent collisions. However, existing catalog and collision avoidance processes have no effective way of dealing with frequent or continuous maneuvers, since they are based on predictions generated days in advance, with no assumption of maneuvers. Thus, if an existing satellite constellation is operating in proximity to one of the SatCom constellations that are frequently automated maneuvering, its current collision avoidance process breaks down. Automated maneuvers may move one vehicle in the constellation out of a conjunction, or it could create a new problematic conjunction [147]. This especially concerns SatCom constellations due to their high number of satellites.

### C. SPACE DEBRIS DETECTION AND AVOIDANCE

Thousands of satellites and rocket bodies (especially in LEO orbits) can break into debris upon collisions, explosions, or degradation in the harsh space environment [43]. These fragmentations increase the collision probability per time and in the worst case these collisions could dominate in-orbit evolution, leading to a situation called the Kessler Syndrome [148]. There are approximately 31 400 cataloged space debris objects, and it is estimated that there are 1,000,000 objects larger than 1 cm in diameter [146]. Debris is travelling in space at a very high velocity where the relative speed of debris objects compared to satellites can be on the level of 15 km/s. Thus, even small pieces can destroy satellites and it is important to develop the means to manage a situation that is becoming more and more complex. The evolving situation is putting space-based services in danger and challenging space traffic in total and could even prevent people to leave Earth in the distant future to look for other inhabitable planets.

The main mitigation approaches to space debris include debris creation avoidance, debris collision avoidance, debris removal, and debris shielding [149]. Provided that the debris is already existing in space, a key enabler in defending the satellites is the ability to detect the time-varying positions of the debris. Several approaches have been developed in the past for debris detection including ground-based radars (GBRs) and space-based radars (SBRs), using either radio frequencies or optical measurement techniques [150]. While the former remains the basis of technology for larger objects, say larger than 10 cm, the latter approach is often preferred for smaller objects. A separate stand-alone space-borne debris detection system would not, however, be a cost- or spectral-efficient solution. To overcome this, complimentary space-borne radar systems that would be more deeply integrated into the emerging satellite communication infrastructure have been recently proposed [54]. Smaller satellites are being launched into the lower orbits by several satellite operators (e.g., SpaceX), and therefore, this solution is now more timely than ever before. Next, we look at this option more carefully.

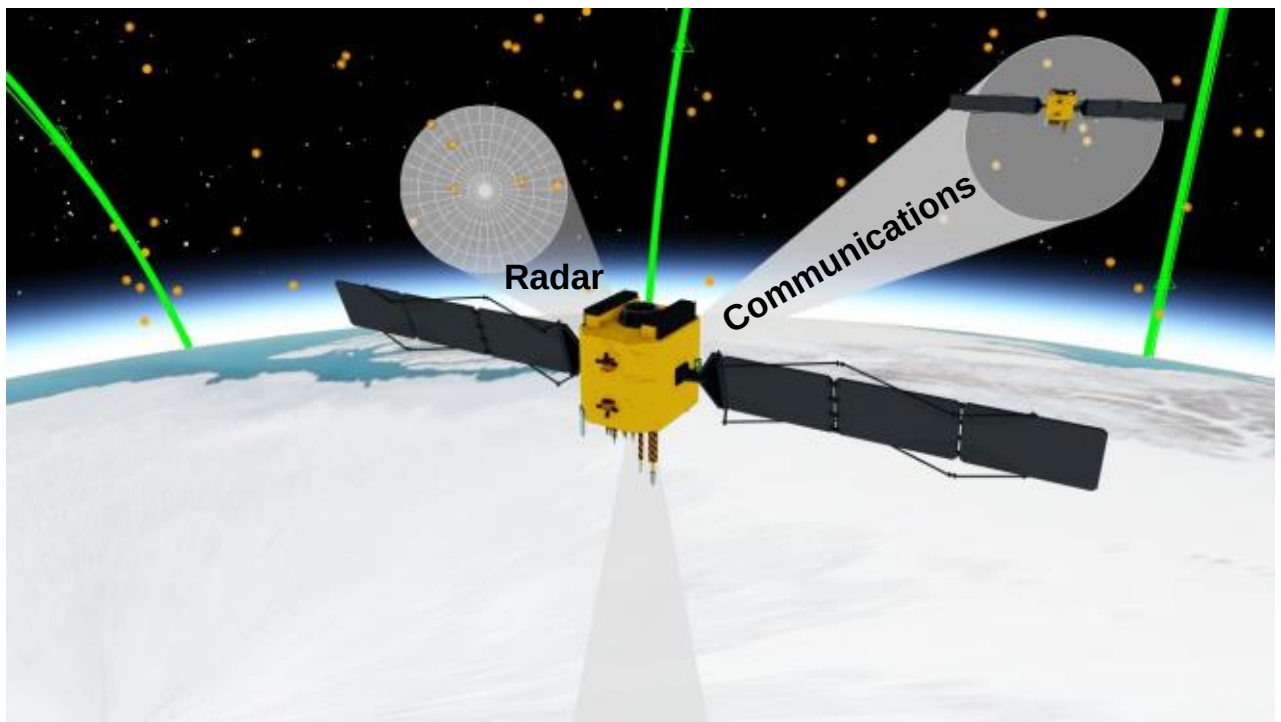

Figure 14. Illustration of the SBRC concept where SatCom signal is used as radar. (Illustration created with STK software.)

*1) SPACE-BORNE RADAR AND COMMUNICATIONS CONCEPT*

A simplified illustration of the 5G space-borne radar and communication (5G-SBRC) concept is shown in Figure 14. Space debris is detected over signals that are designed solely for communication purposes between satellites. Specifically, while the communication incentive is to exchange information between two collaborating entities, the radar incentive is to extract information by sensing radio signals reflected from a non-collaborating target. In general, there are several alternative strategies to develop an SBRC system with different trade-offs regarding the achievable integration gain and compatibility with existing legacy systems. These strategies include (1) independent signals of communication and radar subsystems operated in a single platform, (2) communication over existing radar signals, (3) radar sensing over existing communication signals, and (4) jointly optimized waveforms by redesigning both domains. Clearly, the first option represents the smallest effort for integration with the smallest integration gains, while the last option provides the most advanced integration solutions maximizing the benefits jointly for both communications and radar tasks at the cost of greater integration efforts as both domains must be redesigned. The two options in the middle represent design approaches where some compromises are made between achievable gains and compatibility.

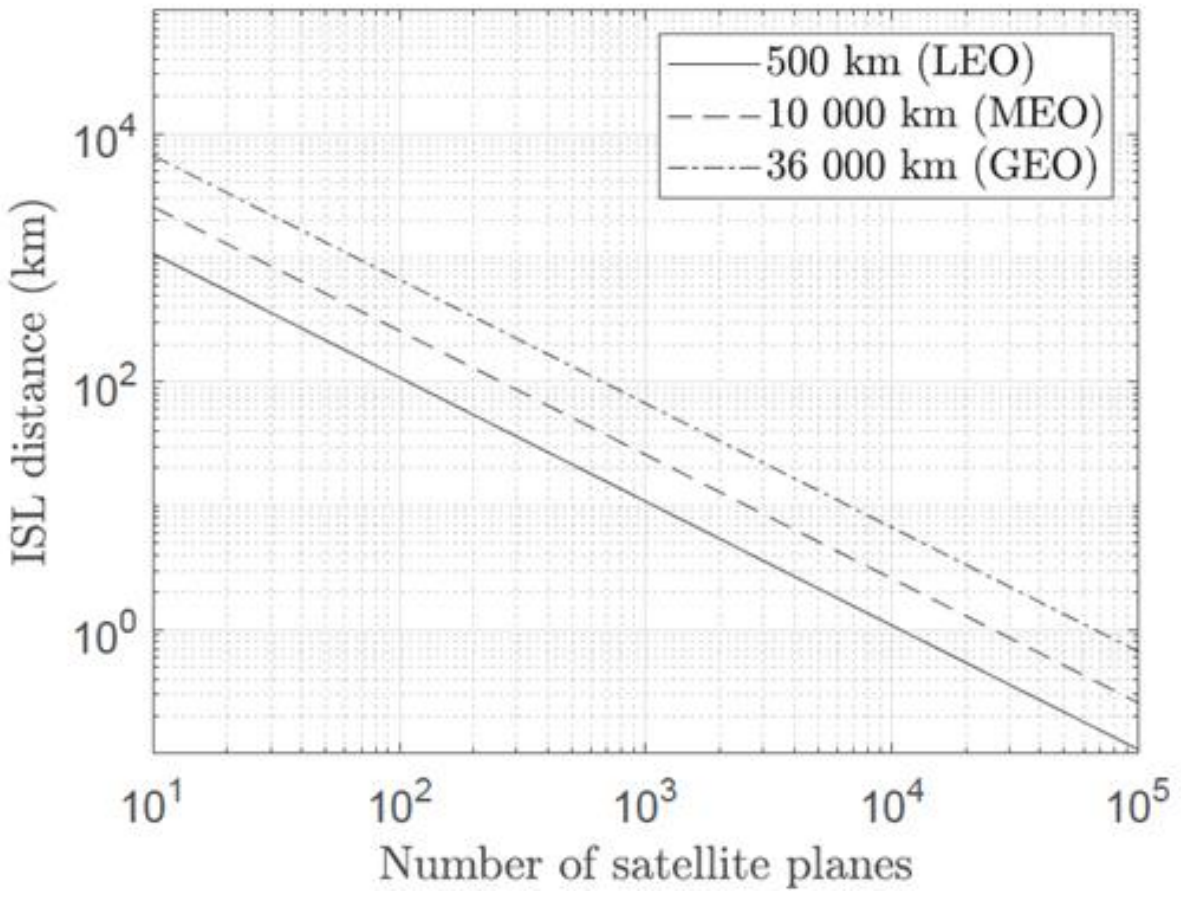

Figure 15. ISL distance versus constellation inter-plane density for different orbits.

The main benefits of the approach are that a separate space-borne radar infrastructure would not be needed, the payload of a satellite can be reduced, and spectrum efficiency is improved. However, the transmission power and antenna size of satellites must also be reduced compared to that of the GBRs. In practice, the satellites cannot be brought arbitrarily close to each other due to cost, safety, and interference regulations. To enable reliable detection of very small objects less than a few centimeters, the detection distance should be reduced to less than a few kilometers when using smaller satellites. To obtain a rough understanding of the required constellation complexity, we plot the ISL distance versus the constellation inter-plane density in Figure 15, which is derived analytically in [151]. For instance, for LEO, we would need about 1000 satellite planes in a constellation to reduce the ISL to less than 10 km. One possible way to overcome this difficulty is to allow debris detection via constellation-dependent cooperation. This has been evaluated in [54].

Terrestrial 5G networks are now being integrated into satellite communications, and it is therefore a good signaling candidate for the proposed SBRC concept. However, the 5G signal structure possesses severe limitations in estimating the velocity of debris objects up to 15 km/s, which is 100 times faster relative speed than that of typical terrestrial vehicular radar applications. We discuss this aspect next in more detail.

*2) CASE EXAMPLE: 5G-SBRC*

We apply the SBRC approach with the *radar sensing over existing 5G communication signal* strategy and provide a numerical example of the estimation accuracy of the sensing part. One main limiting factor is to ensure unambiguity in range and velocity estimations of the high-speed debris objects.

Extracting information from high-speed non-collaborating debris objects is challenging with 5G communication signals, which are optimized for communication purposes.

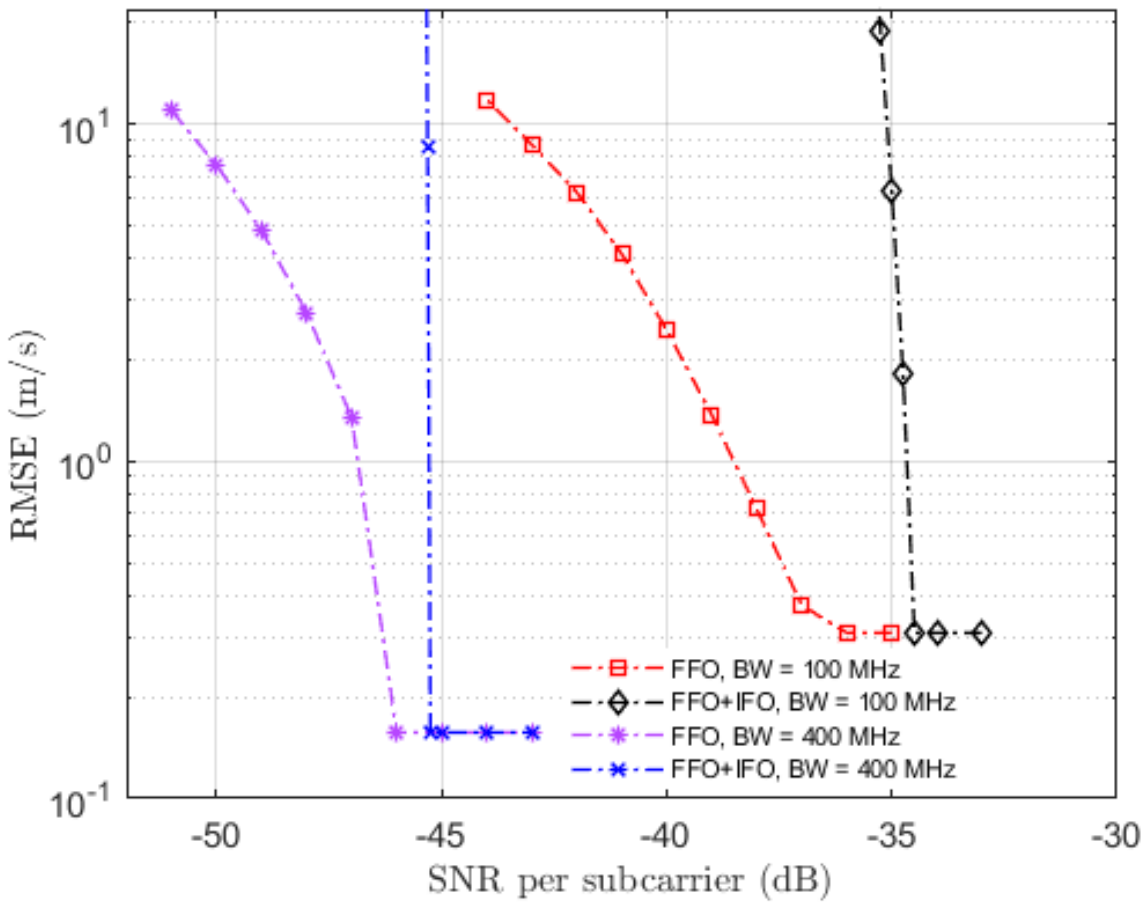

Figure 16. Advantage of 400 MHz 5G on the velocity estimation accuracy of debris objects.

Among other signal characteristics, the 5G bandwidth (BW) has an important role in radar estimation accuracy of the millimeter-sized debris objects. Using our velocity estimation techniques and parameters proposed in [54], we show the processing gain advantage of increasing the BW from 100 MHz to 400 MHz in Figure 16. In space, high velocities also cause an integer frequency offset (IFO) in addition to a fractional frequency offset (FFO). With the IFO compensation up to the signal-to-noise ratio (SNR) threshold, root mean square error (RMSE) performance is comparable to the case where there is only FFO distortion.

*3) FUTURE OUTLOOK*

There are several directions for future work under space debris mitigation. Clearly, signals that optimize jointly the different objectives of communications and debris detection would provide some interesting opportunities for both satellite communications and space debris detection. Furthermore, tighter integration between space-based and ground-based radars could boost overall detection reliability. Obviously, it is not enough to aim at merely improving debris detection approaches. To this end, different types of active space debris removal methods are being proposed including lasers, tethers, sails, and satellites (see the survey from [152]). Commercial debris removal technologies are being developed, e.g., by Astroscale [153].

### D. ENVIRONMENTAL IMPACTS

A very positive development towards more sustainable space operations has included smaller launchers and reusable rockets that can carry satellites into orbit multiple times, such as the ones used by SpaceX. However, in their rockets the second stages are also usually controlled through re-entry and deposited in remote areas of an ocean [43]. The cumulative impact of the rocket bodies can cause environmental damages to the fragile ocean environment. In addition, a large number of launches produces black carbon into the atmosphere. If the number of launches per year is more than 1000, it can lead to a persistent layer of black carbon particles in the northern stratosphere [154]. That could potentially cause changes in the global atmospheric circulation and distribution of ozone and temperature. Over years of active launching this may lead to radiative forcing comparable to the effects of current subsonic aviation.

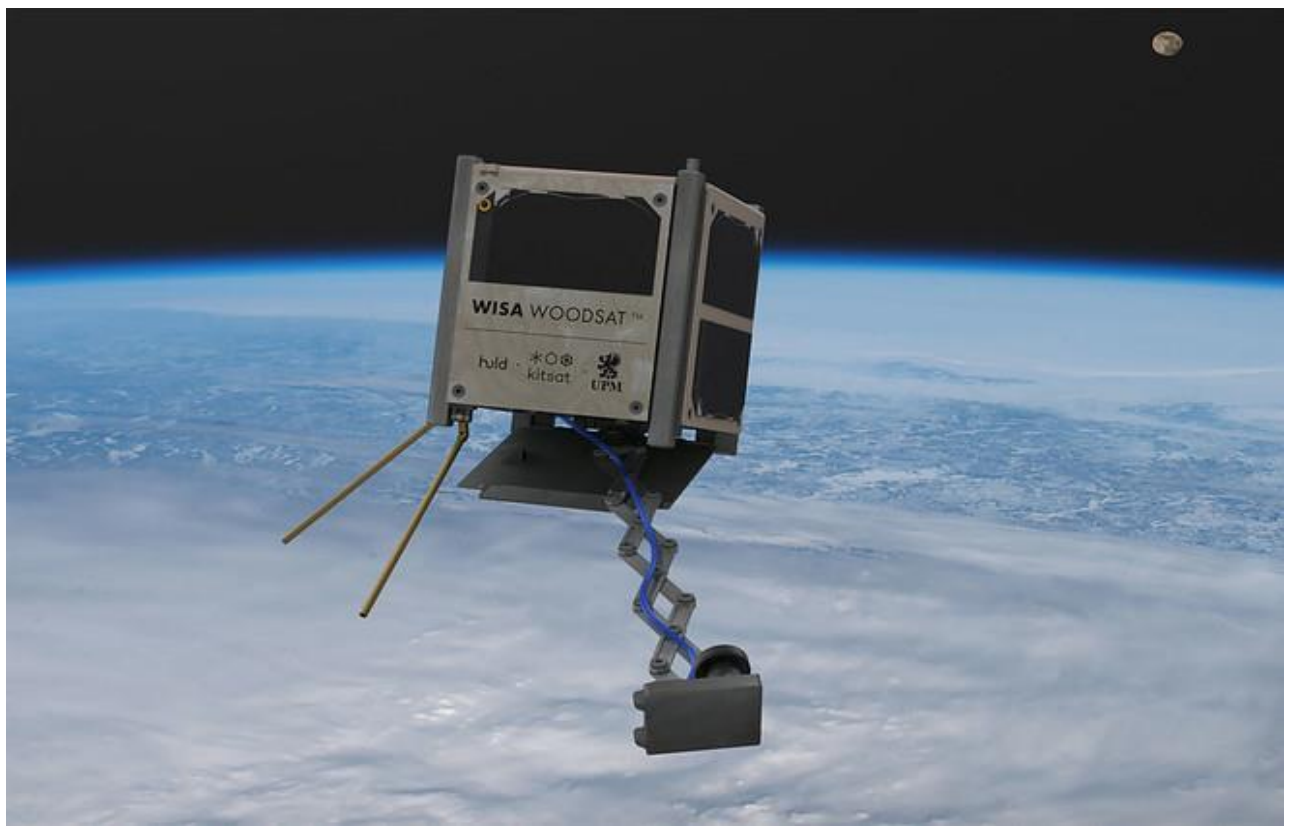

Figure 17. WISA WoodSat, a wooden satellite. Credit: Arctic Astronautics.

Another significant effect of the launches required to keep megaconstellations functional is that the satellites have a limited lifecycle, and over the update cycle, there can be several tons of satellites re-entering Earth's atmosphere every day. As explained in [43], satellites that include aluminum and re-entry will create fine particles that could greatly exceed natural forms of high-altitude atmospheric aluminum deposition.

The sustainable use of space has led to inventions related to satellite structures as well. A recent Finnish initiative called WISA WoodSat is developing the world's first wooden satellite that uses plywood panels in its structure [155] [156]. One of the aims of the mission is to understand how well the wooden structure can be applied to the spacecraft including long-term radiation and harsh space conditions. An interesting feature of this design is that during re-entry into the atmosphere, the satellite will burn more rapidly, causing less risk to humans on the ground and fewer aluminum particles in the atmosphere. The satellite has passed vibrations tests on the premises of ESA and is scheduled to be launched in the second half of 2022. The first stratospheric test flight up to 30 km to test its communication capabilities, command response, and selfie stick camera was successful. The satellite is depicted in Figure 17 showing the wooden panels on the exterior. The satellite is also equipped with a selfie stick that enables inspection of the satellite condition in orbit. In addition to materials research, the mission initially aimed to provide IoT connections using LoRa technology from space, and to support amateur radio communications via space. However, this was not supported by the International Radio Amateur Union. Therefore, there is a requirement to build, test and license a different radio system before the launch can be made.

*In-orbit servicing* is another solution to extend the life of satellites. There have been few human-assisted in-orbit servicing missions and few others for testing autonomous servicing, like the DARPA Phoenix program. Commercial satellite service models are still not fully developed. One company, SpaceLogistics [157], successfully docked its mission extension vehicle-1 (MEV-1) to a client satellite in February 2020, making it the first commercial in-orbit servicing operation. Other companies working towards in-orbit servicing include Altius Space Machines and Orbit Fab. The Consortium for Execution of Rendezvous and Servicing Operations (CONFERS), a satellite servicing industry group, is working on satellite service standards, ranging from sets of principles and best practices to fiducials and refueling interfaces on spacecraft.

### *E. SPACE WEATHER*

Space weather describes phenomena that can cause significant impacts on systems and technologies both in orbit and on Earth [158] [159]. It is caused by solar wind and solar flares in near-Earth space and the upper part of the Earth's atmosphere. Currently, the baseline approach is to collect as much of the required measurement data as possible using ground-based instruments, because they are usually less expensive and easier to maintain and upgrade than space-borne instruments onboard satellites. For example, the Finnish Meteorological Institute provides space weather data using automatic magnetometer stations located in Finland.

Space weather affects 6G services via their impact on systems such as global navigation satellite systems (GNSS) signals [160] [161] and electronics on satellites. Ionospheric scintillations are one of the earliest known effects of space weather, causing interruptions and degradations to the GNSS receivers. The effects of space weather on signal propagation can be mitigated through engineering design solutions.

However, space weather can lead to a total loss of communication due to attenuation and/or severe scintillation when the broadcast signals in certain frequencies cross the ionosphere. Thus, to make the future system resilient, one has to take space weather effects into account. A recent example of the impact was a geomagnetic storm in February 2022, which caused atmospheric drag during the launch of the Starlink satellites – and 40 satellites had to be deorbited back to the atmosphere [162].

In order to make electronics and shielding of satellite stand against harsh space weather, standardized quality procedures and tests defined by the European Cooperation for Space Standardization (ECSS) must be met. This means that the New Space players should not rush too quickly to orbit but rather ensure that even the smallest and cheapest satellites are engineered in a sustainable and reliable way. This is a good action for avoiding the creation of debris.

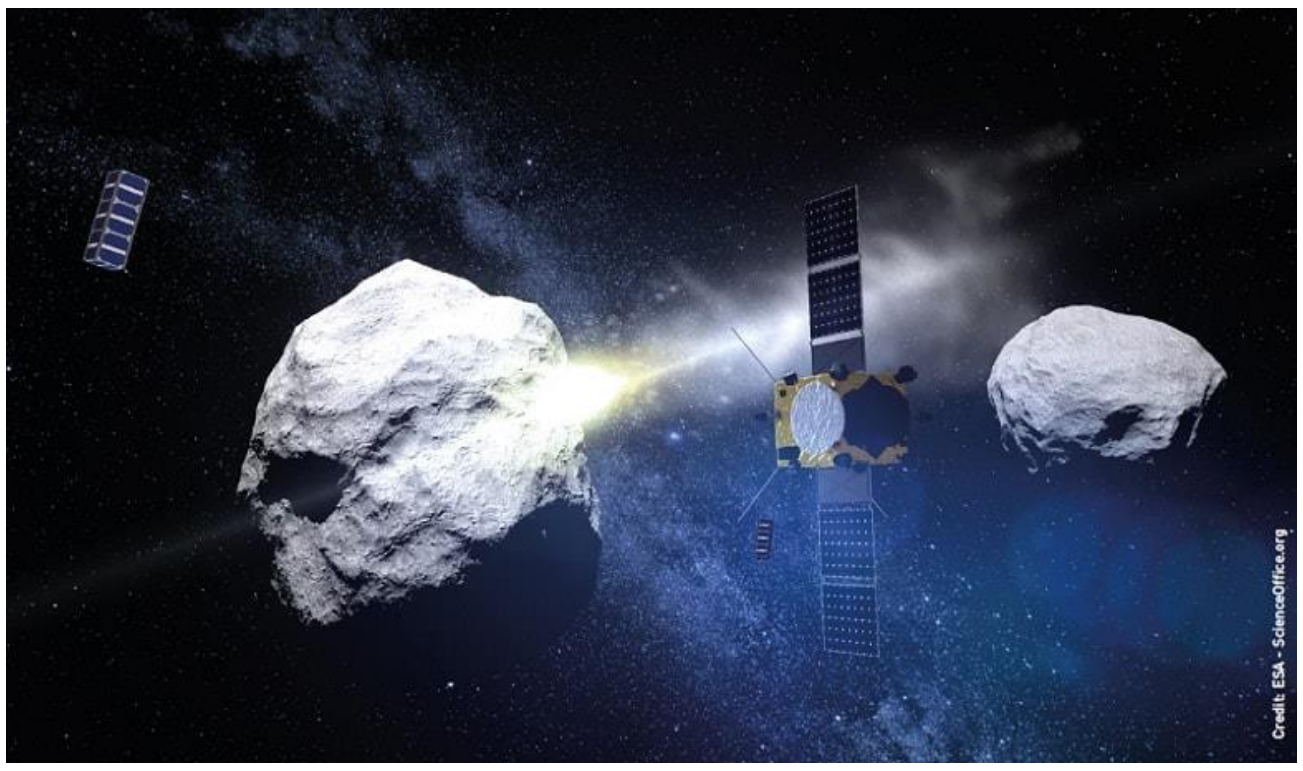

Figure 18. Joint HERA and DART mission for asteroid study and planetary defense mechanisms. Credit: ESA.

### *F. NEAR EARTH OBJECTS AND DEEP SPACE COMMUNICATIONS*

A near-earth object (NEO) is an asteroid or comet that passes close to the Earth's orbit (more accurately, within 45 million kilometers of it). They are classified based on their size and an NEO, which is more than 140 meters in diameter, is considered potentially hazardous since it can create significant damage to the Earth's surface. There are many powerful radars on Earth to detect NEOs, and space-borne missions are also used to complement the information for cataloging NEOs, and they help in preparing counteractions to ease and even avoid hazardous effects. So how does this relate to SatCom and New Space developments?

As an example, in the planned *HERA mission* depicted in Figure 18 the aim is to send a large satellite to the vicinity of the Didymos asteroid [163]. Then, that satellite will launch deep-space CubeSats to study the composition of the asteroid in detail. Also, deep-space intersatellite links will be tested between the large and small satellites. HERA will also demonstrate autonomous navigation around the asteroid (similarly to modern autonomous cars or ships on Earth), and gather crucial scientific data, to help scientists and future mission planners better understand asteroid compositions and structures. In addition, jointly with the DART mission [164]**,** the aim is to study planetary defense mechanisms and the possibility to shift the asteroid orbit with a kinetic impact.

In addition, there is ongoing active research towards deep space communication networks to provide connectivity anywhere in the solar system and to support exploration of the universe. As described in [165], "The deep space exploration missions require high quality of communication performance between the Earth stations and various deep space explorers, such as Mars orbiters and rovers." The paper describes a structured solar system satellite constellation network where several relays are used to create a topology that can support deep space operations and connect objects from deep space to each other and to Earth.

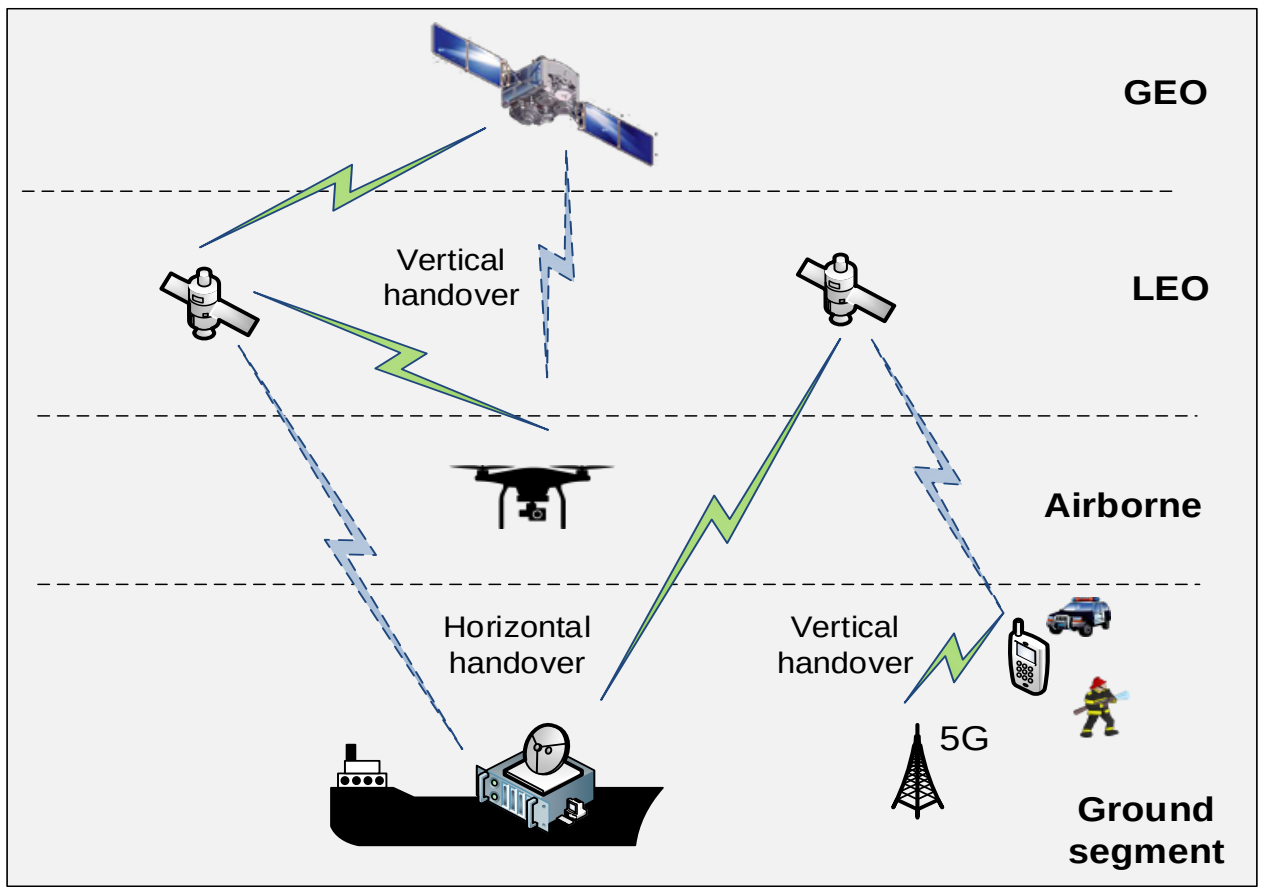

Figure 19. Handover scenarios in a layered network.

Many innovative missions are developed annually, and it is foreseen that the development of a visionary 3D network around the globe and a deep space communication network would enable unforeseen growth in New Space missions. When the connectivity is robust and working anywhere, innovative missions for science and remote sensing can also be served at an unprecedented level.

### G. CYBERSECURITY

There are many emerging cybersecurity challenges due to the development of complex multi-layer systems [166] and the rapid increase of platforms and interfaces. The infrastructure both in space and on the ground must be protected. In addition, application areas such as autonomous and remote-controlled shipping mean opening up previously closed systems due to external control interfaces [48]. Therefore, before the design phase for the whole architecture can begin, it is important to first define relevant threats and assess the impact of those threats on the system. Secondly, map out risk scenarios and finally understand the trade-offs between acceptable and unacceptable risks. After this evaluation, it is possible to proceed in defining the architecture, i.e., using the security-by-design principle. Thus, fine-tuning the multi-layer architecture for future networks requires a new type of approach. In order to make this happen across different countries and stakeholders, effective coordination of actions and organizing training for space actors is required. Space agencies such as ESA and NASA will play an important role in coordinating the activities [167] [168].

Both payload and control communication need to be tested before launching satellites into orbit. It is important to cover the whole end-to-end path when creating reliable, secure communications. Cybersecurity issues have recently been studied from different viewpoints including general issues in space networks [169], integrated satellite-terrestrial systems, [170], 5G and beyond networks [171], looking at the effects of machine learning on security [172], and ensuring secure telemetry and command links [173].

An example of a security challenge is the handover situation that can happen quite frequently in dynamic 3D networks [169]. The situation is depicted in Figure 19. Key management during handover situations is challenging. For example, when a police officer is changing his connection from terrestrial 5G to an LEO satellite, handover information includes both previously accessed networks and newly accessed satellites. Signaling is exchanged between different entities and might be eavesdropped, falsified, or fabricated.

Finally, it should be ensured that the system to use for critical communications is available. For Europe and Finland, it is good to have systems that are under their own control. This was described by a Finnish organization as: “Can we trust systems outside Europe during a time of crisis? It is much better to have a European constellation.”

### H. SPECTRUM MANAGEMENT

**Enabling services**: The number of satellites increases but the spectrum resource is naturally limited [174]. We covered some regulatory aspects related to spectrum allocations already in Section II. The potential for new allocations is shrinking because the expansion of wireless communications is continuing and new systems are emerging faster than the aging systems currently in use are becoming extinct. There are two main ways to cope with the spectrum scarcity problem [32]. (1) Use higher frequencies that are not yet allocated (or optical links [175]) and (2) Use currently allocated frequencies more efficiently. The latter leads to the concepts of dynamic spectrum sharing and spectrum coexistence.

Spectrum sharing means that two or more systems are operating in the same frequency band. In the 6G SatCom this may include spectrum coexistence (1) between different satellite systems, (2) between satellite and terrestrial systems, and (3) between systems in different layers of the multi-layer network. This is a very challenging field and leads to the use of techniques such as spectrum sensing [176] [177], adaptive beam control and multi-beam satellites [178] [179], predictive frequency allocations [180], and licensed spectrum management [33]. The latter includes database-assisted techniques [32] that enable controlled spectrum sharing with guaranteed QoS for the sharing parties. The basic principle of a spectrum database approach is that the secondary user of the spectrum is not allowed to access the band until it has successfully received information from the database that the channel it intends to operate on is free at the location of the device for the time period needed.

**Astronomy:** It is important to use spectrum efficiently to enable an interference-free environment and reliable delivery of wireless services to end users. Another crucial point is to allocate and use frequencies without endangering other space safety services including astronomy [181] [182]. The astronomical radio signals arriving on Earth are extremely weak compared to signals from communication systems and require large radio antennas (called radio telescopes) to detect them. Even a cellular phone on the moon would produce a signal on Earth that radio astronomers consider quite strong. Thus, cosmic radio sources are easily masked if this is not taken carefully into account. Regulations are made to set power and frequency limits in order to keep services operational. Technical studies on aggregate interference effects from new wireless services have to be done in order to support regulatory activities.

Radio astronomy is used to increase our knowledge of space, particularly deep space topics such as pulsars, black holes, radio galaxies, and cosmic microwave background radiation. Thus, it helps to answer questions such as how the universe and planets are born or trying to find extra-terrestrial intelligence [183]. In addition, radio astronomy provides useful information on the sun and solar activities, and thus we can learn about stars in general. Astronomy uses a wide range of frequency bands from a few kHz to tens of GHz, and there have been many interference incidents caused by ground-based broadcast systems, GNSS signals, and cellular phones. New allocations in any layer of the 6G system should not produce harmful interference in a co-channel or in adjacent bands. In addition, large constellations are challenging optical astronomy since satellites can reflect sunlight and appear as bright streaks in telescopes [184].

In Finland, the Sodankylä Geophysical Observatory has already been operating for more than 100 years. For example, they will be using the European Incoherent Scatter Scientific Association (EISCAT) 3D research infrastructure, consisting of thousands of phased array antenna elements operating in 233 MHz band [185]. It uses radar observations and the incoherent scatter technique to cover the near-Earth space environment for space weather forecasts and space debris detection.

**Lessons learned**: The development of new space technologies and the launch of an increasing number of satellites create a lot of potential for business and novel applications. However, there are many safety and security-related issues that need to be taken into account in the development to keep it sustainable and also have those services functional in the future. Avoidance of debris, cyber-native design, and minimization of environmental impact are examples of good practices to follow.

## VII. POTENTIAL RESEARCH TOPICS

In this section, we identify a few promising research directions. In addition to the described technical topics, and space safety topics presented in TABLE VIII, we foresee a plethora of new application areas that can be supported and enabled by multi-layer networks in the future. These include the so-called metaverse, i.e., a virtual environment that blends the physical and digital worlds, facilitated by the convergence between the Internet and extended reality (XR) technologies [187]. Another example is telemedicine [188] and advanced global Internet-based services including the least developed countries.

### A. MULTI-LAYER CONSTELLATION DESIGN

Traditional constellation designs have considered satellites at a single orbit, and updating the designs to multiple layers is not a simple task. There is a need to develop new design methodologies and simulation tools for flexible operations and capacity estimations across layers. Future designs might require the use of machine learning frameworks such as the reinforcement learning-based capacity estimation derived in [189]. Also, stochastic geometry analysis [108] [109] can provide tools for capacity optimization with a minimum number of satellites in different orbits. In addition to space layers, new designs are needed for the ground segment.

Thousands of SatCom and EO satellites will provide a tremendous amount of data in the future. To access the data, customers need to either build their own ground stations and antennas or lease them from ground station providers [28]. It would be useful and more sustainable to create a ground station network that can be shared among constellations. For example, in the Amazon AWS initiative [190], the data can be collected from the numerous satellites orbiting the Earth and stored in a central cloud. In such a case, interested customers will only need to access the cloud without the need to invest in their own infrastructure. Another recent example is the NorthBase in Finland providing secure ground station services to satellite operators [191]. The remaining challenges for the operation are where to locate ground stations to support real-time requirements, how to connect them to the national networks, and setting clear regulatory foundations for their use.

### B. EFFICIENT SPECTRUM USE IN 3D NETWORKS

Dynamic spectrum management needs to be updated to the 6G era. There are many topics to be addressed to make this successful. First, defining the most suitable frequency bands for systems and links. Second, developing spectrum sharing mechanisms to manage the complexity of a dynamic and mobile 3D network. Most probably, AI-based solutions are required [192]. When a massive number of devices are involved in 6G networks and require spectrum assignment, AI-enabled spectrum management is capable of intelligently supporting a massive number of connections and diverse services. It is good to note that the inclusion of non-terrestrial base stations (BSs) brings new challenges to network controller design because both the serving and interfering BSs can move at the same time in the 3D plane [15].

A concept for spectrum sharing in a multi-layer satellite system is seen in Figure 20. There are GSO and NGSO satellites operating in the same frequency bands. In order to coordinate spectrum use and enable predictable QoS for all users, information must be shared between the entities controlling the involved layers or towards a third-party entity controlling spectrum use. The system could use AI-based optimized spot beam allocation and resource management for aggressive frequency reuse and beamforming/precoding techniques [193] to cover the needs of mobile and fixed users. Thus, spot beam settings are dynamically adjusted to support capacity needs and avoid interference. Suitable database approaches and AI technologies are to be developed to realize the approach.

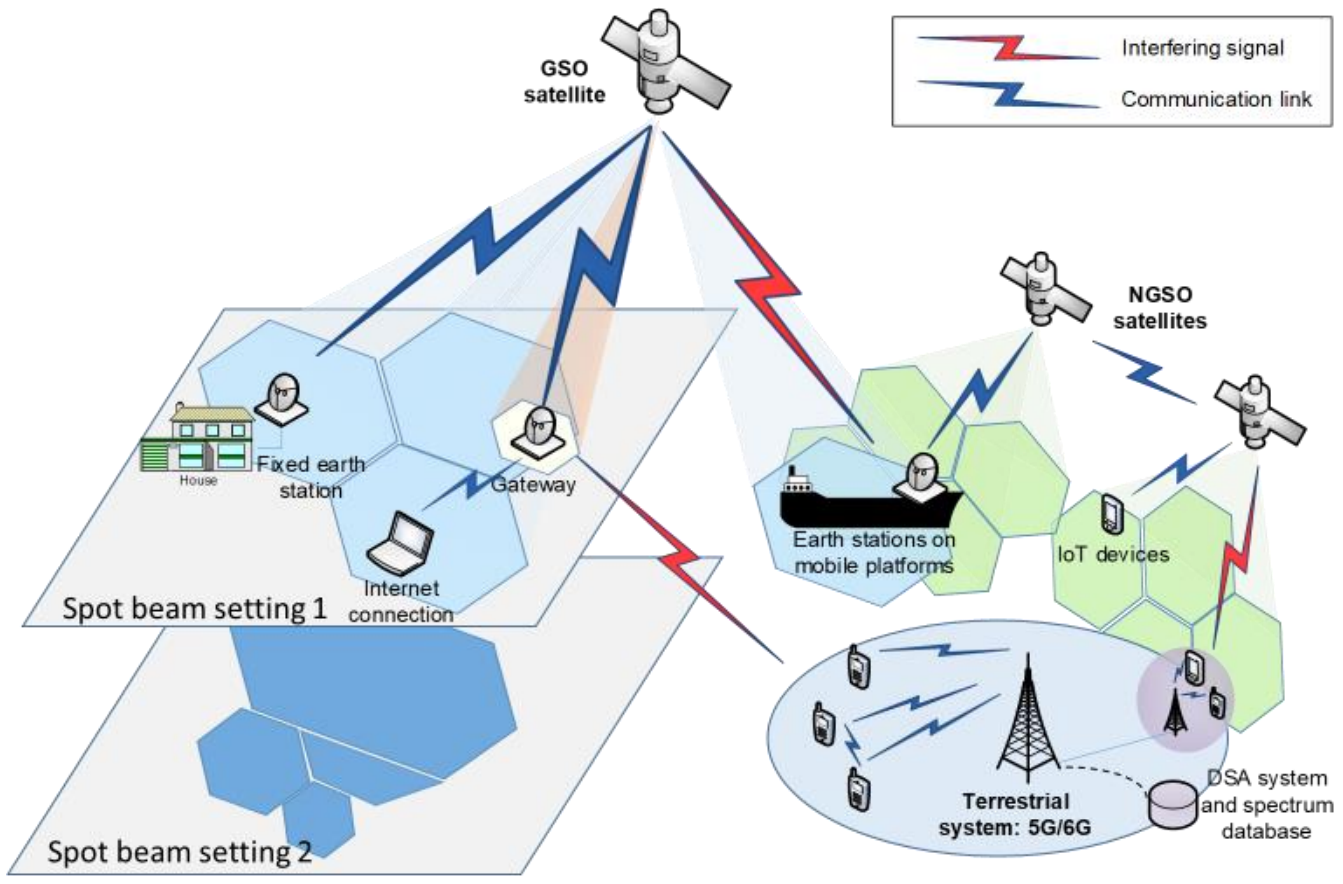

Figure 20. Multi-layer spectrum sharing with dynamic spot beam technology.

New types of antenna solutions are being developed including holographic radio, where electronically active surfaces are designed and utilized to receive, transmit, and reflect arbitrary waveforms [194] [195]. This can potentially lead to energy savings, especially in aerial and terrestrial layers. 5G and 6G SatCom with multibeam technology will enable even more advanced services [196] and possibilities for more efficient use of spectrum in time, frequency, and spatial domains.

### C. CYBER SECURITY IN THE QUANTUM ERA

In 6G communications, post-quantum cryptography will be essential to secure information [40] [197]. We can assume that in the near future an adversary may have access to quantum algorithms that will break the commonly used cryptosystems. Therefore, 6G systems should be quantum proof or at least prepared for a fast transition to post-quantum encryption.

Quantum key distribution (QKD) with an elegant practical solution has already been demonstrated by the Chinese Academy of Sciences and their Micius satellite in the LEO orbit [198]. The experiment used the satellite to establish a secure key between itself and a location in Xinglong, China, and another key between itself and Graz, Austria. The secured connection was used for a video conference.

European plans related to secure space-based connectivity include QKD transmission to securely connect ministries and other organizations across the continent. However, there are also challenges related to the use of optical connections, especially in northern Europe during the winter time due to cloudy weather. Thus, it is not yet clear how suitable the concept is for connecting Helsinki to Brussels, for example. Therefore, a careful analysis of the availability and reliability of the concept for local conditions is required before its wide-scale use.

### D. AI-ENABLED FLEXIBLE NETWORKING

6G will enable innovative services with different capacity and latency requirements, and AI technologies can be used to tailor the network and its functionalities on-the-fly to support those services in the best way [199]. Deep learning can be used to optimize 3D networks for configuration, routing, and computation [200]. Machine learning provides the means to identify and exploit repetitive patterns in the constellation geometry and minimize routing computations [201]. This could finally lead to dynamic network architecture and management with different optimization goals.

Satellite operators aim to reduce the costs of future satellite connectivity by using software-defined satellites that enable radio updates with the latest standard features, such as on-the-fly and 3D-layered architecture control and management based on extensions of software-defined network concepts that are already used for the terrestrial components [202]–[204]. The ability to change coverage areas, power and frequency allocations, and architecture on-demand would mean that a satellite can be manufactured first and tailored to the operator’s needs later. AI-empowered SDN technology and MEC paradigms enable anytime anywhere communications.

Dynamic network slicing provides the means to support a variety of services [131]. Already, a static network slice enables reserving resources ahead of time in a coarse-grained manner for end-to-end services instead of per session. Since static slicing consumes resources, dynamic slicing techniques will be needed in future multi-layer networks to quickly create, adapt, and manage slices according to the needs of users and applications while taking into account dynamically varying network architectures. These technologies will not only facilitate automated network management and increase network flexibility without human interventions, but also provide improved QoS for global multimedia services. This will also improve the longevity and sustainability of the satellites.

### E. DEEP SPACE OPERATIONS AND AUTONOMOUS SATELLITES

Autonomous satellites can use deep learning, expert systems, and intelligent agents to process spacecraft data (telemetry, payload) and make decisions autonomously during the mission. Artificial intelligence enables autonomous re-planning, detection of internal and external events, and reaction accordingly, ensuring fulfillment of mission objectives without the delays introduced by the decision-making loops on the ground [205]. However, an increase in autonomy creates challenges for space traffic management, as described in Section VI and thus, new approaches for collision avoidance and information sharing need to be developed.

TABLE IX. SUMMARY TABLE OF FUTURE RESEARCH DIRECTIONS.

| Technology topics | Challenges and research directions | References |
|---|---|---|
| Multi-layer constellation design | 3D network architecture will include satellites at different orbits. There will be small and large satellites simultaneously used, each having different capabilities. How to optimize the number of satellites in different orbits and their inter-connections? New types of simulation tools to support design and operations are needed. | [106],[114], [107], [189] |
| Spectrum use over horizontal and vertical dimensions | Dynamic spectrum management needs to be updated to the 6G era, taking into account the use of very high frequencies such as terahertz communications. What frequencies can be used for dynamic links and how to utilize AI technologies for flexible operations? Development should cover joint communication and sensing approaches and multi-purpose payloads. | [32], [133], [140], [192] |
| Cyber security in the quantum era | Quantum computing can break current cryptosystems and therefore fast transition for post-quantum encryption might be required in near future. In addition, secure links e.g. in governmental communications will use QKD technology. What is required to ensure the protection of space and ground infrastructure internationally and what will be the role of optical satellite communications? | [44], [142], [197], [198] |
| Flexible satellite architectures and networking | In contrast to traditional satellite systems, the future multi-layer systems are designed to be operated in a dynamic, flexible way. Machine learning and software-defined technology will enable automated operations and updates of operations on-the-fly. How to create sustainable business models for the “payload-as-a service” type of operations while simultaneously ensuring predictable service quality? | [200], [201], [204] |
| Space operations and people as multi-planetary species | Future space operations have to create improved means for space debris detection and removal while enabling reliable space traffic management and autonomous collision avoidance procedures. In addition to providing services to Earth we have to develop deep space communications networks using optical and RF technologies to support space missions across the solar system. | [205], [209] |
| Multi-purpose payloads for advanced space traffic management | Software-defined radios and networking will enable tighter integration of terrestrial and satellite networks and the development of multi-purpose payloads. How to harness small satellites and satellite constellations for real-time space situational awareness? | [203], [210] |

There are ambitious plans for humanity to see us as a multi-planetary species in the future. Countries such as the USA, China, and the UAE have their own plans for the settlement of Mars during the next century, see e.g. [206]. To make this a reality, one needs to also create a supporting infrastructure that includes food and water supply and production, mining, construction, and connectivity. Thus, studies aiming at adding even more layers to the network to also support inter-planetary connections are ongoing. A practical example is NASA’s Mars Cube One (MarCo) mission – relaying data from the lander to Earth with 6U CubeSats [207]. Another recent example is Nokia’s plan to create a cellular network on the moon [208]. The network will provide critical communication capabilities for applications such as vital command and control functions, remote control of lunar rovers, real-time navigation, and streaming of high definition video. The mentioned applications and connectivity capabilities are vital for a long-term human presence on the moon and on Mars. The key challenge related to deep space operations is crystallized in [209] as “Determining the most cost-effective combination of RF and optical assets for communicating with the postulated human Mars assets while still providing for the needs of all the other missions across the solar system.”

### *F. MULTI-PURPOSE PAYLOADS FOR ADVANCED SPACE TRAFFIC MANAGEMENT*

Traditional satellite payloads have been designed to fulfill the requirements of a single mission and different services have required separate satellites. The customer base for satellite services is increasing and very heterogeneous, and future needs for them are not well known. Thus, one needs to be able to adapt and update the satellites and satellite constellations over their lifetime to fulfill the requirements. It is possible to launch satellites with better functionalities as part of the constellation every time new needs arise. However, this is not the most sustainable way. The better option is to update the satellites themselves with multi-purpose payloads using software-defined technologies [203]. Then, the satellite can be updated over-the-air to fulfill new requirements.

However, there are still many open challenges to be tackled. How to use software-defined networking capabilities to integrate satellites tightly as part of 5G/6G networks [210]? Can very small satellites be equipped with advanced multi-purpose payloads to support both SatCom and EO or navigation needs? We envision that joint communication and sensing payloads can be used in the future to gather needed information about the space environment. This provides the means for advanced space traffic management since it is possible to harness satellite constellations to provide accurate up-to-date space situational awareness. Researchers need to identify the most suitable frequency bands, signal formats, and data-sharing mechanisms to enable reliable information.

We have summarized the research directions in Table IX,

complementing the space safety specific table presented in Section VI. Thus, there are definitely interesting times ahead and a lot of challenges for researchers and industries to tackle.

## VIII. CONCLUSIONS

The world is currently experiencing strong developments in satellite communications and the integration of 5G/6G technology into satellite systems. This is leading to a major turning point globally in the satellite-enabled service business. In this paper we have looked at the development of 6G networks and satellite megaconstellations. Unlike other related review papers, we address the importance of sustainability, highlighting the importance of space safety related aspects in the development. We provide a systematic classification of space safety topics related to SatCom development and define promising research directions. As a specific example, we investigate joint communication and sensing related to space debris development, showing how satellite constellations can also help in debris detection and management.

So, what is the role that Europe will play? The USA and China are rapidly advancing with their plans and constellations. After a slower start, Europe is now investing billions of euros in developing our own constellation. Multiple companies and start-ups are growing significantly in the SatCom field. Due to a very strong background and position in 3GPP standardization, it is assumed that Europe will play a very important role in the development of interoperable global space networks. There are ambitious plans for multi-layer networks, and multiple R&D projects are developing future-proof solutions for secure connectivity. Finally, ESA promotes the sustainable use of space for all services in 6G in a sense that supports SDGs via telecom, EO, and navigation. Trustworthy and reliable SatCom is seen as an essential part of sharing EO and navigation data.

## ACKNOWLEDGMENT

The authors would like to acknowledge the interviewed organizations for their contributions: European Space Agency; companies such as Airbus, Nokia, Ericsson, Reaktor Innovations, KNL Networks, Magister Solutions, Northbase, Huld, Telia, Harp Technologies, Missing-Link and nationally relevant organizations such as Erillisverkot, Defence Forces, Finnish Meteorological Institute, Sodankylä Geophysical Observatory, Traficom, Business Finland; and ministries including the Ministry of Interior and the Ministry of Finance. However, the views expressed in this paper are those of the authors.

## REFERENCES

[1] European Space Agency, "From 5G to 6G: Space connecting planet Earth for a sustainable future," White Paper, 2021.

[2] P. Haines, "Overview of satcom market evolution," presentation [restricted access], ESA JCB meeting, 9 Sept. 2020.

[3] Morgan Stanley, "New Space Economy," [Online]. Available: https://www.morganstanley.com/Themes/global-space-economy.

[4] C. Pomeroy, A. C.-Diaz, and D. Bielicki, "Fund me to the Moon: Crowdfunding and the New Space Economy," *Space Policy*, vol. 47, pp. 44-50, Feb. 2019.

[5] G. Denis et al., "From new space to big space: How commercial space dream is becoming a reality," *Acta Astronautica*, vol. 166, pp. 431–443, Jan. 2020.

[6] J. C. McDowell, "The edge of space: Revisiting the Karman line," *Acta Astronautica*, vol. 151, pp. 668–677, Oct. 2018.

[7] M. N. Sweeting, "Modern small satellites - Changing the economics of space," *Proc. IEEE*, vol. 106, pp. 343-361, Mar. 2018.

[8] N. Saeed, A. Elzanaty, H. Almorad, H. Dahrouj, T. Y. Al-Naffouri, and M.-S. Alouni, "CubeSat Communications: Recent advances and future challenges," *IEEE Commun. Surveys Tuts*, vol. 22, pp. 1839–1862, Third Quarter 2020.

[9] D. Roddy, *Satellite Communications*, 4th edition, McGraw-Hill, 2006.

[10] V. Ignatenko, P. Laurila, A. Radius, L. Lamentowski, O. Antropov and D. Muff, "ICEYE Microsatellite SAR Constellation Status Update: Evaluation of First Commercial Imaging Modes," in *Proc. IGARSS*, Sept.-Oct. 2020.

[11] Iceye SAR satellite constellation, [Online]. Available: https://www.iceye.com/sar-data/satellite-capabilities.

[12] S. Dang, O. Amin, B. Shihada, and M.-S. Alouini, "What should 6G be," *Nature Electronics*, vol. 3, pp. 20–29, Jan. 2020.

[13] M. Höyhtyä, M. Corici, M. Covaci, and M. Guta, "5G and beyond for New Space: Vision and research challenges," in *Proc. ICSSC*, Oct. 2019.

[14] J. Lia, Y. Shi, Z. M. Fadlullah and N. Kato, "Space-air-ground integrated networks: A survey," *IEEE Commun. Surveys Tuts.*, vol. 20, pp. 2714-2741, Fourthquarter 2018.

[15] E. C. Strinati et al., "6G in the sky: On-demand intelligence at the edge of 3D networks," *ETRI Journal*, vol. 42, pp. 643–657, Oct. 2020.

[16] M. Mozaffari, W. Saad, M. Bennis, Y.-H. Nam, and M. Debbah, "A tutorial on UAVs for wireless networks: Applications, challenges, and open problems," *IEEE Commun. Surveys Tuts*, vol. 21, pp. 2334–2360, Thirdquarter 2019.

[17] Y. Huo. et al., "Distributed and multi-layer UAV network for the next-generation wireless communication and power transfer," *IEEE Internet Things J.*, vol. 6, pp. 7103–7115, Aug. 2019.

[18] GSMA White Paper, "High altitude platform systems: Towers in the skies," June 2021. [Online]. Available: https://www.gsma.com/futurenetworks/wp-content/uploads/2021/06/GSMA-HAPS-Towers-in-the-skies-Whitepaper-2021.pdf.

[19] Alén Space, "Successful launch of the first satellite of the Sateliot constellation," news announcement, March 2021. [Online]. Available: https://alen.space/successful-launch-of-the-first-satellite-of-the-sateliot-constellation/.

[20] B. Soret, I. Leyva-Marga, S. Cioni, and P. Popovski, "5G satellite networks for Internet of Things: Offloading and backhauling," *Int. J. Satell. Commun. Netw.*, vol. 39, pp. 431–444, July/Aug. 2021.

[21] D. Palma and R. Birkeland, "Enabling the Internet of Arctic things with freely-drifting small-satellite swarms," *IEEE Access*, vol. 6, pp. 71435–71443, Nov. 2018.

[22] ITU Broadband Commission, "The state of broadband: Tackling digital inequalities, A decade of action," September 2020. [Online]. Available: https://www.broadbandcommission.org/publication/the-state-of-broadband-2020/.

[23] International Labour Organization, "Teleworking during the Covid-19 pandemic and beyond: A practical guide," [Online]. Available: https://www.ilo.org/wcmsp5/groups/public/---ed_protect/---protrav/---travail/documents/instructionalmaterial/wcms_751232.pdf.

[24] H. Saarnisaari, S. Dixit, M.-S. Alouini, A. Chaoub, M. Giordani, A. Kliks, M. Matinmikko-Blue, and N. Zhang (Eds.), "6G white paper on connectivity for remote areas," white paper, 6G Research Visions, No.

5, University of Oulu, 2020. [Online]. Available: http://urn.fi/urn:isbn:9789526226750
[25] M. Matinmikko-Blue et al., "White paper on 6G drivers and the UN SDGs," arXiv preprint arXiv:2004.14695, April 2020.
[26] European Space Agency, "ESA and the sustainable development goals." [Online]. Available: https://www.esa.int/Enabling_Support/Preparing_for_the_Future/Space_for_Earth/ESA_and_the_Sustainable_Development_Goals.
[27] M. Palmroth et al., "Toward sustainable use of space: Economic, technological, and legal perspectives," *Space Policy*, vol. 49, 12 pp, Aug. 2021.
[28] O. Kodheli et al., "Satellite communications in the New Space era: A survey and future challenges," *IEEE Commun. Surveys Tuts*, vol. 23, pp. 70–109, Firstquarter 2021.
[29] K. Liolis et al., "Use cases and scenarios of 5G integrated satellite-terrestrial networks for enhanced mobile broadband: The SaT5G approach," *Int. J. Satell. Commun. Netw*., vol. 37, pp. 91–112, Mar. 2019.
[30] F. Rinaldi et al., "Non-terrestrial networks in 5G & beyond: A survey," *IEEE Access*, vol. 8, pp. 165178–165200, Sep. 2020.
[31] X. Lin et al., "On the path to 6G: Embracing the next wave of low Earth orbit satellite access," [Online]. Available: https://arxiv.org/ftp/arxiv/papers/2104/2104.10533.pdf.
[32] M. Höyhtyä et al., "Database-assisted spectrum sharing in satellite communications: A survey," *IEEE Access*, vol. 5, pp. 25322–25341, Dec. 2017.
[33] R. H. Tehrani, S. Vahid, D. Triantafyllopolou, H. Lee, and K. Moessner, "Licensed Spectrum Sharing Schemes for Mobile Operators: A Survey and Outlook," *IEEE Commun. Surveys Tuts*, vol. 18, pp. 2591–2623, Fourth Quarter 2016.
[34] M. Giordani and M. Zorzi, "Non-terrestrial networks in the 6G Era: Challenges and opportunities," *IEEE Network*, vol. 35 pp. 244–251, March/april 2021.
[35] X. Fang et al., "5G embraces satellites for 6G ubiquitous IoT: Basic models for integrated satellite terrestrial networks," *IEEE Internet Things J.*, vol. 8, pp. 14399–14417, Sep. 2021.
[36] F. Tariq et al., "A speculative study on 6G," *IEEE Wireless Commun.*, vol 27, pp. 118–125, Aug. 2020.
[37] A. Yastrebova, R. Kirichek, Y. Koucheryavy, A. Borodin and A. Koucheryavy, "Future networks 2030: Architecture & requirements", *Proc. ICUMT*, pp. 1-8, Nov. 2018.
[38] T. Huang et al., "A survey on green 6G networks: Architecture and technologies," *IEEE Access*, vol. 7, pp. 175758–175768, Dec. 2019.
[39] IMT-2030 2021. IMT-2030 (6G) Promotion Group, "White paper on 6G vision and candidate technologies," June 2021.
[40] P. Raatikainen et al., "Holistic approach to 6G networks," *VTT White Paper*, Sept. 2021. [Online]. Available: https://info.vttresearch.com/download-beyond-5g.
[41] J. Radtke, C. Kebschull, and E. Stoll, "Interactions of the space debris environment with mega constellations—Using the example of the OneWeb constellation," *Acta Astronautica*, vol. 131, pp. 55–68, Feb. 2017.
[42] A. H. Sanchez, T. Soares, and A. Wolahan, "Reliability aspects of mega-constellation satellites and their impact on the space debris environment," in *Proc. RAMS*, Jan. 2017.
[43] A. C. Boley and M. Byers, "Satellite mega-constellations create risks in Low Earth Orbit, the atmosphere and on Earth," *Nature Sci. Rep.*, Vol. 11, May 2021.
[44] D. Housen-Couriel, "Cybersecurity threats to satellite communications: Towards a typology of state actor responses," *Acta Astronautica*, vol. 128, pp. 409–415, Nov./Dec. 2016.
[45] T. Breton, Speech at the workshop on space based secure connectivity project, [Online]. Available: https://ec.europa.eu/commission/commissioners/2019-2024/breton/announcements/workshop-space-based-secure-connectivity-project_en
[46] R. Radhakrishnan, W. W. Edmonson, F. Afghah, R. M. Rodriguez-Osorio, F. Pinto, and S. C. Burleigh, "Survey of inter-satellite communication for small satellite systems: Physical layer to network layer view," *IEEE Commun. Surveys Tuts*, vol. 18, pp. 2442–2473, May 2016.
[47] Ministry of Finance, Finland, "Sustainable growth programme for Finland," 2021. [Online]. Available: https://julkaisut.valtioneuvosto.fi/bitstream/handle/10024/163363/VN_2021_69.pdf?sequence=1&isAllowed=y.
[48] M. Höyhtyä and J. Martio, "Integrated satellite-terrestrial connectivity for autonomous ships: Survey and future research directions," *Remote Sens.*, vol. 12, 24 pp., Aug. 2020.
[49] 3GPP, "Feasibility study on maritime communication services over 3GPP system," TR 22.819 V1.0.0, Technical report, Jan. 2018.
[50] H. Lueschow and R. Pelaez, "Satellite communications for security and defence," in *Handbook of Space Security*, pp. 779–796, Springer, 2020.
[51] M. Heikkilä et al., "Field trial with tactical bubbles for mission critical communications," *Trans. Emerg. Telecommun. Tech.*, 2021.
[52] A. Anttonen and M. Höyhtyä, "Emerging 5G satellite-aided networks for mission-critical services: A survey and feasibility study," research report, VTT-R-01049-19, Oct. 2019. [Online]. Available: https://www.researchgate.net/publication/340173872_Emerging_5G_satellite-aided_networks_for_mission-critical_services_A_survey_and_feasibility_study.
[53] ITU/UNESCO Broadband Commission, "The state of broadband: People-centered approaches for universal broadband," Sept. 2021. [Online]. Available: https://www.itu.int/dms_pub/itu-s/opb/pol/S-POL-BROADBAND.23-2021-PDF-E.pdf
[54] A. Anttonen, M. Kiviranta, and M. Höyhtyä, "Space debris detection over intersatellite communication signals," *Acta Astronautica*, vol. 187, pp. 156–166, Oct. 2021.
[55] ESA - Reprogrammable satellite shipped to launch site. [Online]. Available: https://www.esa.int/Applications/Telecommunications_Integrated_Applications/Reprogrammable_satellite_shipped_to_launch_site.
[56] SpaceNews, "Lockheed Martin + USC To Build Smart Smallsats," [Online]. Available: https://news.satnews.com/2020/08/06/lockheed-martin-usc-to-build-smart-smallsats/.
[57] Space Safety Coalition, "Best practices for the sustainability of space operations," Sep. 2019. [Online]. Available: https://spacesafety.org/best-practices/.
[58] SpaceNews, "From space traffic awareness to space traffic management," [Online]. Available: https://spacenews.com/from-space-traffic-awareness-to-space-traffic-management/
[59] Space Safety Coalition website. Available: https://spacesafety.org/.
[60] World Economic Forum, "Space Sustainability Rating." [Online]. Available: https://www.weforum.org/projects/space-sustainability-rating.
[61] Ministry of Economic Affairs and Employment, Finland, "Act on Space Activities," 2018 . [Online]. Available: https://tem.fi/documents/1410877/3227301/Act+on+Space+Activities/a3f9c6c9-18fd-4504-8ea9-bff1986fff28/Act+on+Space+Activities.pdf?t=1517303831000.
[62] 3GPP, "Technical Specification Group Services and System Aspects; Guidelines for Extraterritorial 5G Systems," TR 22.926 V18.0.0, Technical Report, Dec. 2021.
[63] C. Stöcker, R. Bennett, F. Nex, M. Gerke and J. Zevenbergen, "Review of the current state of UAV regulations", *Remote Sens.*, vol. 9, no. 5, pp. 459, 2017.
[64] Ministry of Defence, "Finland's Cyber Security Strategy", [Online]. Available: https://www.defmin.fi/files/2378/Finland_s_Cyber_Security_Strategy.pdf , 2013.
[65] A. Al-Hourani, S. Kandeepan and S. Lardner, "Optimal LAP altitude for maximum coverage," *IEEE Lett. Wireless Commun.*, vol. 3, pp. 569–572, Dec. 2014.
[66] M. Kishk, A. Bader, and M.-S. Alouni, "Aerial base station deployment in 6G cellular networks using tethered drones: The

mobility and endurance tradeoff," *IEEE Veh. Technol.Mag.*, vol. 15, pp. 103-111, Dec. 2020.

[67] Li B., Fei Z., and Zhang Y., "UAV communications for 5G and beyond: Recent advances and future trends," *IEEE Internet Things J.*, vol. 6, pp. 2241-2263, Apr. 2019.

[68] A. W. Mast and J. P. Bruckmeyer, "Radios, payloads, & onboard processing made easy," in *Proc. IEEE Aerosp. Conf.*, Mar. 2017.

[69] N. Efthymiou, Y. F. Hu, R. E. Sheriff, and A. Properzi, "Inter-segment handover algorithm for an integrated terrestrial/satellite-UMTS environment," in *Proc. PIMRC*, pp. 993–998, Sep. 1998.

[70] E. Falletti, M. Mondin, F. Dovis, and D. Grace, "Integration of a HAP within a terrestrial UMTS network: Interference analysis and cell dimensioning," *Wireless Pers. Commun.*, vol. 24, pp. 291–325, Feb. 2003.

[71] B. Evans, M. Werner, E. Lutz, M. Bousquet, G. E. Corazza, G. Maral, and R. Rumeau, "Integration of satellite and terrestrial systems in future multimedia communications," *IEEE Wireless Commun*., vol. 12, pp. 72–80, Oct. 2005.

[72] E. Corbel, I. Buret, J.-D. Gayrard, G. E. Corazza, and A. Bolea-Alamanac, "Hybrid satellite & terrestrial mobile network for 4G: Candidate architecture and space segment dimensioning," in *Proc. ASMS*, pp. 162–166, Aug. 2008.

[73] J. Ylitalo et al., "Hybrid satellite systems: Extending terrestrial networks using satellites," in *Cooperative and Cognitive Satellite Systems*, edited by S. Chatzinotas, B. Ottersten, and R. De Gaudenzi, pp. 337–371, Academic Press, Elsevier, 2015.

[74] J. Praks et al., "Aalto-1, multi-payload CubeSat: Design, integration and launch," *Acta Astronautica*, vol. 187, pp. 370–383, Oct. 2021.

[75] Satnews, W-Cube smallsat, Available: https://news.satnews.com/2021/09/01/the-first-75-ghz-signals-are-sent-from-space-by-the-w-cube-smallsat/ .

[76] P.-D. Arapoglou et al., "DVB-S2X-enabled precoding for high throughput satellite systems," *Int. J. Satell. Commun. Netw*., vol. 34, pp. 439–455, May/June 2016.

[77] N. Mazzali et al., "Enhancing mobile services with DVB-S2X superframing," *Int. J. Satell. Commun. Netw*., vol. 36, pp. 503–527, Nov./Dec. 2018.

[78] K. Liolis, A. Franchi, and B. Evans, "Editorial for Wiley IJSCN's special issue "Satellite networks integration with 5G"," *Int. J. Satell. Commun. Netw.*, vol. 39, pp. 319–321, July/Aug. 2021.

[79] 3GPP, "3rd Generation Partnership Project; Technical Specification Group Services and System Aspects; Study on architecture aspects for using satellite access in 5G," TR 23.737 V17.2.0, Technical Report, Mar. 2021.

[80] 3GPP, "Technical Specification Group Radio Access Network; Study on New Radio (NR) to support non terrestrial networks," TR 38.811 V15.0.0, Technical Report, June 2018.

[81] 3GPP, "Technical Specification Group Radio Access Network; Solutions for NR to support Non-Terrestrial Networks (NTN)", TR38.821 V16.1.0, Technical Report, May 2021.

[82] 3GPP, "Technical Specification Group Services and System Aspects; Study on using satellite access in 5G", TR22.822 V16.0.0, Technical Report, June 2018.

[83] 3GPP, "Technical Specification Group Services and System Aspects; Study on management and orchestration aspects of integrated satellite components in a 5G network", TR28.808 V17.0.0, Technical Report, Mar. 2021.

[84] 3GPP, "Technical Specification Group Radio Access Network; Study on Narrow-Band Internet of Things (NB-IoT) / enhanced Machine Type Communication (eMTC) support for Non-Terrestrial Networks (NTN)," TR 36.763 V17.0.0, Technical Report, June 2021.

[85] 3GPP, "Technical Specification Group Core Network and Terminals; Study on PLMN selection for satellite access," TR 24.821 V17.0.0, Technical Report, Sept. 2021.

[86] European Space Agency. ESA's Annual Space Environment Report. Ref GEN-DB-LOG-00288-OPS-SD. [Online]. Available: https://www.sdo.esoc.esa.int/environment_report/Space_Environment_Report_latest.pdf. (2020).

[87] I. del Portillo, B. Cameron, and E. Crawley, "A technical comparison of three low earth orbit satellite constellation systems to provide global broadband,"*Acta Astronautica*, vol. 159, pp. 123–135, June 2019.

[88] C. Ravishankar, R. Gopal, N. BenAmmar, G. Zakaria and X. Huang, "Next-generation global satellite system with mega-constellations," *Int. J. Satell. Commun. Netw.*, vol. 39, pp. 6–28, Jan./Feb. 2021.

[89] A. Yastrebova, M. Höyhtyä, and M. Majanen "Mega-constellations as enabler of autonomous shipping," in *Proc. ICSSC*, Oct. 2019.

[90] A. Jones, "China is developing plans for a 13,000-satellite megaconstellation" Spacenews, April 2021. [Online]. Available: https://spacenews.com/china-is-developing-plans-for-a-13000-satellite-communications-megaconstellation/.

[91] J. Rainbow, "Inmarsat unveils multi-orbit Orchestra constellation," Spacenews, July 2021. [Online]. Available: https://spacenews.com/inmarsat-unveils-multi-orbit-orchestra-constellation/.

[92] Inmarsat Orchestra system, [Online]. Available: https://www.inmarsat.com/en/about/technology/orchestra.html

[93] SES O3B system, [Online]. Available: https://www.ses.com/our-coverage/o3b-mpower.

[94] Kepler Commmunications, [Online], Available: https://kepler.space/.

[95] European Commission, "Secure Space initiative," [Online]. Available: https://ec.europa.eu/defence-industry-space/major-space-breakthrough-secure-digital-connections-future-2021-01-12_en.

[96] Airbus, "European study on EUs space connectivity", press release https://www.airbus.com/newsroom/press-releases/en/2020/12/european-space-and-digital-players-to-study-build-of-eus-satellitebased-connectivity-system.html.

[97] European Commission, "Space: EU initiates a satellite-based connectivity system and boosts action on management of space traffic for a more digital and resilient Europe," press release, [Online], Available: https://ec.europa.eu/commission/presscorner/detail/en/ip_22_921.

[98] European space agency, 4S program. [Online]. Available: https://artes.esa.int/safety-and-security-4s.

[99] D. C. Beste, "Design of satellite constellations for optimal continuous coverage," *IEEE Trans. Aerosp. Electron. Syst*., vol. 14, pp. 466-473, May 1978.

[100] R. D. Lüders, "Satellite networks for continuous zonal coverage," *Am. Rocket Soc. J.*, vol. 31, pp. 179-184, Feb. 1961.

[101] J. G. Walker, "Satellite constellations," *J. Brit. Interpl. Soc.*, vol. 37, pp. 559–572, 1984.

[102] C. J. Wang, "Structural properties of a low Earth orbit satellite constellation the Walker delta network," in *Proc. MILCOM*, pp. 968-972 Oct. 1993.

[103] J. Jiang, S. Yan, and M. Peng, "Regional LEO satellite constellation design based on user requirements," in *Proc. ICCC*, Aug. 2018.

[104] T. Savitri, Y. Kim, S. Jo and H. Bang, "Satellite constellation orbit design optimization with combined genetic algorithm and semianalytical approach," *Int. J. Aerosp. Eng.*, vol. 2017, Article ID 1235692, 17 pages, 2017.

[105] S. Boumard et al., "Constellation and data link design for a nanosatellite Earth Observation system," Proc. ICSSC, Oct. 2022.

[106] H. Nishiyama, Y. Tada, N. Kato, N. Yoshimura, M. Toyoshima, and N. Kadowaki, "Toward optimized traffic distribution for efficient network capacity utilization in two-layered satellite networks," *IEEE Trans. Veh. Technol.*, vol. 62, pp. 1303-1313, Mar. 2013.

[107] J. Qi, Z. Li, and G. Liu, "Research on coverage and link of multi-layer satellite network based on STK," in *Proc. ChinaCom*, Aug. 2015.

[108] N. Okati, T. Riihonen, D. Korpi, I. Angervuori, and R. Wichman, "Downlink coverage and rate analysis of low Earth orbit satellite constellations using stochastic geometry," *IEEE Trans. Commun.*, vol. 68, pp. 5120-5134, Aug. 2020.

[109] A. Yastrebova et al., "Theoretical and simulation-based analysis of terrestrial interference to LEO satellite uplinks," in *Proc. GLOBECOM*, Dec. 2020.

[110] Satellite Network Simulator 3 (SNS3), [Online]. Available: https://www.sns3.org/content/home.php.

[111]J. Puttonen, L. Sormunen, H. Martikainen, S. Rantanen and J. Kurjenniemi, “A system simulator for 5G non-terrestrial network evaluations,” in *Proc. WoWMoM*, June 2021.

[112]H. Saarnisaari and C. de Lima, “Application of 5G new radio for satellite links with low peak-to-average power ratios,” *Int. J. Satell. Commun. Netw*, vol. 39, pp. 445–454, July/Aug. 2021.

[113]A. Yastrebova, A. Anttonen, M. Lasanen, M. Vehkaperä and M. Höyhtyä, “Interoperable simulation tools for satellite networks,” in *Proc. WoWMoM*, June 2021.

[114]Satellite Constellation Network Emulator, [Online]. Available: https://artes.esa.int/projects/realtime-satellite-network-emulator.

[115]5G Test Network Finland (5GTNF), [Online]. Available: https://5gtnf.fi/.

[116]M. A. Uusitalo et al., “Ultra-reliable and low-latency systems for port automation,” *IEEE Wireless Commun.*, vol. 28, pp. 114–120, Aug. 2021.

[117]H. Vuojala et al., “Spectrum access options for vertical network service providers in 5G,” *Telecomm. Policy*, vol. 44, May 2020.

[118]M. Asad Ullah, K. Mikhaylov and H. Alves, “Massive machine-type communication and satellite integration for remote areas,” *IEEE Wireless Commun.*, vol. 28, pp. 74–80, Aug. 2021.

[119]M. Asad Ullah et al., “Situational awareness for autonomous ships in the Arctic: mMTC direct-to-satellite connectivity,” *IEEE Commun. Mag.*,vol. 60, pp.32-38, June 2022.

[120]M. Höyhtyä, J. Huusko, M. Kiviranta, K. Solberg and J. Rokka, "Connectivity for Autonomous Ships: Architecture, use cases and research challenges", in *Proc. ICTC*, pp. 345-350, Oct. 2017.

[121]T. Ojanperä et al., “5G-enabled road safety and cybersecurity services for connected and automated vehicles,” in *Proc. VTC Spring*, May 2021.

[122]A. Yastrebova, T. Ojanperä, J. Mäkelä and M. Höyhtyä, “Hybrid connectivity for autonomous vehicles: Conceptual view & Initial results,” in Proc. *VTC Spring*, May 2021.

[123]M. Vehkaperä, M. Hoppari, J. Suomalainen, J. Roivainen, and S. Rantala, “Testbed for local area private network with satellite-terrestrial backhauling,” in *Proc. ICECCE*, June 2021.

[124]J. F. Hayes, “Adaptive feedback communications,” *IEEE Trans. Commun. Technol.*, vol. CT-16, pp. 29–34, Feb. 1968.

[125]S. Haykin, “Cognitive radio: Brain-empowered wireless communications,” *IEEE J. Sel. Areas Commun.*, vol. 25, pp. 201–220, February 2005.

[126]A. D. Panagopoulos, P.-D. M. Arapoglou, G. E. Chatzarakis, J. D. Kanellopoulos, and P. G. Cottis, “Coexistence of the broadcasting satellite service with fixed service systems in frequency bands above 10GHz,” *IEEE Trans. Broadcast.*, vol. 52, pp. 100-107, Mar. 2006.

[127]M. Höyhtyä, J. Kyröläinen, A. Hulkkonen, J. Ylitalo, and A. Roivainen, “Application of cognitive radio techniques to satellite communication,” in *Proc. DySPAN*, pp. 540–551, Oct. 2012.

[128]M. Matinmikko et al., “Cognitive radio trial environment: First live authorized shared access-based spectrum-sharing demonstration,” *IEEE Veh. Tech. Mag.*, vol. 8, pp. 30–37, Sept. 2013.

[129]M. M. Sohul, Y. Miao, T. Yang, and J. H. Reed, “Spectrum access system for the citizen broadband radio service,” *IEEE Commun. Mag.*, vol. 53, pp. 18–25, Jul. 2015.

[130]M. Palola et al., “Field trial of the 3.5 GHz citizens broadband radio service governed by a spectrum access system (SAS),” in Proc. DySPAN, Mar. 2017.

[131]M. Höyhtyä et al., “Critical communications over mobile operators’ networks: 5G use cases enabled by licensed spectrum sharing, network slicing, and QoS control,” *IEEE Access*, vol. 6, pp. 73572–73582, December 2018.

[132]M. Höyhtyä et al., “Licensed shared access field trial and a testbed for integrated satellite-terrestrial communications including research directions for 5G and beyond” *Int. J. Satell. Commun. Netw*, vol. 39, pp. 455–472, July/Aug. 2021.

[133]M. Matinmikko-Blue, S. Yrjölä and P. Ahokangas, “Spectrum management in the 6G era: The role of regulation and spectrum sharing,” in *Proc. 6G SUMMIT*, Mar. 2020.

[134]M. Höyhtyä, A. Mämmelä, A. Chiumento, S. Pollin, M. Forsell, and D. Cabric, “Database-assisted spectrum prediction in 5G networks and beyond: A review and future challenges,” *IEEE Circuits Syst. Mag.*, vol. 19, pp. 34–45, Third Quarter 2019.

[135]T. S. Rappaport et al., “Millimeter wave mobile communications for 5G cellular: It will work!,” *IEEE Access*, vol. 1, pp. 335–349, May 2013.

[136]J. G. Andrews *et al.*, “Modeling and analyzing millimeter wave cellular systems,” *IEEE Trans. Commun.* vol. 65, pp. 403–430, Jan. 2017.

[137]M. Giordani and M. Zorzi, “Satellite Communication at Millimeter Waves: a Key Enabler of the 6G Era,” in *proc. ICNC*, pp. 383-388, Feb. 2020.

[138]First W-band transmission from space, [Online], Available: https://artes.esa.int/news/first-wband-transmission-space

[139]A. Lamminen, J. Säily, J. Ala-Laurinaho, J. de Cos and V. Ermolov, “Patch antenna and antenna array on multilayer high-frequency PCB for D-Band,” *IEEE Open J. Antennas Propag.*, vol. 1, pp. 396–403, June 2020.

[140]J. Kokkoniemi, J. M. Jornet, V. Petrov, Y. Koucheryavy and M. Juntti, “Channel modeling and performance analysis of airplane-satellite terahertz band communications,” *IEEE Trans. Veh. Tech.*, vol. 70, pp. 2047–2061, Feb. 2021.

[141]Y. Rahmat-Samii, V. Manogar and J. M. Kovitz, “For satellites, think small, dream big – A review of recent antenna developments for CubeSats,” *IEEE Antennas Propag. Mag.*, vol. 59, pp. 22-30, Feb. 2017.

[142]H. Hauschildt, C. Elia, H. L. Moeller and J. M. P. Armengol, “HydRON: High throughput optical network,” in *Proc. SPIE Free-Space Laser Commun. XXXI*, Mar. 2019.

[143]C. Preyssl, “Safety risk assessment and management - The ESA approach,” *Reliab. Eng. Syst.*, vol. 49, pp. 303-309, 1995.

[144]G. Musgrave, A. Larsen and T. Sgobba (Eds.), *Safety Design for Space Systems*, Elsevier, 2009.

[145]UK Space Agency, press release [Online]. Available: https://www.gov.uk/government/news/g7-nations-commit-to-the-safe-and-sustainable-use-of-space

[146]ESA, Space Debris by the Numbers, [Online]. Available: https://www.esa.int/Safety_Security/Space_Debris/Space_debris_by_the_numbers

[147]T. J. Muelhaupt, M. E. Sorge, J. Morin and R. S. Wilson, “Space traffic management in the new space era,” *J. Space Saf. Eng.*, vol. 6, pp. 80–87, June 2019.

[148]D. Kessler and B. Cour-Palais, “Collision frequency of artificial satellites: The creation of a debris belt,” *J. Geophys. Res.* vol. 83, pp. 2637–2646, June 1978.

[149]A. Murtaza, S. Pirzada, T. Xu, and L. Jianwei, “Orbital debris threat for space sustainability and way forward,” *IEEE Access*, vol. 8, pp. 61000–61019, 2020.

[150]H. Klinkrad, *Space Debris: Models and Risk Analysis*, Springer, 2006.

[151]E. Ekici, I. Akyildiz, and M. Bender, “A distributed routing algorithm for datagram traffic in LEO satellite networks,” *IEEE/ACM Trans. Netw.*, vol. 9, pp 137–147, Apr. 2001.

[152]C. Mark and S. Kamath, “Review of active space debris removal methods,” *Space Policy*, vol. 47, pp. 194-206, Feb. 2019.

[153]Astroscale, “Active debris removal,” [Online]. Available: https://astroscale.com/services/active-debris-removal-adr/

[154]M. Ross, M. Mills and D. Toohey, “Potential climate impact of black carbon emitted by rockets,” *Geophys. Res. Lett.*, vol. 27, Dec. 2010.

[155]T. Pultarova, “The World’s first wooden satellite will launch this year,” https://www.space.com/first-wooden-satellite-will-launch-in-2021.

[156]Wisa WoodSat, [Online]. Available: https://www.wisaplywood.com/wisawoodsat/.

[157]Space Logistics, [Online]. Available: https://www.northropgrumman.com/space/space-logistics-services/.

[158]H. Koskinen et al. “Space weather effects catalogue,” ESA Space Weather Study (ESWS), 2001.

[159]R. Schwenn, “Space Weather: The Solar Perspective.” *Living Rev. Sol. Phys.* **3,** 2 (2006).

[160]G. W. Hein, “Status, perspectives and trends of satellite navigation,” *Satellite Navigation*, Article 22, 2020.

[161]A. Yastrebova, M. Höyhtyä, S. Boumard, E. Lohan, and A. Ometov, “Positioning in the Arctic Region: State-of-the art and future perspectives,” *IEEE Access*, vol. 9, pp. 53964–53978, Mar. 2021.

[162]Geomagnetic storm brings down Starlink satellites, [Online]. Available: https://spaceweather.com/archive.php?view=1&day=09&month=02&year=2022

[163]‘HERA mission’, https://www.esa.int/Our_Activities/Space_Safety/Hera/Hera.

[164]‘Double Asteroid Redirection Test (DART) mission’, [Online]. Available: https://www.nasa.gov/planetarydefense/dart.

[165]P. Wan and Y. Zhan, “A structured Solar System satellite relay constellation network topology design for Earth-Mars deep space communications,” *Int. J. Satell. Commun. Netw.*, vol. 37, pp. 292-313, May/June 2019.

[166]M. Kivelä, A. Arenas, M. Barthelemy, J. P. Gleeson, Y. Moreno and M. A. Porter, “Multilayer networks,” *J. Complex Netw.*, vol. 2, pp. 203-271, Sept. 2014.

[167]NASA Cybersecurity services, [Online]. Available: https://www.nasa.gov/centers/ivv/cybersecurity.html .

[168]European Space Security and Education Centre, [Online]. Available: https://www.esa.int/About_Us/Corporate_news/ESA_ESEC.

[169]C. Jiang, X. Wang, J. Wang, H.-H. Chen and Y. Ren, “Security in space information networks,” *IEEE Commun. Mag.*, vol. 53, pp. 82-88, Aug. 2015.

[170]I. Ahmad, J. Suomalainen, P. Porambage, A. Gurtov, J. Huuskoand M. Höyhtyä,“Security of Satellite-Terrestrial Communications: Challenges and Potential Solutions”.TechRxiv,11-Jul-2022,doi:10.36227/techrxiv.20254605.v1.

[171]I. Ahmad, S. Shahabuddin, T. Kumar, J. Okwuibe, A. Gurtov and M. Ylianttila, “Security for 5G and beyond”, *IEEE Commun. Surveys Tuts.*, vol. 21, no. 4, pp. 3682-3722, 4th Quart. 2019.

[172]J. Suomalainen, A. Juhola, S. Shahabuddin, A. Mämmelä and I. Ahmad, “Machine learning threatens 5G security,” *IEEE Access*, vol. 8, pp. 190822-190842, Oct. 2020.

[173]S. S. Saha, S. Rahman, M. U. Ahmed and S. K. Aditya, “Ensuring cybersecure telemetry and telecommand in small satellites: Recent trends and empirical propositions,” *IEEE Aerosp. Electr. Syst. Mag.*, vol. 34, pp. 34–49, Aug. 2019.

[174]M. Höyhtyä et al., “Spectrum occupancy measurements: Survey and use of interference maps,” *IEEE Commun. Surveys Tuts.*, vol. 18, pp. 2386–2414, Fourth Quarter 2016.

[175]P. H. Pathak, X. Feng, P. Hu and P. Mohapatra, "Visible light communication networking and sensing: A survey potential and challenges", *IEEE Commun. Surveys Tuts.*, vol. 17, no. 4, pp. 2047-2077, 4th Quart. 2015.

[176]M. Jia, X. Liu, X. Gu, and Q. Guo, “Joint cooperative spectrum sensing and channel selection optimization for satellite communication system based on cognitive radio,” *Int. J. Satell. Commun.*, vol. 35, pp. 139–150, Mar./Apr. 2017.

[177]F. Dimc, G. Baldini, and S. Kandeepan, “Experimental detection of mobile satellite transmissions with cyclostationary features,” *Int. J. Satell. Commun.*, vol. 33, pp. 163–183, Mar.-Apr. 2015.

[178]S. Maleki, S. Chatzinotas, B. Evans, K. Liolis, J. Grotz, A. Vanelli-Coralli, and N. Chuberre, “Cognitive spectrum utilization in Ka band multibeam satellite communications,” *IEEE Commun. Mag.*, vol. 53, pp. 24–29, Mar. 2015.

[179]S. Tani et al., “An adaptive beam control technique for Q band satellite to maximize diversity gain and mitigate interference to terrestrial networks,” *IEEE Trans. Emerg. Topics Comput.*, vol. 7, pp. 115–122, Jan./Feb. 2019.

[180]M. Höyhtyä, S. Pollin, and A. Mämmelä, “Improving the performance of cognitive radios through classification, learning, and predictive channel selection,” *Adv. Electron. Telecommun.*, vol. 2, pp. 28–38, December 2011.

[181]J. Hall et al., “Light pollution, radio interference and space debris: Threats and opportunities in the 2020s,” White paper, *Bull. Am. Astron. Soc.*, Vol. 51, Sept. 2019.

[182]A. Venkatesan, J. Lowenthal, P. Prem, and M. Vidaurri, “The impact of satellite constellations on space as an ancestral global commons,” *Nature Astron.*, vol. 4, pp. 1043–1048, Nov. 2020.

[183]J. Tarter, “The search for extraterrestrial intelligence (SETI),” *Annu. Rev. Astron. Astr.*, vol. 39, pp. 511–548, Sept. 2001.

[184]A. Witze, “How satellite ‘megaconstellations’ will photobomb astronomy images,” *Nature News*, Aug. 2020. https://doi.org/10.1038/d41586-020-02480-5.

[185]D. Kastinen, T. Tveito, J. Vierinen, and M. Granvik, “Radar observability of near-Earth objects using EISCAT 3D”, *Ann. Geophys*., vol. 38, pp. 861–879, Jul. 2020.

[186]D. J. Israel and H. Shaw, “Next-generation NASA Earth-orbiting relay satellites: Fusing optical and microwave communications,” in *Proc. IEEE Aerosp. Conf.*, Mar. 2018.

[187]L.-H. Lee et al., “All one needs to know about metaverse: A complete survey on technological singularity, virtual ecosystem, and research agenda,” https://arxiv.org/pdf/2110.05352.pdf.

[188]R. Bashshur, C. R. Doarn, J. M. Frenk, J. C. Kvedar, and J. O. Woolliscroft, “Telemedicine and the COVID-19 pandemic, lessons for the future,” *Telemedicine and e-Health*, vol. 26, pp. 571–573, May 2020.

[189]C. Jiang and X. Zhu, “Reinforcement Learning Based Capacity Management in Multi-Layer Satellite Networks,” *IEEE Trans. Wireless Commun.*, vol. 19, pp. 4685–4699, July 2020.

[190]AWS Ground Station: Easily Control Satellites and Ingest Data With Fully Managed Ground Station as a Service, [Online]. Available: https://aws.amazon.com/ground-station/.

[191]Finnish NorthBase groundstations, [Online]. Available: http://www.northbase.fi/.

[192]M. Jia, X. Zhang, J. Sun, X. Gu, and Q. Guo, “Intelligent resource management for satellite and terrestrial spectrum shared networking towards B5G,” *IEEE Wireless Commun.*, vol. 27, pp. 54–61, Feb. 2020.

[193]M. A. Vazquez, A. Perez-Neira, D. Christopoulos, S. Chatzinotas, B. Ottersten, P.-D. Arapoglou, A. Ginesi, and G. Tarocco, “Precoding in multibeam satellite communications: Present and future challenges,” *IEEE Wireless Commun.*, vol. 23, pp. 88–95, Dec. 2016.

[194]E. Björnson, L. Sanguinetti, H. Wymeersch, J. Hoydis, and T. L. Marzetta, “Massive MIMO is a reality— What is next? Five promising research directions for antenna arrays,” *Digit. Signal Process.*, vol. 94, pp. 3–20, Nov. 2019.

[195]H. Lu, Y. Zeng, S. Jin, and R. Zhang, “Aerial intelligent reflecting surface: Joint placement and passive beamforming design with 3D beam flattening,” *IEEE Trans. Wireless Commun.*, pp. 1–1, 2021.

[196]F. Rinaldi, H.-L. Määttänen, J. Torsner, S. Pizzi, S. Andreev, A. Iera, Y. Koucheryavy, and G. Araniti, “Broadcasting services over 5G NR enabled multi-beam non-terrestrial networks,” *IEEE Trans. Broadcast.*, vol. 67, pp. 33–45, Mar. 2021.

[197]D. J. Bernstein and T. Lange, “Post-quantum cryptography,” *Nature*, vol. 549, pp. 188–194, Sept. 2017.

[198]S.-K. Liao et al., “Satellite-relayed intercontinental quantum network,” *Physical Review Letters*, vol. 120, Jan. 2018.

[199]J. Hoydis, F. A. Aoudia, A. Valcarce and H. Viswanathan, “Toward a 6G AI-native air interface,” *IEEE Commun. Mag.*, vol. 59, pp. 76–81, May 2021.

[200]N. Kato, Z. M. Fadlullah, F. Tang, B. Mao, S. Tani, A. Okamura, and J. Liu, “Optimizing space-air-ground integrated networks by artificial intelligence,” *IEEE Wireless Commun.*, vol. 26, pp. 140–147, Aug. 2019.

[201]I. Leyva-Mayorga et al., "LEO small-satellite constellations for 5G and Beyond-5G communications," *IEEE Access*, vol. 8, pp. 184955–184964, Oct. 2020.

[202]D. Kreutz et al., "Software-defined networking: A comprehensive survey," *Proc. IEEE*, vol. 103, pp. 14-76, Jan. 2015.

[203]S. Xu, X.-W. Wang, and M. Huang, "Software-defined next-generation satellite networks: Architecture, challenges, and solutions," *IEEE Access*, vol. 6, pp. 4027-4041, Jan. 2018.

[204]Y. Bi et al., "Software defined space-terrestrial integrated networks: architecture, challenges, and solutions," *IEEE Network*, vol. 33, pp. 22–28, Jan./Feb. 2019.

[205]European Space Agency, "AIKO: Autonomous operations thanks to artificial intelligence," [Online]. Available: https://www.esa.int/Our_Activities/Telecommunications_Integrated_Applications/TTP2/AIKO_Autonomous_satellite_operations_thanks_to_Artificial_Intelligence.

[206]SpaceWatch. "UAE aims to establish settlement on mars by 2117," [Online] Available: https://spacewatch.global/2017/02/uae-aims-establish-human-settlement-mars-2117/.

[207]NASA, "Mars Cube One (MarCo): Mission overview," [Online]. Available: https://www.jpl.nasa.gov/cubesat/missions/marco.php.

[208]Nokia, Press release, [Online]. Available: https://www.nokia.com/about-us/news/releases/2020/10/19/nokia-selected-by-nasa-to-build-first-ever-cellular-network-on-the-moon/.

[209]D. S. Abraham et al., "Recommendations emerging from an analysis of NASA's deep space communications capacity," in *Space Operations: Inspiring Humankind's Future*, edited by H. Pasquier C. A. Cruzen, M. Schmidhuber, and Y. H. Lee, pp. 475–511, Springer, 2019.

[210]Kratos, "Software-defined satellites meet software-defined ground,", https://www.kratosdefense.com/constellations-podcast/articles/software-defined-satellites-meet-software-defined-ground, 2021.

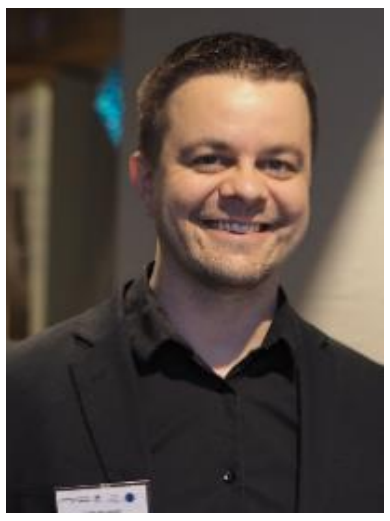

**Marko Höyhtyä** (Senior Member, IEEE) received the D.Sc. (Tech.) degree in telecommunication engineering from the University of Oulu, where he currently holds an associate professor position. He is an associate professor at the National Defence University as well. Since 2005, he has been with VTT Technical Research Centre of Finland Ltd. in various researcher and team leader positions. He is currently working as a research professor, focusing on satellite communications and situational awareness technologies. He was a Visiting Researcher at the Berkeley Wireless Research Center, CA, from 2007 to 2008, and a Visiting Research Fellow with the European Space Research and Technology Centre, the Netherlands, in 2019. His research interests include critical communications, autonomous systems, and resource management in terrestrial and satellite communication systems.

**Sandrine Boumard** received her Master of Science and Master of Advanced Studies degrees in Electronics, Systems, Radar and Radio-communication from the "Institut National des Sciences Appliquées," INSA, in Rennes, France, in 1998. She has since been working at VTT. She is currently a senior scientist on the autonomous systems connectivity team in the mobility and transport research area of the carbon neutral solutions business area at VTT. Her research focus is on physical layer algorithms, e.g., synchronization and OFDM systems, but her experience ranges from channel modeling to system-level analysis and simulation as well as VHDL modeling. She has been involved in several research and development projects, national, and international projects. She has co-authored several conference and journal papers as well as book chapters.

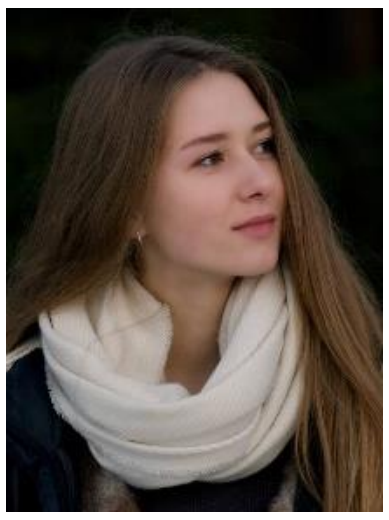

**Anastasia Yastrebova** received her M.Sc. in Information Technology from Tampere University (former Tampere University of Technology), Finland, in 2019. Currently, she is a Research Scientist at the Technical Research Centre of Finland, VTT Ltd in the Autonomous Systems Connectivity team. She is also working towards a Ph.D. degree in Communications Engineering at the University of Oulu, Finland. Her research focus is on heterogeneous wireless communication networks, including satellite-terrestrial systems and next-generation communication systems for remote monitoring, and autonomous operation of on-land and maritime systems.

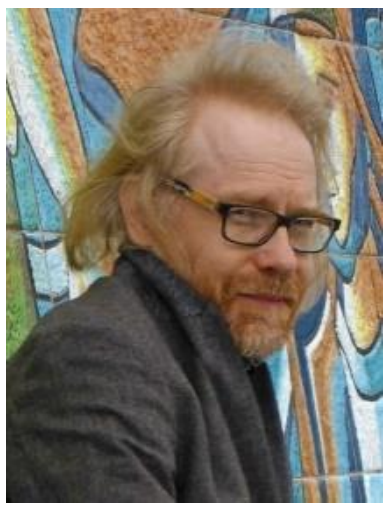

**Pertti Järvensivu** received the degree of M.Sc. (Tech.) in electrical engineering from the University of Oulu in 1992. From June 1988 to December 1999, he was with the Telecommunication Laboratory and the Mathematics Division at the University of Oulu in various teaching and research positions. He joined VTT in January 2000 and is currently a Principal Scientist in the Antennas and RF technologies team. He was the leader of the Wireless Access team from 2003 to 2011 and the leader of the Cognitive Radios and Networks team from 2011 to 2013. Since 2000, he has been a project manager in several publicly funded and private customer projects, for example, on mm-wave radio technologies. He is currently managing ESA Contract No. 4000135759/21/NL/MGu, "Automotive FMCW radar technology for space applications," in ESA's Technology Development Element (TDE) program. His interests include millimeter wave radio communication, satellite communication systems, and space applications of radio techniques.

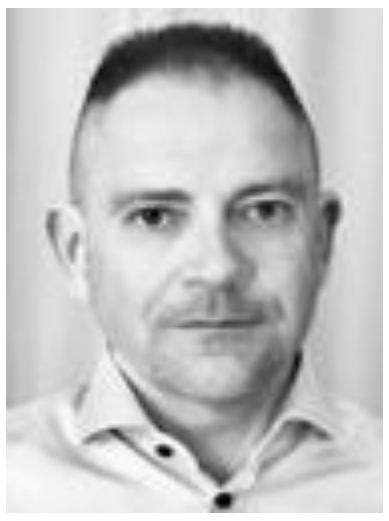

**Markku Kiviranta** (Senior Member, IEEE) received the degrees of the MSc (Tech.) and Lic.Sc. (Tech.) from the University of Oulu, Finland, in 1994 and 2000, respectively. In 2017, he received the D.Sc. (Tech.) degree from the School of Electrical Engineering, Aalto University, Finland. He started his professional career in 1993 at VTT, and since then, he has had a strong impact on ramping up and performing PHY layer algorithm and performance activities for microwave and satellite links and 3G/4G/5G-capable base stations and terminals as well as for high data rate and short link modems. In 2005, he spent a year at the Berkeley Wireless Research Centre in California. He has led the Digital Transceiver and Cognitive Radios and Networks teams at VTT for 10 years, after which he has worked as a principal scientist of radio systems.

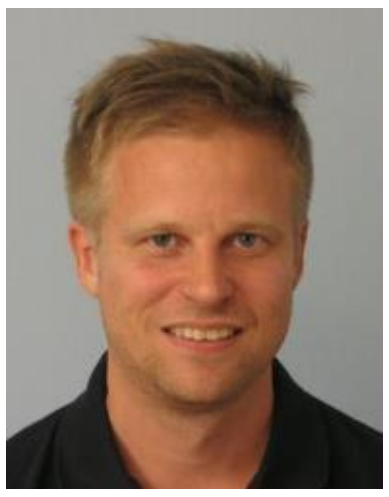

**Antti Anttonen** (Senior Member, IEEE) received the D.Sc. (Tech.) degree from the Department of Electrical and Information Engineering, University of Oulu, Finland, in 2011. He works as a Senior Scientist at VTT Technical Research Centre of Finland Ltd. He has visited Lucent Technologies, USA in 2000, the University of Hannover, Germany in 2008, and the University of Leuven, Belgium in 2014. He has participated in numerous national and European research projects related to wireless communications. His main interests include smart resource management, data traffic control, and the application of statistical theories for heterogeneous wireless networks.